\documentclass{aa}  

\usepackage{graphicx}
\usepackage{natbib,twoopt}
\usepackage[breaklinks=true]{hyperref} 
\usepackage{xcolor}
\usepackage{color}
\usepackage{amsmath}    
\usepackage{txfonts}
\usepackage{hyperref}
\usepackage{comment}
\usepackage{tablefootnote}
\usepackage{booktabs, multirow}
\usepackage[normalem]{ulem}

\usepackage{hyperref}
\hypersetup{colorlinks=true,linkcolor=[rgb]{1.,0.2,0.2},citecolor=[rgb]{0.1,0.4,1.},filecolor=[rgb]{0.7,0.2,0.2},urlcolor=[rgb]{0.7,0.2,0.2}}
\def\kms{\mbox{km~s$^{-1}$}}
\definecolor{darkspringgreen}{rgb}{0.09, 0.6, 0.27}

\begin{document} 

   \title{The Fornax Cluster VLT Spectroscopic Survey IV -- Cold kinematical substructures in the Fornax core from COSTA}
   \titlerunning{FVSS IV. Cold kinematical substructures in Fornax core from COSTA}


\author{N. R. Napolitano \inst{1,3} \thanks{E-mail: napolitano@mail.sysu.edu.cn},
M. Gatto \inst{2,3}, C. Spiniello \inst{3, 4}, M. Cantiello \inst{5}, M. Hilker \inst{6}, M. Arnaboldi\inst{6}, C. Tortora\inst{3}, A. Chaturvedi\inst{6}, R. D'Abrusco\inst{7}, R. Li\inst{1}, M. Paolillo\inst{2,8}, R. Peletier\inst{9}, T. Saifollahi\inst{9}, M. Spavone\inst{3}, A. Venhola\inst{10}, V. Pota
, M. Capaccioli\inst{2}, G. Longo\inst{2}}
\authorrunning{Napolitano, N.R., et al.}
\institute{School of Physics and Astronomy, Sun Yat-sen University Zhuhai Campus, 2 Daxue Road, Tangjia, Zhuhai, Guangdong 519082, P.R. China
\and
Department of Physics E. Pancini, University Federico II, Via Cinthia 6, 80126-I, Naples, Italy
\and
INAF, Osservatorio Astronomico di Capodimonte, Via Moiariello  16, 80131, Naples, Italy
\and 
Sub-Department of Astrophysics, Department of Physics, University of Oxford, Denys Wilkinson Building, Keble Road, Oxford OX1 3RH, UK
\and
INAF Osservatorio Astr. di Teramo, via Maggini, I-64100, Teramo, Italy
\and
European Southern Observatory, Karl-Schwarzschild-Str. 2, D-85748, Garching bei M\"unchen, Germany
\and
Center for Astrophysics $|$ Harvard \& Smithsonian, 60 Garden Street, 02138 Cambridge (MA)
\and
INFN, Sez. di Napoli, via Cintia, 80126, Napoli, Italy
\and
Kapteyn Astronomical Institute, University of Groningen, P.O. Box 72, 9700 AV Groningen, The Netherlands
\and
Astronomy Research Unit, University of Oulu, Pentti Kaiteran katu 1, 90014 Oulu, Finland
}

\date{}

 
\abstract
{Substructures in stellar haloes are a strong prediction of galaxy formation models in $\Lambda$CDM. Cold streams, e.g. from small satellite galaxies, are extremely difficult to detect and kinematically characterize. 
The COld STream finder Algorithm (COSTA), is a novel algorithm to find streams in the phase space of planetary nebulae (PNe) and globular cluster (GCs) populations.
COSTA isolates groups of ($N$) particles with small velocity dispersion (between 10 \kms~and $\sim120$ \kms), using an iterative ($n$) sigma-clipping over a defined number of ($k$) neighbor particles. 
} 
{We have applied COSTA to a catalog of PNe and GCs from the Fornax Cluster VLT Spectroscopic Survey (FVSS), within $\sim$200 kpc from the cluster core, to detect 
cold substructures and characterize their kinematics (mean velocity and velocity dispersion).} 
{We have selected more than 2000 PNe and GCs from the FVSS catalogs and we have adopted a series of optimized set-up of the COSTA parameters, based on Montecarlo simulations of the PN and GC populations, to search for
realistic stream candidates. We have found 13 cold substructures, with velocity dispersion ranging from $\sim 20$ to $\sim100$ kms$^{-1}$, which are likely associated either to large galaxies or to ultra-compact dwarf (UCD) galaxies in the Fornax core.} 
{These streams show a clear correlation of their luminosity with the internal velocity dispersion, and their surface brightness with size and distance from the cluster center that are compatible with dissipative processes producing them. However, we cannot exclude that some of these substructures have formed by violent relaxation of massive satellites finally merged into the central galaxy. Among these substructures we have: 1) a stream connecting NGC 1387 to the central galaxy, NGC 1399, previously reported in literature; 2) a new giant stream produced by the interaction of NGC 1382 with NGC 1380 and (possibly) NGC 1381;  3) a series of streams kinematically connected to nearby ultra compact dwarf galaxies (UCDs); 4) clumps of tracers with no clear kinematical association to close cluster members.}
{ 
We show evidence for a variety of cold substructure predicted in simulations. Most of the streams 
are kinematically connected to UCDs, supporting the scenario that they can be remnants of disrupted dwarf systems. However we also show the presence of long coherent sub-structures connecting cluster members and isolated clumps of tracers possibly left behind by their parent systems before these merged into the central galaxy. Unfortunately, the estimated low-surface brightness of these streams does not allow us to find their signatures in the current imaging data and deeper observations are needed to confirm them.  
}
   \keywords{Galaxies: clusters: intracluster medium -- Galaxies: interactions -- Galaxy: formation -- Galaxy: kinematics and dynamics
  }

   \maketitle
%



\section{Introduction}
\label{sec:introduction}
In the context of the hierarchical structure formation scenario, galaxy clusters have formed through highly non-linear growth \citep[e.g.][]{blumenthal-1986}. During their assembly a variety of physical processes, like tidal interactions, ram pressure stripping, and gas accretion take place, all contributing to shaping the luminous and dark matter component of galaxies in their cores \citep[e.g.][]{Genel-2014,Vogelsberger-2014,Schaye-2015}. 
These physical mechanisms are expected to leave signatures in the kinematics of the galaxy components, from their stellar haloes \citep{Duc-2011, amorisco-2019}, out to the the intracluster regions \citep{napolitano-2003,Murante-2007}, e.g. in the form of cold substructures which survive after stripping events due to the long dynamical times in the outskirts of galaxies.
Semi-analytic models combined with cosmological N-body simulations, as well as hydrodynamical simulations, have shown that the amount of substructures in stellar haloes, their stellar populations, and their dynamics, directly probe fundamental aspects of galaxy formation in $\Lambda$CDM. In particular, they can provide insight in the hierarchical assembly of massive galaxies, e.g. the mix between their ``in situ'' and ``accreted'' components
\citep{Cooper-2013,Cooper-2015a,Cooper-2015b,Pillepich-2015,pulsoni-2020,pulsoni-2021}.
One of the classical approaches to look for such substructures in galaxy halos is to use deep photometric observations \citep{Mihos-2005,Martinez-Delgado-2010,Montes-2019,Iodice-2017,Iodice-2019,Spavone-2017,Spavone-2018,Spavone-2020}. However, this approach is challenging due to the faint surface brightness of the tidal streams and remnants, typically below $\mu \sim 27$ mag/arcsec$^2$, which implies that only the brightest substructures are generally detected, while most of the accreted mass provided by the fainter tidal substructures, having generally surface brightness of the order of 30 mag/arcsec$^2$ or below in the $V-$band \citep{cooper-2010, Cooper-2015a}, remain hidden in the central galaxy background. 
To go beyond the purely photometric studies, in the last few years deep spectroscopy programs have provided kinematic information of the tidal debris around galaxies and allowed us look into the phase-space (projected positions and line-of-sight velocities) to search for the typical signatures expected for galaxy interactions \citep[e.g.][]{johnston-1998,romanowsky-2012,longobardi-2015a,hartke-2018}.
However, the use of kinematic tracers like planetary nebulae (PNe) or globular clusters (GCs), which are observable out to large distances from the galaxy centers \citep{durrell-2003,merrett-2003,shih&mendez-2010,cortesi-2011,richtler-2011}, 
is often a viable alternative to the standard kinematical measurements based on the 
integrated stellar light in the faint galaxy haloes
(PNe; \citealt{napolitano-2002,romanowsky-2003,douglas-2007,delorenzi-2009,coccato-2009,napolitano-2009,pota-2013,longobardi-2015a,hartke-2018, pulsoni-2018} -- GCs; \citealt{cote-2003,romanowsky-2009,schuberth-2010,woodley&harris-2011,richtler-2011,forbes-2011,romanowsky-2012,foster-2014,veljanoski-2016,longobardi-2018}). 
These discrete tracers allow us to probe the dynamics and the kinematics further out in galaxy clusters, where the potential of the cluster begins to dominate over that of individual galaxies (e.g. \citealt{spiniello-2018,pota-2018}).
First attempts to use GCs and PNe to find signatures of substructures from minor merger or accretion events have led to a fair number of claims based on the assumption that shells and tidal streams are located in chevron-like substructures in the position-velocity diagram (but see also \citealt{2013MNRAS.436.1322C}). This could be explained in terms of a near radial infall of objects with almost the same initial potential energy \citep{cote-2003,romanowsky-2009,mcneil-2010,shih&mendez-2010,schuberth-2010,woodley&harris-2011,romanowsky-2012,foster-2014,longobardi-2015a}. 
However, these patterns do not represent low dispersion streams made by a handful of particles and originated from dwarf galaxies in a recent encounter with a massive galaxy. 

Recently, we have developed an optimized stream finding algorithm, named COld STream finder Algorithm \citep[COSTA][G+20 hereafter]{gatto-2020}, that is able to spot tidal debris in the phase-space by detecting cold kinematics substructures moving in a warm/hot environment background of relaxed particles. COSTA is the first algorithm of this kind, although a similar concept was proposed to detect large substructures in the phase-space of galaxies in rich clusters (\citealt{1988AJ.....95..985D}).

In particular, COSTA relies on a deep friend-of-friend procedure that, through an iterative sigma clipping, detaches groups of neighbors particles with a cold kinematics (tens of \kms). 
This procedure allows to find small samples of tens of low velocity dispersion particles, as expected for low surface brightness streams originated by the disruption of dwarf galaxies orbiting in the diffuse stellar halos of giant galaxies or cluster dominant (cD) galaxies. 
{In G+20, COSTA}  
has been fully tested on hydro-simulations of galaxy encounters and Montecarlo simulations of realistic cluster-like velocity fields.
COSTA can efficiently work on samples of few hundreds to thousands discrete tracers in cluster cores, like the one we have collected in the multi-instrument observational program 
{\it Fornax Cluster VLT Spectroscopic Survey }(FVSS, see \S\ref{sec:datasets}).  Within this program we have {assembled catalogs }
of $\sim 1200$ GCs \citep[][P+18 hereafter]{pota-2018} and $\sim 1200$ PNe \citep[][S+18 hereafter]{spiniello-2018}, out to $\sim 200$kpc from the cluster center.
The {velocity dispersion ($\sigma$) radial } 
profiles from {both} PNe and GCs have shown a clear signature of an intracluster population with a sharp $\sigma$ increase, from $\sim$200 
km s$^{-1}$ to 
$\sim$350
km s$^{-1}$, at a radius of $\sim$10 arcmin (60 kpc) from the centre of NGC1399 \citep[see also][]{napolitano-2002}. In P+18 and S+18 we have discussed that this velocity dispersion raise is compatible with the scenario that both PN and GC populations at this distance start to feel the cluster potential, rather than the one of the central galaxy. This scenario has been recently confirmed with updated FVSS measurements of the PN and GC population \citep{Chaturvedi-2021}. An alternative scenario of a mix of populations producing an inflated profile as observed in Hydra cluster (\citealt{2018A&A...619A..70H}) cannot be excluded in Fornax, although does not seem to be supported by dynamical arguments (see S+18 and P+18).
In these regions, the dynamical timescales are long enough to preserve kinematical substructures for a longer time \citep{napolitano-2003,arnaboldi-2004,bullock-johnston-2005,arnaboldi-2012,coccato-2013,longobardi-2015a}, hence the FVSS represents the ideal dataset to use COSTA for searching stream candidates.
The Fornax cluster is the most massive galaxy overdensity after the Virgo cluster within 20 Mpc and, as such, it is an ideal target to search for cold substructures produced by the interaction of the large population of dwarf galaxies \citep[see e.g.][]{munoz-2015,venhola-2017,Ordenes-Briceno2018,Venhola-2019} with the cluster environment and investigate the assembly of the diffuse halo and the intracluster component in its core \citep{Arnaboldi-1996,napolitano-2002,Iodice-2016,Spavone-2020}. 
Despite its regular appearance, recent investigations have found that the assembly of Fornax is still ongoing, as shown by deep photometry with the ESO/VST, that found signatures of stellar and GC tidal streams (e.g. \citealt{Iodice-2016}, I+16 hereafter; \citealt{Iodice-2017}; \citealt{dabrusco-2016}, DA+16; \citealt{cantiello-2020}, C+20; see also \citealt{Chaturvedi-2021}).
Due to its proximity, Fornax gives us a unique opportunity to kinematically map the complexity of its core out to at least 200 kpc using discrete kinematical tracers (e.g. GCs and PNe) and finally connect the ``hot'' large scale kinematics down to the ``cold'' scale of dwarf satellite galaxies that are expected to produce most of the kinematical substructure in Fornax.

In this paper we apply COSTA to the full sample of PN and GC radial velocities. 
In \S\ref{sec:datasets} we present the available discrete tracer populations (PNe and GCs) and demonstrate that {their} 
velocity and spatial distributions are statistically consistent with belonging to the same parent population of tracers on cluster scale and, hence, they can be combined together to look for substructures. In \S\ref{sec:COSTA} we briefly introduce COSTA and describe the parameter set-up adopted for the stream finding. In \S\ref{sec:Results} we present the stream candidates and discuss their reliability and general properties. We also identify 
correlations among their observed parameters and discuss possible physical mechanisms behind them.
In \S\ref{sec:discussion} we discuss 
the results and give further details {on the}
stream candidates. In \S\ref{sec:conclusions} we finally draw some conclusions and perspectives.
In this paper, we will assume for the Fornax Cluster a distance modulus of $m-M=31.51$ \citep{Blakeslee-2009}.

\section{Datasets}
\label{sec:datasets}

\subsection{The Fornax Cluster VLT Spectroscopic Survey}
The dataset used in this work is based on the first catalog of GCs and PNe produced within the Fornax Cluster VLT Spectroscopic Survey \citep[FVSS;][]{pota-2018}. This program aims at collecting multi-instrument observations {of the Fornax Cluster } using the VLT@ESO telescopes. {In particular, in FVSS I, we have collected }
multi-object spectroscopy of $387$ new GC and Ultra-compact dwarf (UCD) systems with VIMOS at VLT \citep{pota-2018}. This sample, {added to archival data of } 
further $746$ GCs \citep[][]{bergond-2007,schuberth-2010} makes {a final catalog}
of $1183$ GCs/UCDs \citep{pota-2018}. 
{In FVSS II we obtained} 
dispersed imaging with FORS2 (VLT), with which we detected and measured radial velocities of $\sim 1268$ PNe \citep[][]{spiniello-2018}. Together with 184 PNe previously observed by \citet{mcneil-2010}, they
add up to a total of 1452 kinematical tracers. 
Recently in \citet[][FVSS III]{Chaturvedi-2021}, we have refined the data-reduction of the GC sample and pushed the measurement of GC radial velocities to lower signal-to-noise ratios (SNR$\sim5$). This increased the number of GC measured in the VIMOS data to 777 and the total GC velocities to 2341, after including other unpublished data.  
Hence, within FVSS, we have collected the most extended velocity field from GCs and PNe ever measured in a cluster (out to $\sim 300$kpc).

In this paper, {we use, for the first time, }
a combination of GCs and PNe to reveal cold substructures in the core of the Fornax cluster. 
The main reason {for}
combining radial velocities from different tracers is to maximize the number of test particles populating the streams, hence augmenting the probability to detect small substructures.  
The total size ($>2000$ test particles) and velocity errors of the two datasets ($\Delta_V \simeq$ 37 kms$^{-1}$ for GCs, P+18, and between 30 and 45 kms$^{-1}$ for PNe, S+18 -- we will adopt an average error of 37.5 kms$^{-1}$ in the following), are suitable for the detection and characterization of stream candidates, in hot environments like the stellar halo in cluster cores (see G+20).  
For the GC population we decided to use the P+18 catalog because this is based on a higher SNR sample that allows us to keep the statistical errors under a value (37 kms$^{-1}$) smaller than the one of the new catalog \citep[70--100 \kms, see][]{Chaturvedi-2021} and similar to the one of the PNe\footnote{We do not exclude from the PN+GC catalog the UCD candidates from P+18 because they are a minor fraction of the total sample ($<0.1\%$) and because if they are part of a stream, they can likely be the nuclear core of that stream (see \S\ref{sec:discussion}).}. 

There are 
differences between the GCs and PNe observing strategies which impact their spatial coverage of the Fornax core area. We refer the interested reader to the original papers for a detailed description of the observation techniques, data analysis and sample characterizations. Here below we report only the information that is of main interest for the analysis performed in this paper and the strategies to resolve the spatial disuniformity of the observed samples.
\begin{figure*}
	\centering
	\includegraphics[scale=0.4]{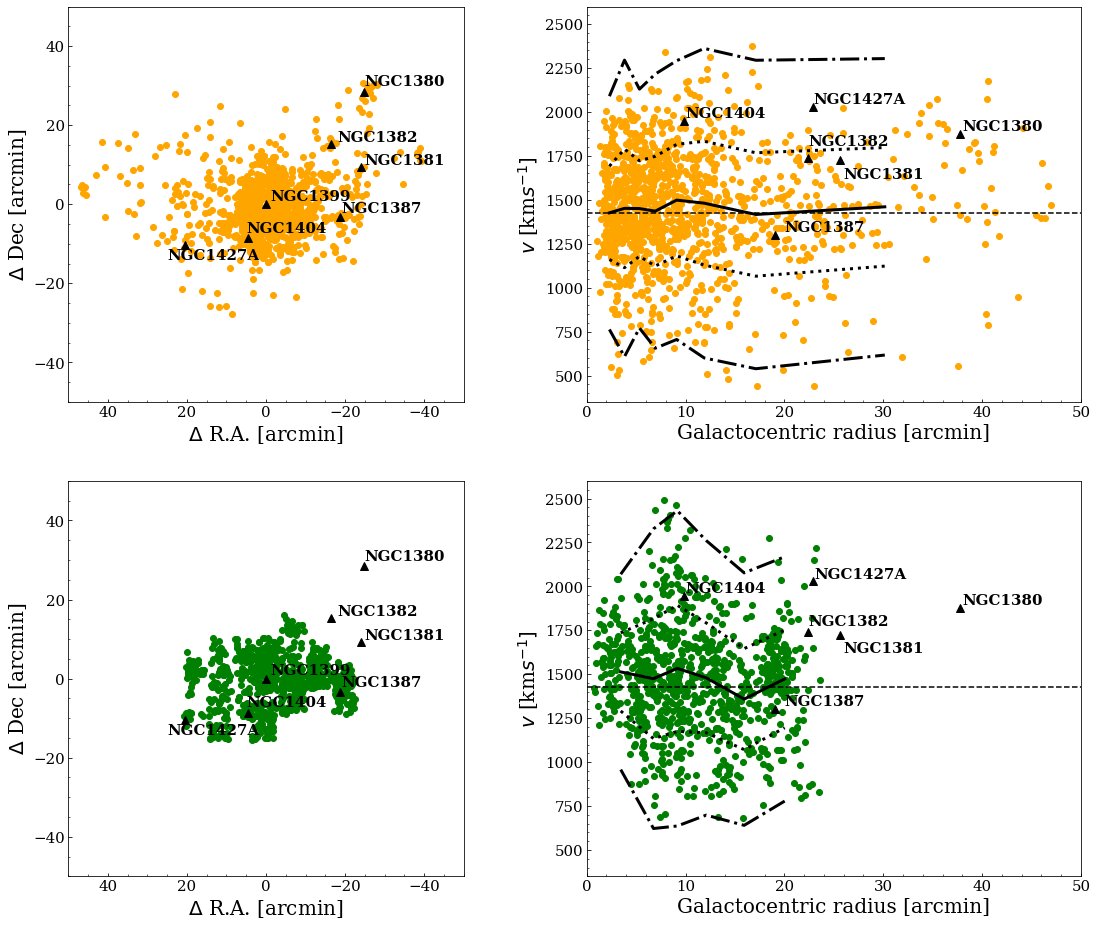}
	\caption{Relative distances, from NGC 1399, of the GCs (top left) and PNe (bottom left) and reduced phase-space (top right: GCs; bottom right: PNe). Overplotted on the phase-space there are the mean velocity of the particles (solid line), the $1\sigma$ (dotted lines) and the $2.5\sigma$ (dash-dotted lines). Black triangles are the main galaxies in the field. The dashed horizontal line in the phase space is the systemic velocity of NGC 1399 (1425 \kms).
	}
	\label{fig:gcs_pne}
\end{figure*}{}

\subsection{The GC sample} 
P+18 have presented a new catalog of GCs observed in an area of about 1 deg$^2$ around the NGC~1399, the bright galaxy in the center of the Fornax cluster, which corresponds to $\sim 175$ kpc galactocentric radius.
Observations consisted of a mosaic of 25 VIMOS pointings where about 2400 slits {were} allocated over a photometrically selected sample of GC candidates{. This sample was defined} using VST/OmegaCAM photometry in the de-reddened $g$ and $i$ bands from the Fornax Deep Survey(FDS) \citep{dabrusco-2016,Iodice-2016} and preliminary
VISTA/VIRCAM photometry in the K\textup{s} band from the Next Generation Fornax Survey \citep[NGFS; see][]{munoz-2015}.
The observed sample had $i-$mag in the range $17.0 \leq i \leq 23.0$ mag{. This} restriction {has} been used in order to avoid contamination by foreground stars at bright magnitudes and too low signal-to-noise spectra at faint magnitudes. 

Spectra have been analysed with \textit{iraf/fxcor} and the derived radial velocities (or GC redshift) showed typical errors of the order of 37 km s$^{-1}$. 
The new catalog has been combined with the literature catalogs from \citet{schuberth-2010}, {including} sources within 18 arcmin from NGC 1399, and \citet{bergond-2007} {covering} a strip of about 1.5 degree in right ascension and 0.5 degree in declination. 

The final sample includes 1183 GCs in total, with a systemic velocity of 1452 $\pm$ 9 km s$^{-1}$ (i.e. fully consistent with the one of NGC~1399).

Fig.~\ref{fig:gcs_pne} shows the position in RA and DEC of the objects and a reduced phase-space made by the radial velocity of GCs (top panels) vs. cluster-centric radius (i.e. the distance from the center of NGC~1399). 
In the reduced phase-space we report ({as black lines}) the mean velocities and the $1\sigma$ and $2.5\sigma$ contours of the velocities vs. radius. These were calculated dividing the sample in distance bins, such that in each bin the number of GCs is about the same ($\sim 150$).

\subsection{The PN sample}
S+18 have assembled a kinematic catalog of PNe out to 200 kpc in the Fornax cluster core, using a counter-dispersed slitless spectroscopic technique \citep[CDI,][]{Douglas-1999}. They obtained the final PNe dataset observing 20 new pointings (for a total of 5 hours exposure time) with FORS2,  
and also supplemented these new data with 180 central PNe velocities presented in \citet{mcneil-2010}. 
The covered total final area is $\sim 50\arcmin \times 33\arcmin$, centered around $\alpha =$3:37:51.8 and $\delta =-$35:26:13.6. 

The CDI technique has been shown to be very successful in PNe analysis since it allows both the detection of PNe and the measurement of their Doppler shift within a single observation. CDI uses two counter dispersed frames with the position angle of the field rotated by 180 deg and with an [OIII] filter (specifically the [OIII]/3000, 51\AA\ wide) to select the light. In this way all the Oxygen emission-line objects (among which PNe) appear as point-like sources, while sources emitting continuum (e.g. stars and background galaxies) show up as strikes or star-trails. PN candidates will thus appear in the two images (at the two rotated angles) at the same y-position, but shifted of a $\delta x$ which is proportional to their line-of-sight velocity.

These line-of-sight velocity measurements were then calibrated using the PNe in common with \citet{mcneil-2010}, which were calibrated against the measurements of \citet{Arnaboldi-1994} obtained with NTT multi-object-spectroscopy. 

The final sample of PNe with measured velocities comprises 1452 objects, it has a mean velocity of $1433$ kms$^{-1}$ with a standard deviation of $312$ kms$^{-1}$, calculated after applying heliocentric correction to each field separately. 
However, for the purpose of this paper, which aims at detecting streams possibly distributed over large areas, we have excluded the tracers strictly around NGC~1379 (i.e. 150 PNe in Field 1 from Fig.~1 of S+18), because this is isolated and not extended enough to probe the intracluster region. Furthermore, according to \citet{gatto-2020}, the number of tracers might be too small to detect cold substructures, if any.

In Fig.~\ref{fig:gcs_pne} (bottom panels) we report the position in RA and DEC of the objects and the reduced phase-space of PNe as done for GCs (upper panels).



\subsection{On the spatial homogeneity of the PN and GC samples}
\begin{figure}
    \centering
    \includegraphics[width=9cm]{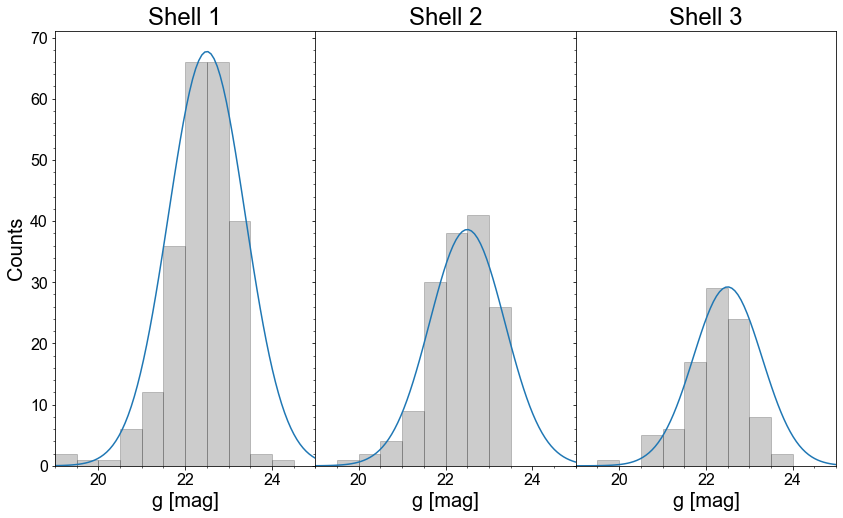}
    \caption{GCLF evaluated in the three different radial bins, along with their fitted Gaussian function whose parameters are described in the text.}
    \label{fig:gclf}
\end{figure}

Looking at the 2D distribution of GCs in Fig.~\ref{fig:gcs_pne} (top left), {one can see that} this is rather spherically uniform, with {the} number density of objects increasing towards the center of the cluster.
This homogeneity come from a fair uniformity in the GC selection for the slit allocation of the VLT observations, both in spatial and luminosity distribution. Spatially-wise, the VLT allocation was meant to collect a complementary dataset to previous existing observations, which allowed us to uniform the overall radial coverage of the spectroscopic sample finally collected by FVSS. Luminosity-wise, FVSS strategy aimed at emulating the depth of previous dataset, in order to minimize the luminosity function in-homogeneity among the different datasets. In particular, we used the \citet{schuberth-2010} as reference sample, because the most copious one. 
The net result is that, looking at  the globular cluster luminosity function (GCLF) in three radial shells (see \S\ref{sec:gc_vs_pn} and Table \ref{tab:K-S test} for their definition) from the cluster center in Fig. \ref{fig:gclf}, we see that the GC sample become incomplete at a magnitude $mag_{\rm g}\sim23.4$ in all bins.

\begin{figure*}
    \centering
    \includegraphics[width=15.5cm]{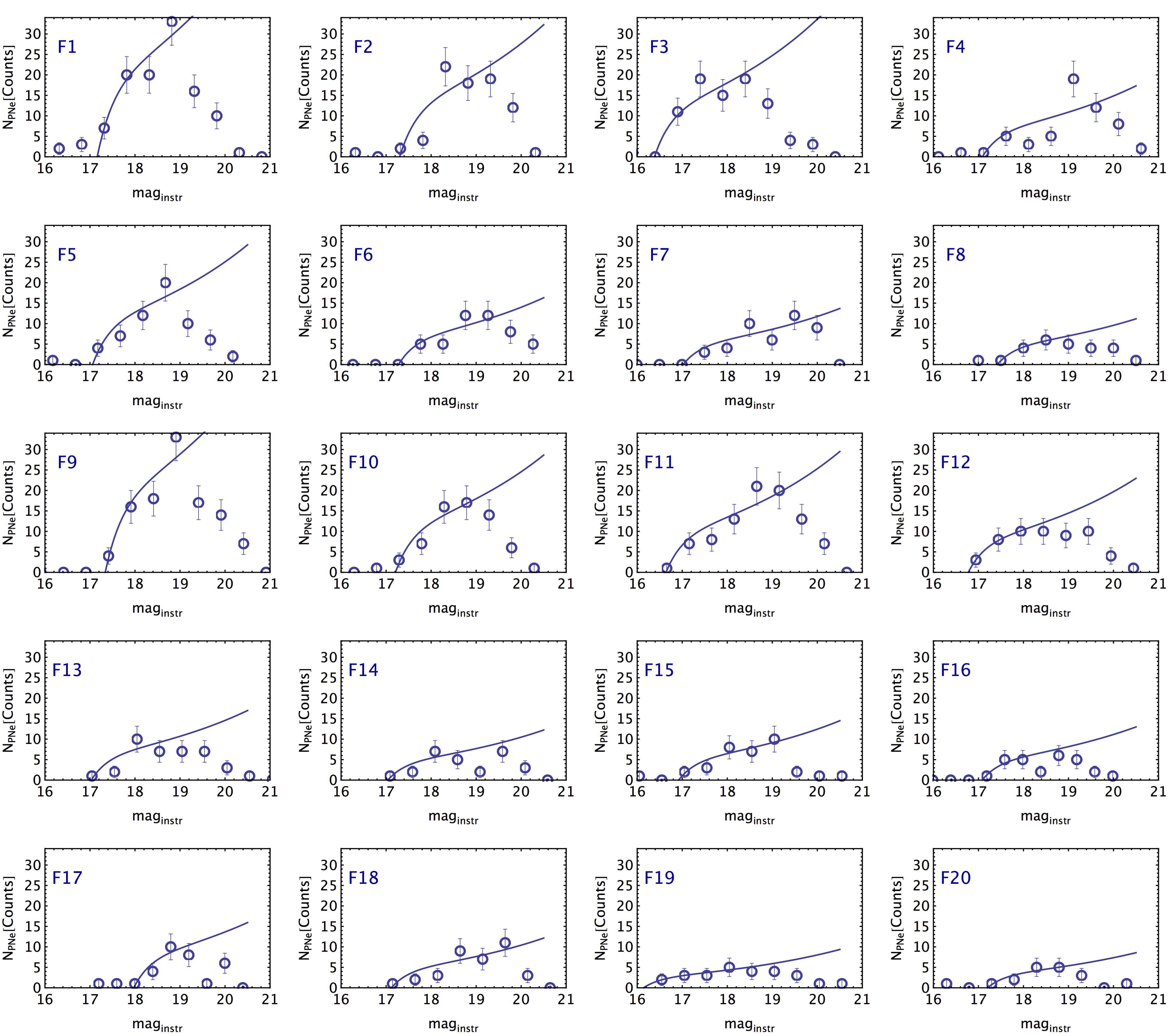}
    \caption{Planetary nebulae luminosity function in all the S+18 fields. The best fit of Eq.~\ref{eq:pnlf} is shown as continuous line. We have excluded from the fit the bins with one single ultra-luminous object, generally brighter than $\sim17$mag. See discussion in the text.
    }
    \label{fig:pnlf}
\end{figure*}
To quantify this, we have first fitted the bright half of the GCLF with a Gaussian distribution \citep[][and references therein]{brodie-2006}: we have obtained a mean equal to $\langle mag_{\rm g}\rangle =22.5$ for the all bins, a variance $\sigma(mag_{\rm g})=0.88,~ 0.87, ~0.79$ and a normalization factor, $A=300,~ 168, ~114$ for the three bins respectively. Then, we have derived the magnitude where the observed interpolated histogram deviates from the best-fit by 50\% of the counts ($mag_{\rm 50}$ in short) and found these to be $mag_{\rm 50}\sim23.5,~23.6,~23.2$, with a mean of 24.4 and scatter of 0.2. Hence, only the most external radial bin shows a slightly brighter limiting magnitude but reasonably within the tentative variance of these measurements.
Due to the high-SNR selection in P+18, there are only few GCs in the magnitude bin fainter than $mag_{\rm g}\sim23.4$, and only in the more external bin. 
Due to the overall regularities discussed above, we have decided to keep the full GC sample for the analysis.

On the other hand, PNe have a smaller spatial extension, i.e. $\sim$25 arcmin, versus $\sim$ 50 arcmin of GCs (see Fig. \ref{fig:gcs_pne}, bottom left), but more importantly they have, overall, a non homogeneous distribution. 
This is due to the tiling of the FORS2 observations, which produces a patchy sky coverage around the cluster center (see Fig.1 in S+18). 
Moreover, since the fields have been observed {under}  different observational conditions (seeing), and {sometimes with different final integration times,} each of them has a different intrinsic depth and consequently also a different limiting magnitude for the PNe detection (e.g. the luminosity at which almost half of the real PNe are detected). 
Hence, the number of identified PNe in each pointing varies not only because of the intrinsic {local} differences of the density of PNe but also due to the inability to find them down to the same magnitude limit. 

This aspect {was not relevant for S+18}, where they were not interested in using the spatial information of PNe for their analysis. 
In our study, instead, the spatial incompleteness can affect the stream detection because artificial overdensities, generated by a different observation depth, can give a higher chance to mimic a stream.
To correct for differences in the field-to-field depth, we need to select PNe which are within the same limiting magnitude (assuming all PNe at the same distance), taken as the mean magnitude where half of the PNe are missed, with respect to their intrinsic luminosity function, as done for GCs.  We first build 
the Planetary Nebula luminosity function (PNLF) in each pointing and best-fit a standard empirical formula to interpolate it. 
Then, assuming that the sample is complete in the brightest bins, we compute the magnitude where the fraction of observed PNe is 50\% of the expected number of PNe from the best-fitted PNLF (see below).

We have derived the photometry of the PNe detected in the FORS2 data where {the} [OIII] filter 
was used to isolate the [OIII] emission from PNe. This gives the so called [OIII] magnitude, which is generally used to construct the PNLF \citep{ciardullo-1998}.
{In particular, we use the software SExtractor \citep{bertin&arnouts-1996} and retrieve the output parameter $MAG_{\rm AUTO}$ which has often been found to provide the best estimate for the PNe [OIII] magnitudes \citep[see also][for a discussion]{arnaboldi-2003}.}

As we are not interested in absolute photometry, but only on the relative photometry of the observed PNe, we did not calibrate the PN fluxes, but arbitrarily assumed $m_{\rm zero}=$ 25 as zeropoint. Hence, the numerical values of $MAG_{\rm AUTO}$ correspond to instrumental magnitude{s}, mag$_{\rm instr}$.

SExtractor was able to measure fluxes of 99\% of all sources detected by eye in S+18, hence allowing us to derive the PNLF in all the S+18 fields, as shown in Fig. \ref{fig:pnlf}. 
These measured PNLFs 
have been best fitted using the fitting formula of \citet{aguerri-2005}: 
\begin{equation}
   N(m) = ce^{0.307m}[1 - e^{3(m^*-m)}]
   \label{eq:pnlf}
\end{equation}
where $c$ is a positive normalization constant 
and $m^*$ represents the apparent magnitude of the bright cutoff. 
The best fits are also shown in Fig.~\ref{fig:pnlf} as continuous line in each field. Note that in most of the fields (i.e. F1, F2, F4, F5, F6, F10, F15, F16, F17, F20) there are some ultra-luminous [OIII] emitters. These are typically found in survey of intracluster PNe and have been associated to either PNe located in the closer side of the cluster (e.g. \citealt{aguerri-2005}) or to other kind of contaminants like [OII] emitting galaxies at $z\sim0.35$ or Lyman$\alpha$ galaxies at $z\sim3.14$ (e.g. \citealt{castro-rodriguez2003}). As discussed in S+18, they have minimized the presence of the high-$z$ outliers, however we cannot exclude some residual contamination. We have dropped these ultra-luminous ``outliers'' in the PNLF fitting procedure by excluding all bright bins with only one count. However, we did not exclude these ultra-luminous objects from the PN sample because 1) they are very few and 2) we cannot securely classify them as non-cluster members. 
Finally, the limiting magnitude has been obtained by comparing a simple interpolating function of the data with respect to the fitted PNLF to find at which magnitude the interpolated counts fall at 50\% from the fitted PNLF ($mag_{\rm 50}$ as for the GCs).
From the distribution of the limiting magnitude from Fig.~\ref{fig:pnlf} 
we have found an average $mag_{\rm 50}=19.3\pm0.1$: hence all PNe with measured $MAG_{\rm AUTO}>19.3$ have been excluded from the final sample, which finally contains 887 PNe. 

\begin{figure*}
	\centering
	\includegraphics[scale = 0.4]{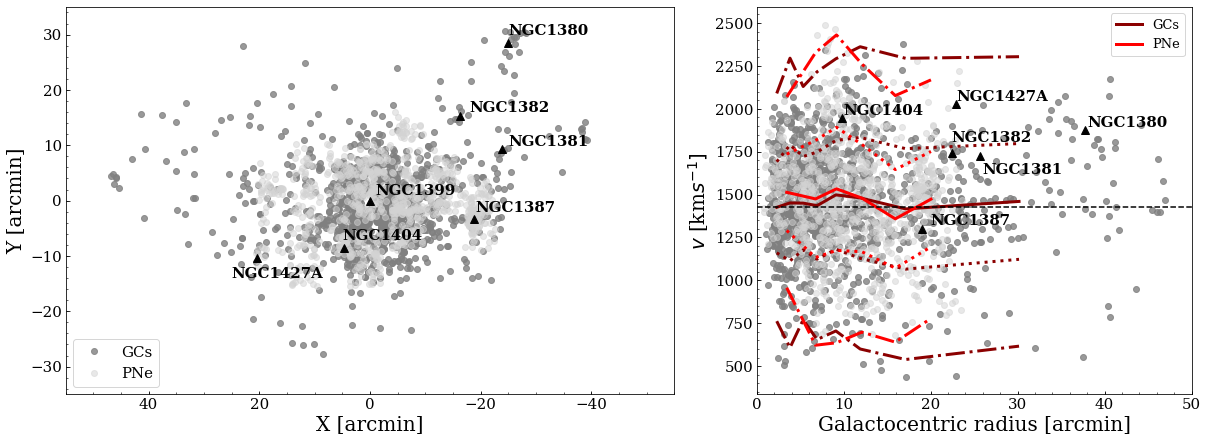}
	\caption{\textit{Left:} image of the whole sample of GCs (dark gray) and 887 bright PNe (light gray) in RA vs DEC {with the galaxies overplotted as black triangles.} \textit{Right}: GCs (dark gray) and PNe (light gray) in the phase-space. The lines correspond to the mean velocity of the particles (solid line, dark red for the GCs and light red for the PNe), to the velocity dispersion at $1\sigma$ (dotted line) and to the velocity dispersion at $2.5\sigma$ (point-dotted line). 
	}
	\label{fig:sample}
\end{figure*}

\subsection{The combined GC+PN sample}
\label{sec:gc+pn}
In this section we discuss the phase space properties of the GC and PN populations, to show that they can be combined to increase the statistics of the stream tracers. 
If we assume that PNe and GCs both respond in the same way to the encounters and they have initially no statistically significant differences in their kinematics, they will keep sharing the same phase space properties when captured in the streams. 
This is a reasonable assumption if their kinematics is (almost) indistinguishable in their parent galaxy. Generally, this is not true, and in fact the two families (GCs and PNe) are often considered dynamically disconnected, with PNe and red, metal-rich GCs sharing more often spatial and velocity distributions (e.g. \citealt{napolitano-2014}). However, as discussed in P+18 and S+18, there are clear similarities between the GC and PN populations in the Fornax core. 
{
In particular, their velocity dispersion profiles show clear and spatially-similar 
signatures of a superposition of the bright central galaxy, that dominates the inner regions up to $\sim25$ kpc (4.3$'$), and the cluster potential, that becomes dominant outside $\sim 100$ kpc (17.2$'$).}

Most of the kinematical differences
are due to the different spatial distribution (i.e. their number density profiles) and intrinsic anisotropy of the two populations 
\citep[see Eq.~3-4 in][]{napolitano-2014}. Given the typical values of these parameters, in massive galaxies the difference between their projected velocity dispersions can be of the order of 10--20\% \citep[see][]{napolitano-2014}, hence 
within the typical errors of individual dispersion values obtained by PNe and GCs. 

In principle, if we could have the information on the number density profile and anisotropy of the tracers in dwarf galaxies (as done for massive systems) we could predict what difference in the PNe and GCs projected velocity dispersion one should expect.
Unfortunately, we know little about the detailed slope of these populations in low-mass systems, while we have sparse information on the metal-rich and metal-poor stellar populations in the Local Group, which might be used as a reference being PNe and GCs possibly tracers of the former and the latter populations respectively. 
For instance, \citet{walker&penarrubia-2011} discuss the dispersion predictions of different metal-rich and -poor sub-populations assuming realistic profile slopes and anisotropy parameters and they show (see e.g. their Fig.~3) that the velocity dispersion difference becomes very small (of the order of $10\%$ or smaller) outside 1--2 $R_e$. If we assume that the PNe and GCs 
follow the kinematics of these sub-components, then we can reasonably expect that the two populations should show differences in their velocity dispersion profiles of the same {order of} magnitude, again within the typical statistical errors. 

Assuming that all these arguments are valid, we use in our analysis a total of 2070 objects, where GCs represent $57\%$ of the whole catalog and the PNe $43\%$. 

\subsubsection{The GC vs. PN phase space properties}
\label{sec:gc_vs_pn}
In order to {further support } 
the assumption that PNe and GCs are a single family of tracers, for stream hunting sake, {here } we 
look in more details {at} their phase space properties.

As already mentioned, Fig.~\ref{fig:gcs_pne} shows the position in RA and DEC of the objects and a reduced phase-space made by the radial velocity (of GCs {top panels, in yellow} and PNe, {bottom panels, in green}) vs. cluster-centric radius (i.e. the distance from the center of NGC~1399). 

The first evident feature is the already mentioned different spatial extension of PNe and GCs, which implies that within 25 arcmin we will rely on the combined sample with a higher tracer density and higher statistics, while outside 25 arcmin the chance of finding streams is considerabl{y }
reduced because of the smaller statistics. Also, we can see that NGC~1387 has an exceptional high density of PNe, while the GC coverage is limited 
(possibly because of the incomplete slit allocation of the GC observations with VIMOS). 
In the phase space, this large overdensity {does not seem to be }
aligned to the galaxy systemic velocity (a discrepancy that has not been solved in S+18). 
{Finally, another visible difference between the two phase-space diagrams} resides very close to the center, i.e. $R<5'$, where the PN sample is numerically smaller and {shows a smaller overall scatter in the radial velocity}. This is {due} to two main factors: first a large spatially incompleteness of the PNe due to the bright NGC 1399 diffuse halo and second, a shallower and lower resolution dataset from FORS1 observations from \citet{mcneil-2010} covering this area. However, for the purpose of this paper we will exclude regions too close to the galaxy center, hence the differences inside 5\arcmin~radius will not affect {any of the }
final results. 


In Fig.~\ref{fig:sample} we show how the combined GC and PN samples (plotted with different gray scales) add up together. 
Looking at the phase-space, we see that, besides some residual inhomogeneities in the PN distribution (left panel), after having cleaned the PN catalog from incompleteness,
the two samples show (in the right panel) a smaller dispersion in the center and an increasing dispersion going toward larger radii. Overall the two samples do not show large differences in terms of mean velocity and velocity dispersion (continuous and dotted lines respectively in the bottom of Fig.~\ref{fig:sample}). We will discuss this in a more quantitative way in the next Section. 

\begin{figure}
\centering
\hspace{-0.4cm}
\includegraphics[width=8.7cm]{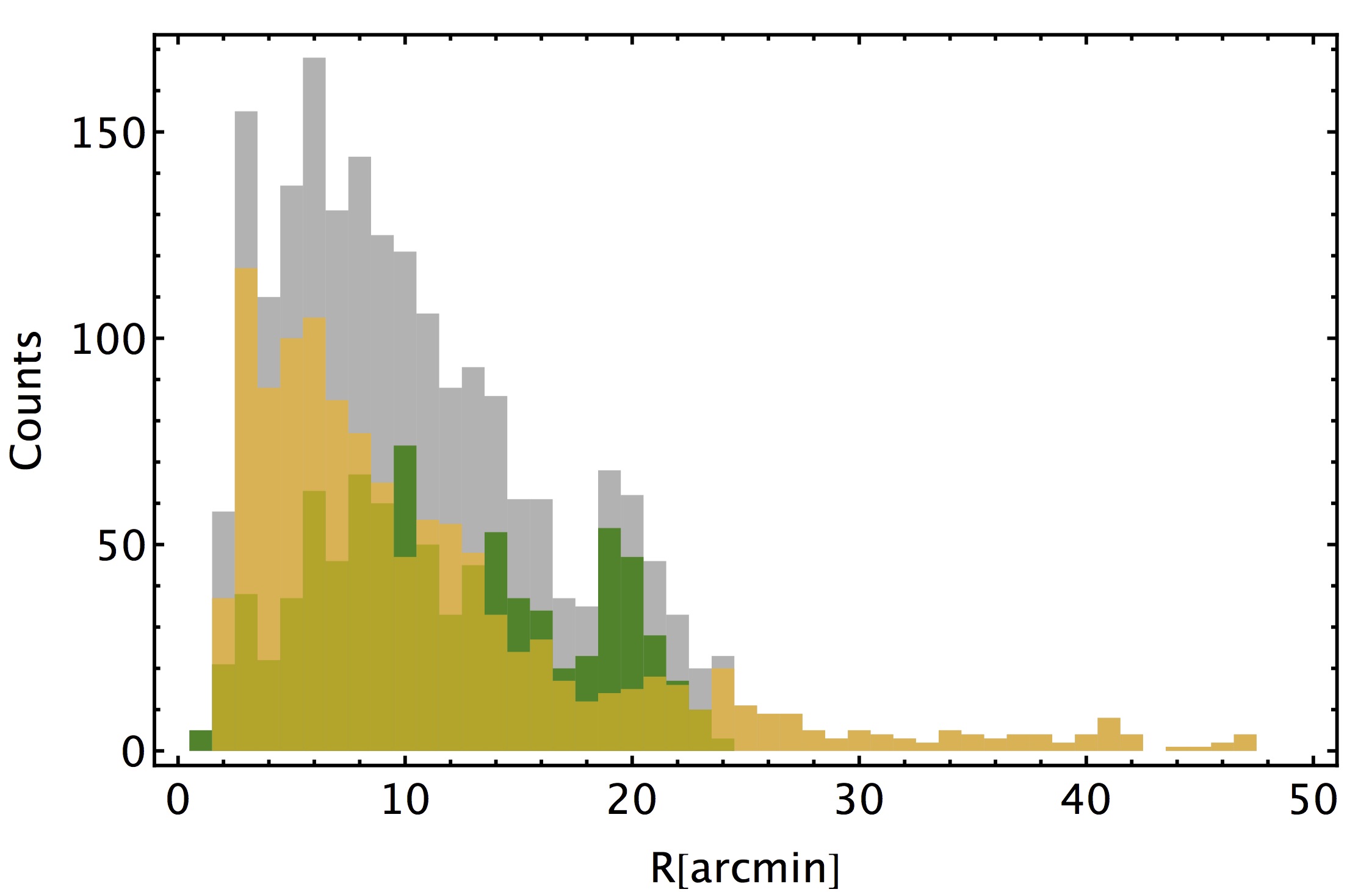}

\includegraphics[width=8.8cm]{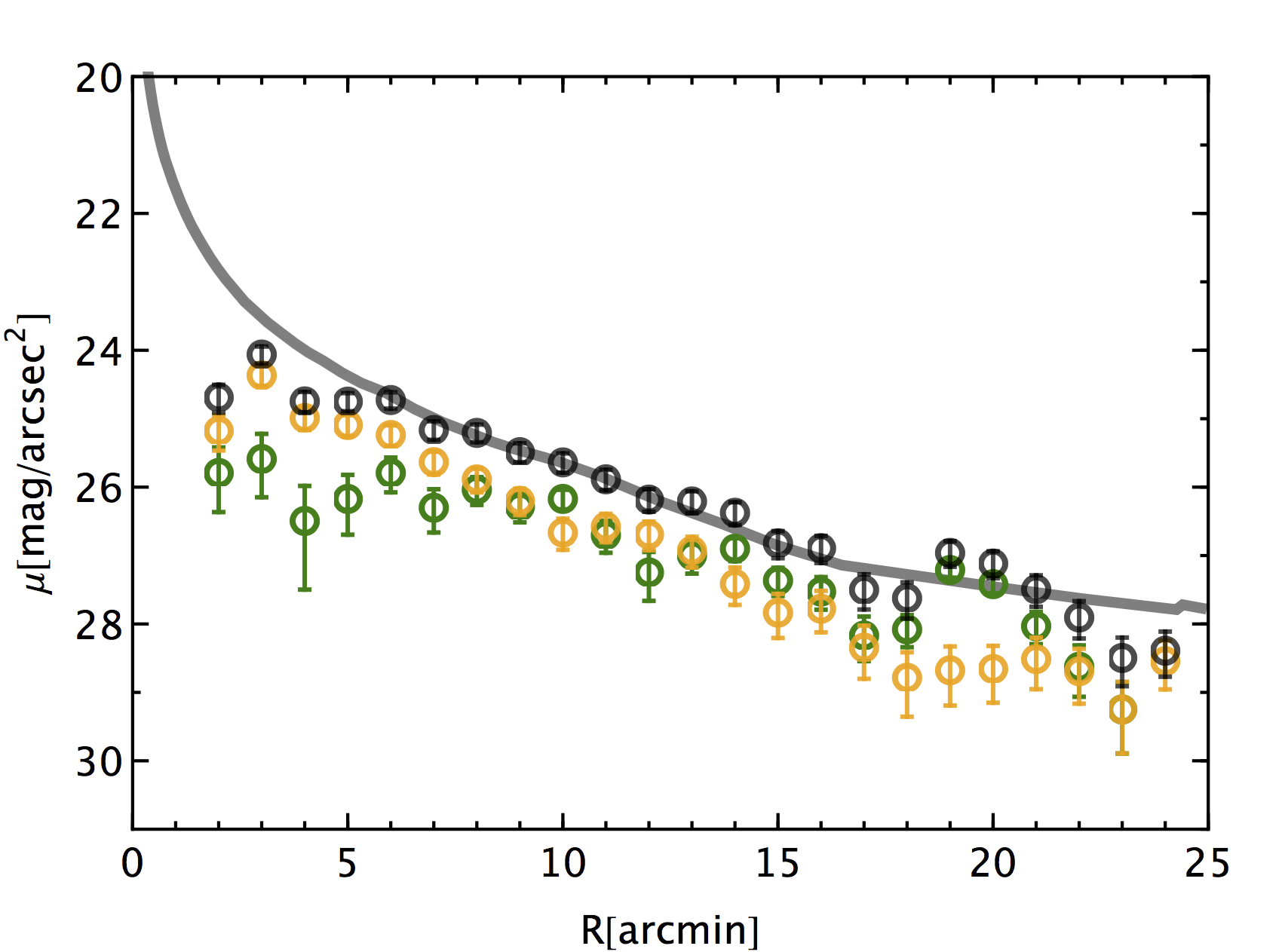}
	\caption{\textit{Top}: number counts as a function of the radius of GCs (in yellow), PNe (in green) and the total population (in gray). The overdensity of PNe around NGC 1387 is clearly seen at $R\sim20''$ (see text).  
	\textit{Bottom}: Radial density distribution of GCs (yellow open dots), PNe (green) and the total population (black), arbitrarily re-scaled to match the surface brightness of the Fornax core from I+16 (gray tick line). Error bars represent Poissonian errors on the tracer counts. 
	}
	\label{fig:histo_sample}
\end{figure}

\begin{figure}
    \hspace{-0.5cm}
	\includegraphics[width=9.3cm]{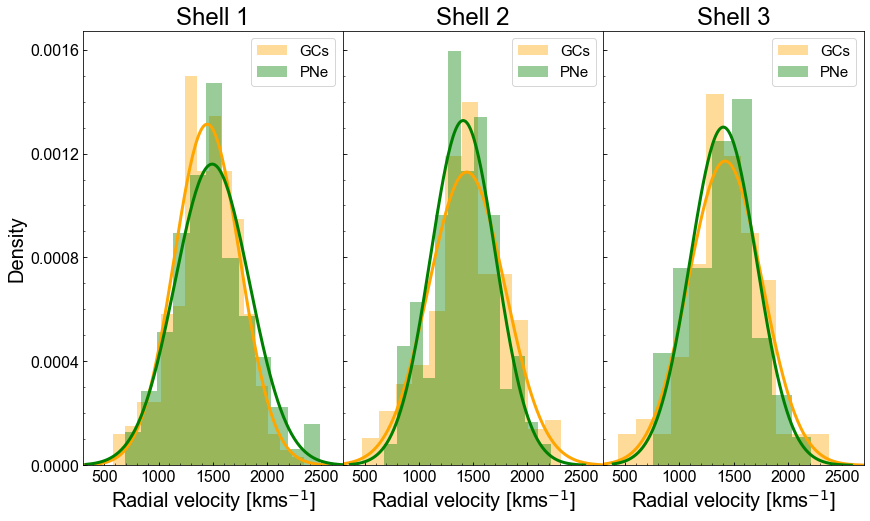}
	\caption{Histograms of the density of GCs (yellow) and of the PNe ({green}) 
	per velocity interval in the three shells at different radial distances. Overlapped to these histograms there are the distributions of the two samples.}
	\label{fig:K-S test}
\end{figure}

\begin{table}
	\centering
	\caption{Parameters and p-values of the Kolmogorov-Smirnov test.}
	\label{tab:K-S test}
	\begin{tabular}{c c c c}
		\hline
		GCs & PNe & Radius (arcmin) & p-value\\
		\hline
		298 & 208 & $5.0 \leq R \leq 8.5$ & 0.37\\ 
		187 & 202 & $11 \leq R \leq 16.0$ & 0.19\\ 
		104 & 101 & $16.0 \leq R \leq 18.0$ \& $20.0 \leq R \leq 25.0$ & 0.51 \\
		\hline
	\end{tabular}
\end{table}


To quantify more the differences in {the} 
spatial distributions {of PNe and GCs}, in Fig.~\ref{fig:histo_sample} we 
plot the distribution of the cluster-centric radii of the two sub-populations. 
In the top panel
we plot number counts as a function of the radius for the separated populations of GCs (yellow) and PNe (green) and the total population (gray). They show some distinctive features, some of them mirroring the ones in the phase-space: 1) PNe extend out to a distance of about 25$'$ from NGC~1399, since no observations were performed outside this radius, while GCs are distributed out to $\sim 45'$; 2) inside $\sim7'$, PNe show a clear incompleteness (i.e. their normalized counts are systematically lower than the ones of GCs), while outside $7'$ they 
nicely align with GC number densities, decreasing  
with a similar slope (e.g. in the range $7'-17'$); 3) there is a PN density peak at about  $20'$ 
around NGC~1387, where we observe a mild, but visible dip in the GC density,
due to the selection effect in the GC spectroscopic sample mentioned above.
On the bottom panel {of the same figure}, we have converted the number counts per radial bin into a surface density (N/arcmin$^2$), where the area is the one of the circular annulus enclosing any bin, and finally arbitrary re-scaled along the vertical axis to match the surface brightness distribution of the central galaxy from I+16. The latter plot shows a strong similarity {between the PNe+GC tracers and the galaxy light, }
except in the {areas }
where there is a clear effect of the spatial incompleteness (i.e. the center below 5$'$ and {around NGC~1387.}
{L}ooking at the individual density distributions, PNe and GCs also show a nice consistency in most of the radial bins, hence confirming that these two population of tracers are fairly representative of the total stellar light in the cluster.    

\subsubsection{Kolmogorov-Smirnov test}
\label{sec:Kolmogorov-Smirnov test}
To check whether the two tracers are statistically representative of a single kinematical population, we performed a Kolmogorov-Smirnoff (K-S) test. 

We have defined three shells at different radial distances from NGC~1399 (see Tab. \ref{tab:K-S test}), and we have performed a K-S test on the GC and PN velocity dispersion distribution separately in each shell.
We excluded from this non-parametric test the inner region of the cluster ($R \leq 5$ arcmin) and the galactocentric distances where NGC~1387 and NGC~1404 are located because these regions will be excluded in the research of the cold substructures (see Sect. \ref{sec:runningCOSTA}).

The inner and outer radii of each shell were selected in order to have about the same number of GCs and PNe, except for the inner shell, which has a larger number of GCs (298 vs. 208 PNe) because of the 
higher number of GCs in the inner regions of the cluster (see also the top of the figure \ref{fig:histo_sample}). 

Because the PNe are selected out to a maximum distance of about $\sim 25$ arcmin, no shells beyond this radius are considered.

In table \ref{tab:K-S test} we show the results of the K-S test in each shell. In Fig.~\ref{fig:K-S test}, we 
plot the histograms of the density of GCs (yellow) and of PNe (green)
per velocity interval for the three shells and their equivalent Gaussian distribution. 
The results obtained from the K-S test are consistent with the null hypothesis that both GCs and PNe follow the same distribution. 
Indeed the p-values for all the shells showed in  Table~\ref{tab:K-S test} are well above the significance level of 5\%. Moreover, it is also evident from Fig.~\ref{fig:K-S test} that the two distributions are very similar, as also shown by the Gaussian fit to their velocity histograms. In Appendix \ref{sec:red_blue_pne} we discuss the further case where we separate blue and red GCs and find that their velocity distributions remain statistically consistent between them and with the one of the PNe.

Having presented several arguments supporting the assumption that GCs and PNe have similar spatial and velocity profiles, from now on we will consider 
them as a single family of test particles, which allows us to increase the statistics and, consequently, the chances of finding cold substructures.

\section{The COSTA algorithm}
\label{sec:COSTA}

In this section we briefly summarize the features and assess reliability of the COSTA algorithm. This has been fully tested on hydro-simulations of galaxy encounters and Montecarlo simulations of realistic cluster-like velocity fields. These latter consist of realizations of the phase space distribution of tracer particles created by random sampling from a smooth model of the observed tracer density profile in equilibrium with an analytic model of the Fornax potential
(see G+20 for details).
Here we report the most relevant properties and definitions that we use throughout the paper.

\subsection{COSTA steps}
\label{sec:steps}
The algorithm looks for coherent structures both in the RA/DEC position-space and in the reduced phase-space (radial velocity vs. cluster-centric radius), with the additional necessary condition to have a low velocity dispersion. 
Indeed, we are particularly interested in dwarf galaxy disruption as 
a major mechanism,
still active in the local universe, contributing to the intracluster stellar population and the assembly of large stellar halos around galaxies. Therefore, we introduce a velocity dispersion threshold to define a ``cold'' substucture, $\sigma_{\rm cut}$, 
to vary in the range of [10-120] \kms, according to typical dwarf dispersion values found in the Coma cluster \citep{coma_dw_FJ}\footnote{Even if most of the dwarf galaxies in Coma from \citet{coma_dw_FJ} have $\sigma<100$ \kms, we decided to use a slightly larger range of $\sigma_{\rm cut}$, to avoid border effects. These could happen, e.g., when taking larger $N$ neighbor particles, as this could increase the chance of measuring a larger velocity dispersion for a group of particles, hence producing a clustering of parameters toward higher $\sigma_{\rm cut}$.}.

The main steps performed by COSTA are summarized here below (see G+20 for a more detailed description):
\begin{itemize}
    \item[i)] for each particle, COSTA finds the first $k$ nearest neighbors in the position space;
    \item[ii)] it estimates the mean velocity and velocity dispersion of these $k$ tracers;
    \item[iii)] it performs an iterated sigma clipping by removing all tracers beyond n standard deviations from the mean velocity of the group. During each iteration, it recalculates the average velocity and the velocity dispersion, until no more outliers are found;
    \item[iv)] it retains all structures with a number of particles greater than $N_{\rm min}$ and with a velocity dispersion lower than $\sigma_{\rm cut}$.
\end{itemize}
\noindent

Finally, since the search of closer neighbors is based on a projected circular distance, while streams tend to lie on elongated structures, COSTA performs a further step:

\begin{itemize}

\item[v)] if there are groups having some particles in common, they are considered belonging to a single stream if their mean velocity and velocity dispersion values are consistent with each other within uncertainties. 

\end{itemize}
\noindent
{Ultimately, COSTA }
depends on four free parameters: the three friend-of-friend parameters (\emph{k, n, N\textsubscript{\rm min}}) and the $\sigma_{\rm cut}$, which need to be properly chosen to maximize the number of real cold substructures (completeness) and minimize the number of spurious detections (purity), caused by the intrinsic stochastic nature of the velocity field of hot systems. 

In G+20, we have shown 
how to obtain a list of parameter combinations that produce an acceptable probability of finding ``false positives'', i.e. spurious cold substructures based on Montecarlo simulations.  
This is done by assigning a maximum reliability to the combination of parameters for which no streams are detected on Montecarlo simulations of the Fornax core where all test particles are in equilibrium with the ``warm''cluster potential (where no cold streams are added, see G+20). Vice versa, a lower reliability is assigned to the parameter combinations that find an increasing number of (false) detection in the same simulations (see next Section).

\begin{figure}
    \centering
    \includegraphics[width=0.48\textwidth]{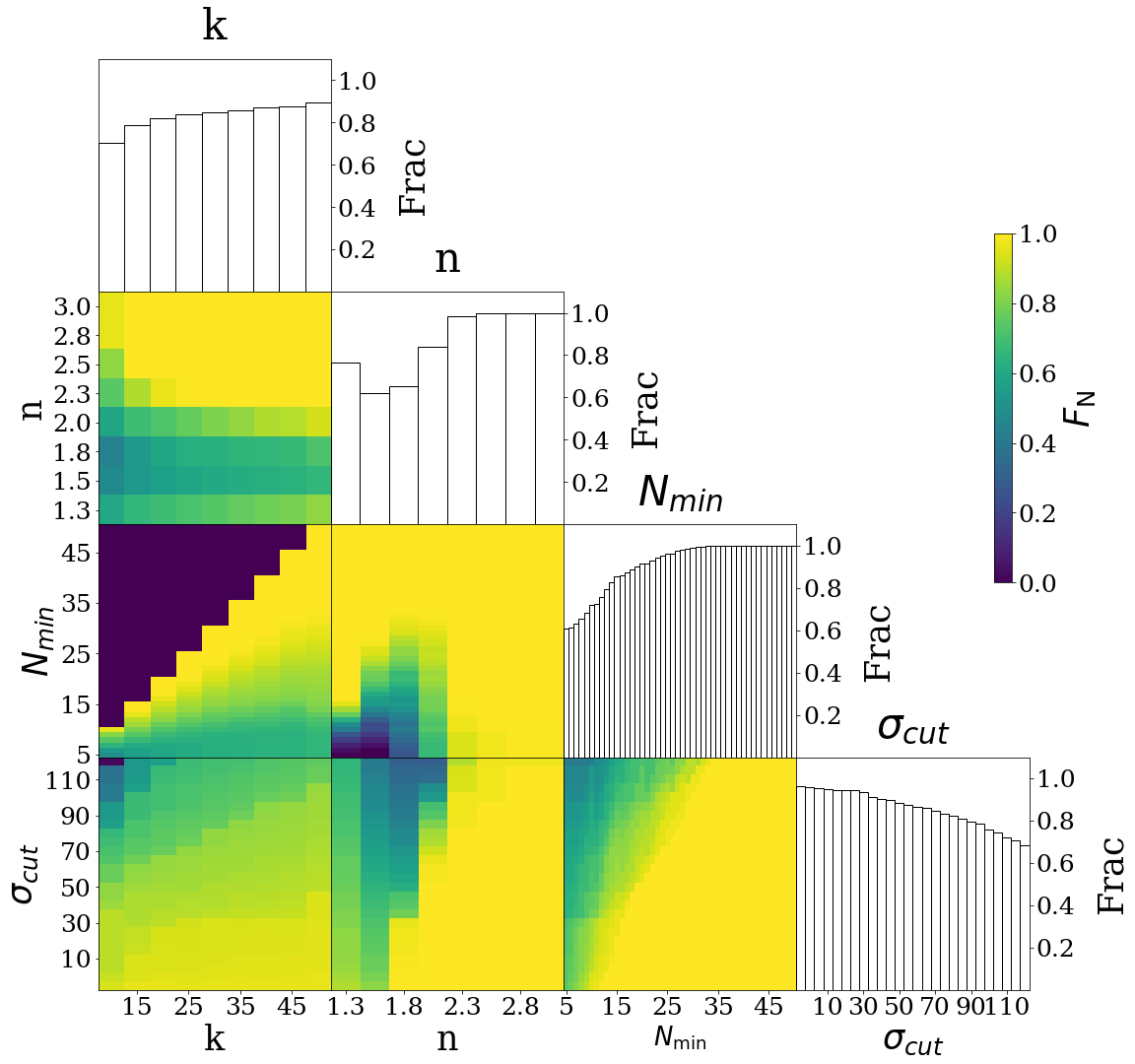}
    \caption{Reliability map in the 4-dimensional parameter space for the {simulated } Fornax cluster. The 2D projected plots for each pair of parameters, are color-coded by the fraction of configurations ($F_{\rm N}$) with reliability~$\geq50\%$ according to the legend. Histograms show the fraction of times the reliability passes the 50\% threshold for each value of the four free parameters of COSTA.}
    \label{fig:rel_map}
\end{figure}

\subsection{Reliability map from Montecarlo simulations of the Fornax cluster core}
\label{sec:simulation of the cluster}

Following G+20, the first step to perform before running COSTA on a given dataset, is to derive the reliability map.
This is obtained by producing Montecarlo  simulations 
of the dataset {to which COSTA will be applied }
(we refer to the original \citealt{napolitano-2001} paper for more
details). 
In G+20 we have obtained such Montecarlo simulations for a Fornax-like cluster, including the presence of the major galaxies in the Fornax core (see G+20 for details). We have also shown that COSTA is able to retrieve a series of artificial streams with different physical sizes and particle numbers, corresponding to the expected surface brightness for streams produced by the interaction of dwarf galaxies with the cluster environment (i.e. of the order of 28-29 mag/arcsec$^2$). 

Here we briefly summarize the main concepts of the simulated sample and how this is used to define the reliability map.
We start by simulating random discrete radial velocity fields of particles:
\begin{itemize}
\item we consider only the region covered by our objects, i.e.  
$1.8$ deg$^2$ around the cD, NGC 1399, where there are two other bright early-type roundish galaxies:
NGC 1404, {located at $\sim9$ arcmin South-East of the cD, and NGC 1387, $\sim 19$ arcmin West of NGC 1399; }
\item following \citet{napolitano-2001}, we consider GCs and PNe at the equilibrium in the gravitational potential of these three galaxies, where we assume for the cluster a total mass of about $10^{14}$ solar masses and a Hernquist \citep{hernquist-1990} density distribution of the stellar-like tracers;
\item we consider a dark matter halo following a Navarro-Frenk-White profile (NFW); hence, the potential of the system at equilibrium is given by the total mass:
\begin{equation}
	M_\textup{tot} = M_\textup{l} + M_\textup{dm}
\end{equation}
where $M_\textup{l}$ and $M_\textup{dm}$ are the total luminous and dark mass respectively;
\item we assume an isotropic velocity dispersion tensor, and solve the radial Jeans equation to derive the 3D velocity dispersion $\sigma^2$ along the three directions in the velocity space, and generate a full 3D phase space;
\item  we 
simulate an observed phase space, first, by projecting the tracer distribution on the sky plane. Then, we derive the line-of-sight velocity of the individual particles by randomly extracting the observed velocities from a Gaussian distribution centered at the true value and having a standard deviation equal to the velocity errors. We assume these latter to be 
37 kms$^{-1}$, i.e. about the average errors of real GCs and PNe (see P+18 and S+18). 
\end{itemize}

We {check carefully }
that the mock catalogs of positions and radial velocities closely resemble the real one,  and also that the simulated cluster {is }
consistent with the mean observable quantities of the galaxies in the area 
(see  Table~2 in G+20). 


Finally, we create a ``white noise sample'' (WNS) as described in G+20 by randomly drawing 100 realizations of the mock catalog, with no substructures, from the model described above.
COSTA is run on the WNS using a grid of different parameters to
to check which combinations produce any (false) detection, due to randomly connected particles.  

The parameters have been uniformly taken in the following ranges:
\begin{itemize}
\item $10 \leq k \leq 50$;
\item $1.3 \leq n \leq 3$;
\item $5 \leq N_{\rm min} \leq k$;
\item $10 \leq \sigma_{\rm cut} \leq 120$ \kms .
\end{itemize}
This choice allowed us to search {both for} 
small substructures, that we {expect }
to find with 
low \emph{k}, low $N_{\rm min}$ and low velocity dispersion {value, and for }
larger and spatially extended groups, with greater values of both $k$ and $N_{\rm min}$ and rather hot velocity dispersion, e.g. of the kind expected from moderate luminosity galaxies like NGC 1387 \citep[see e.g.;][]{Iodice-2016}.

For a given parameter combination, the reliability is defined as the fraction of detections over the 100 realizations in the WNS, i.e. $100 - N_{\rm det}/100$\% (with $N_{\rm det}$ expected to be zero, by definition in the WNS).
In G+20 we have discussed the impact of threshold on the completeness of the candidates and checked that a good compromise between the completeness and the contamination is obtained with a threshold of 70\% in reliability. However, 
for this first test on Fornax, we have decided to apply a less conservative threshold of 50\%, which can both return more candidates, at the cost of a higher probability of spurious detections.  
In case we will gain all real streams, the completeness is expected to be incremented by 20\% (see G+20), while in case the extra candidates are all false positive, the contamination
can raise up to $\sim 58\%$, which is a risk we decided to accept. 
In Fig.~\ref{fig:rel_map} 
we show the reliability distribution in the 4-dimensional parameter space. The 2D projection plots, are color-coded by the fraction of configurations ($F_{\rm N}$) with reliability$\geq50\%$. Yellow regions are those with the highest density of configurations with minimal or no false detections. Histograms, in the same figure, show the fraction of times where the reliability passes the 50\% threshold for each value of the four free parameters of COSTA.

\begin{figure*}
    \centering
    \includegraphics[scale=0.16]{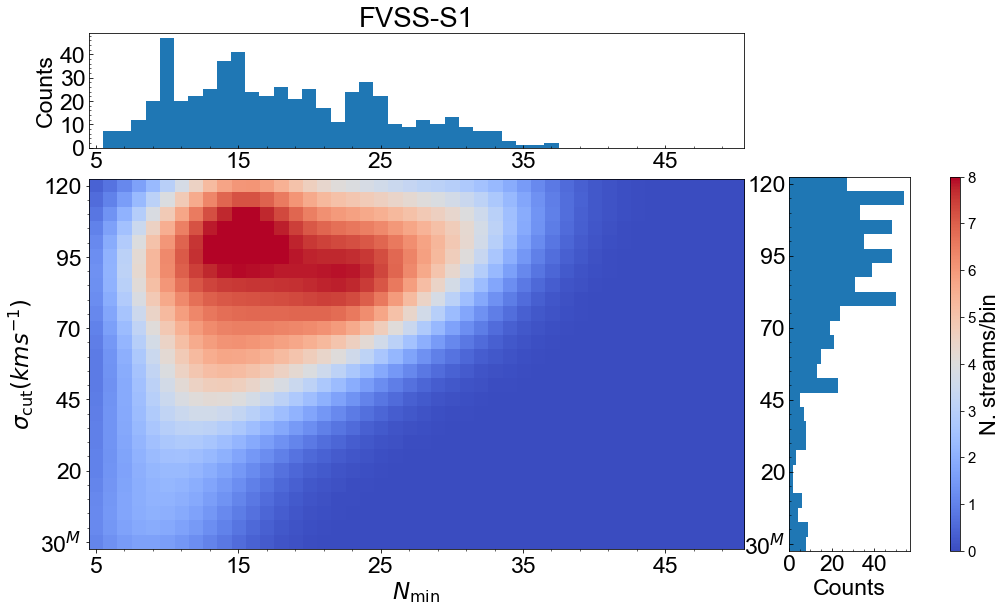}
    \includegraphics[scale=0.16]{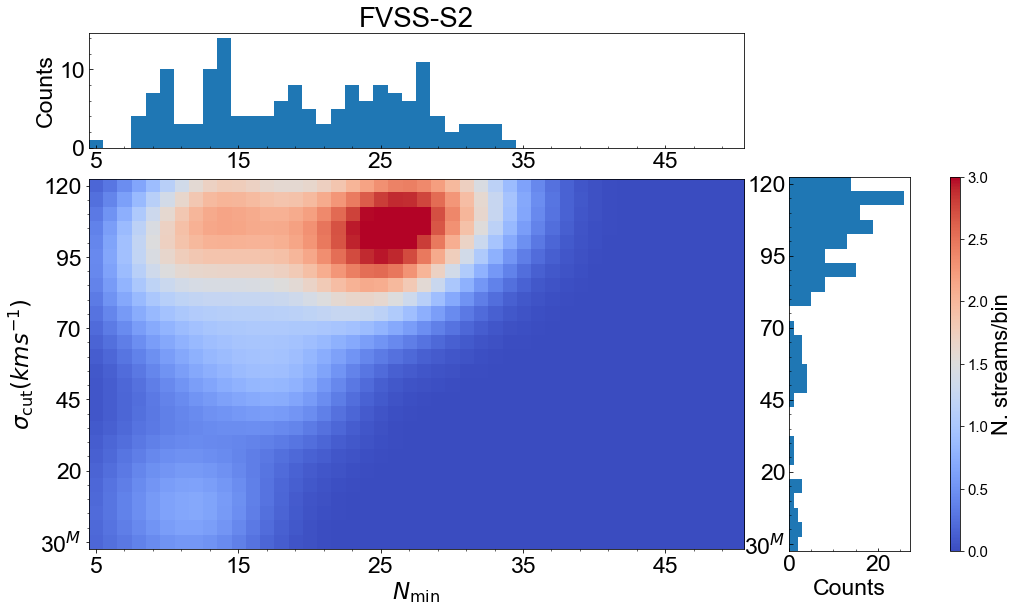}
    \includegraphics[scale=0.16]{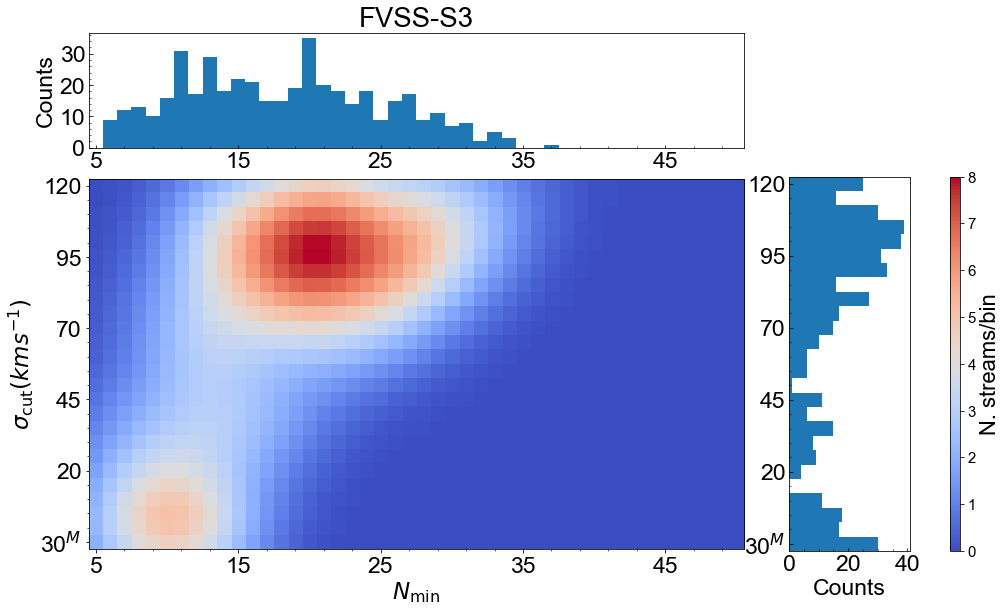}\\
    \includegraphics[scale=0.16]{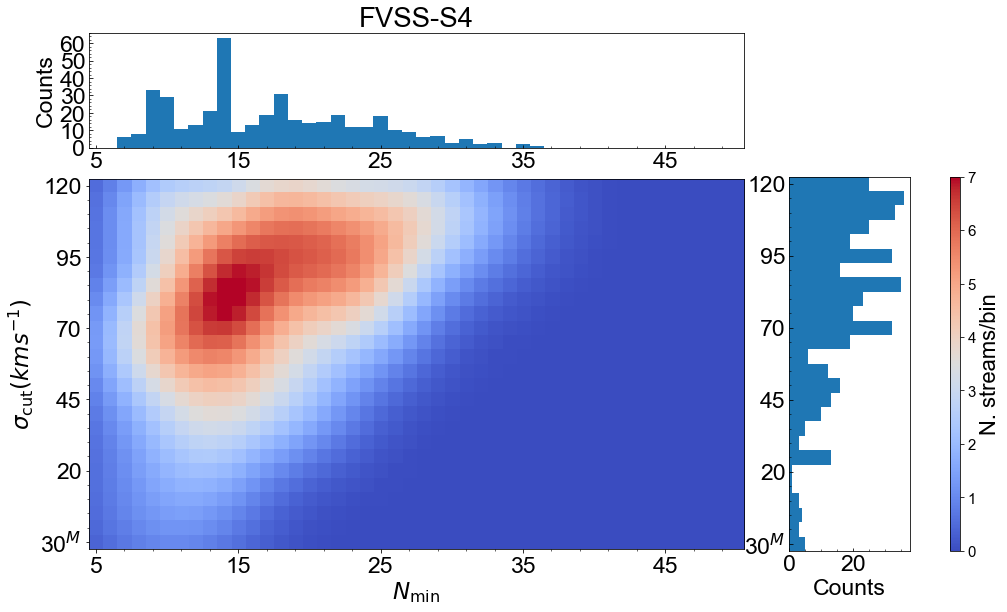}
    \includegraphics[scale=0.16]{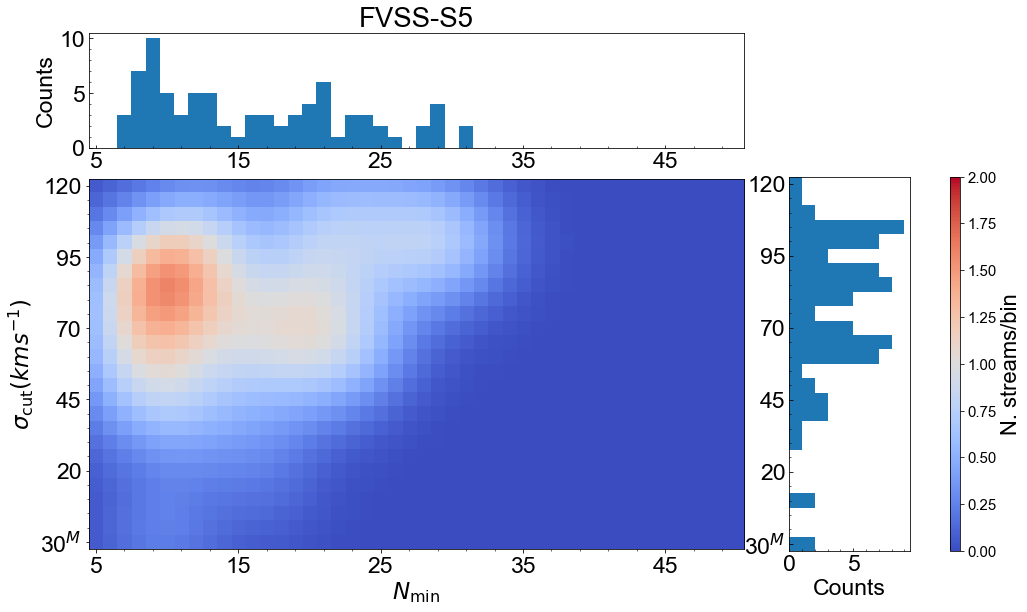}
    \includegraphics[scale=0.16]{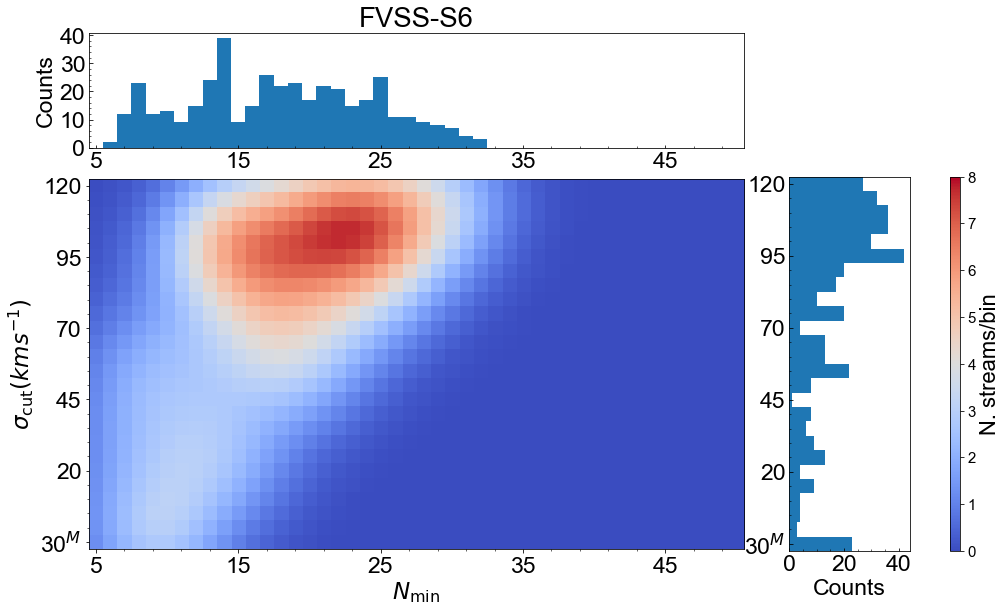}\\
    \includegraphics[scale=0.16]{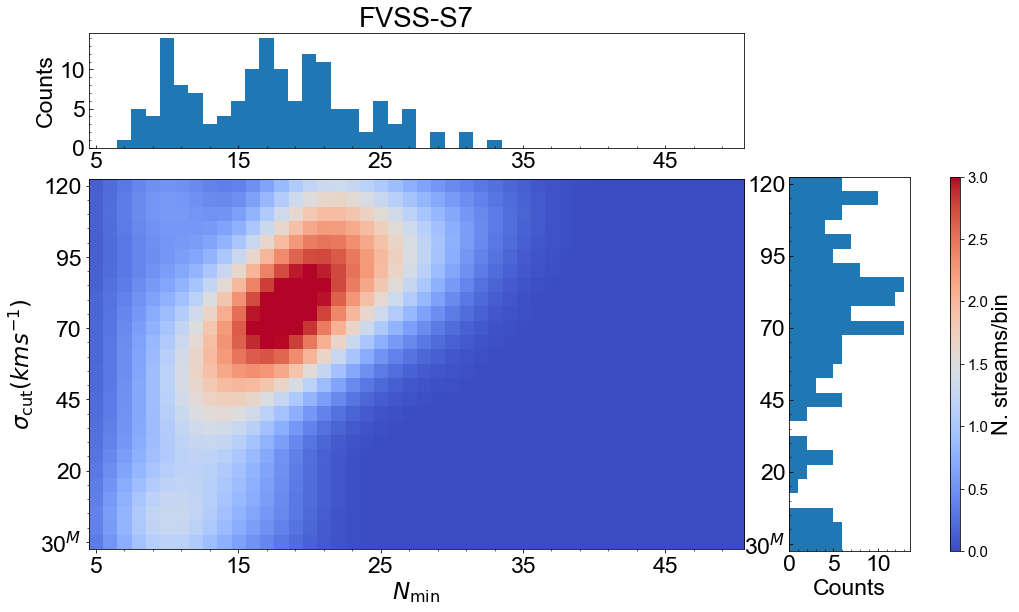}
    \includegraphics[scale=0.16]{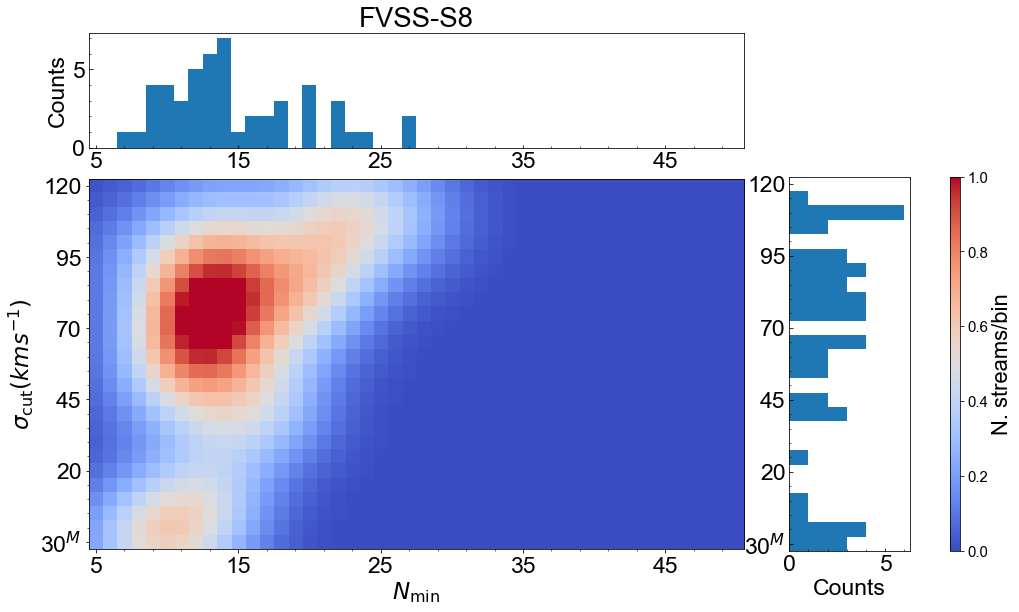}
    \includegraphics[scale=0.16]{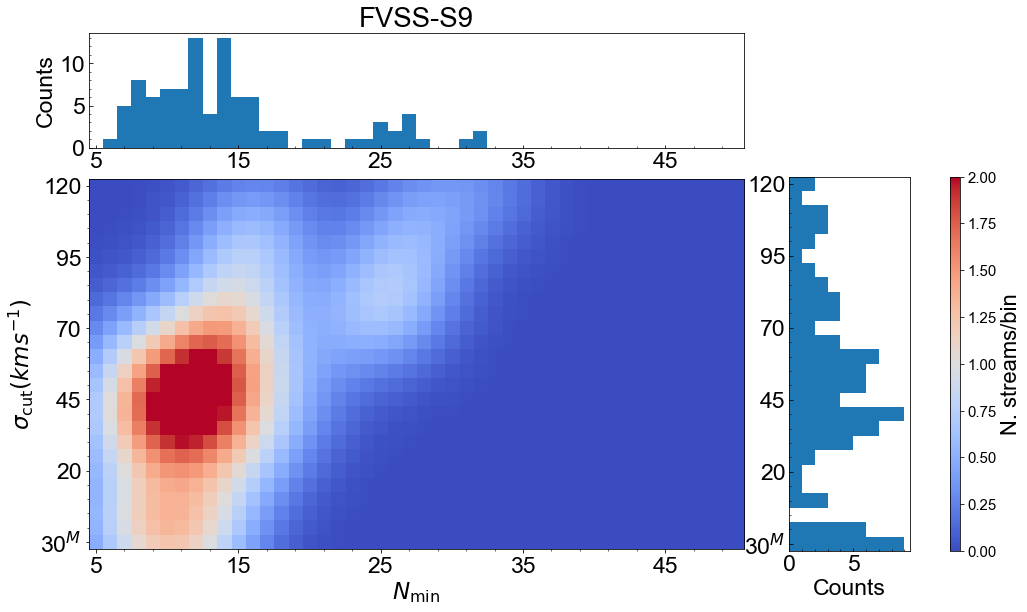}\\
    \includegraphics[scale=0.16]{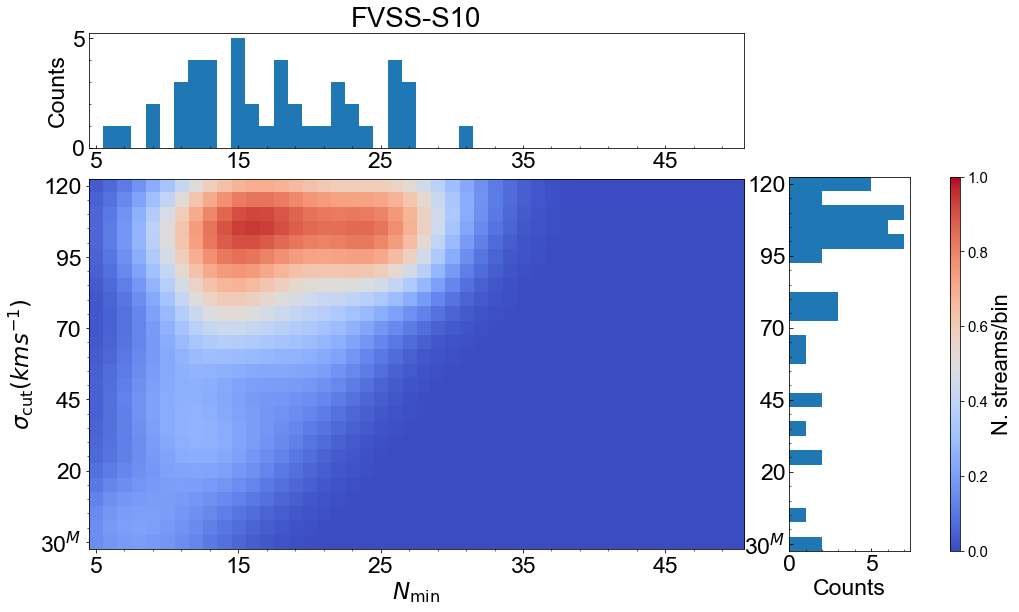}
    \includegraphics[scale=0.16]{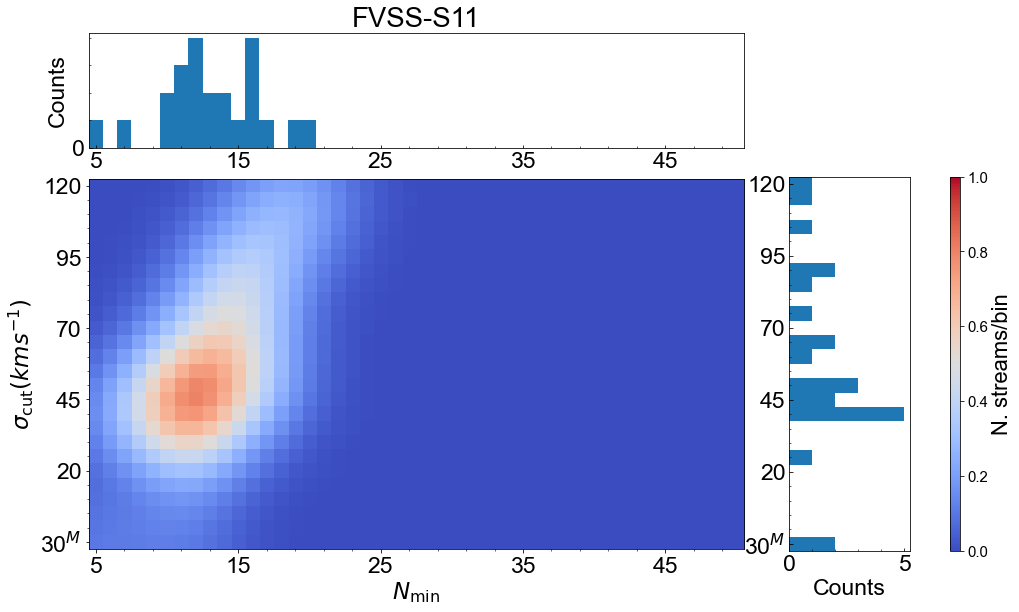}
    \includegraphics[scale=0.16]{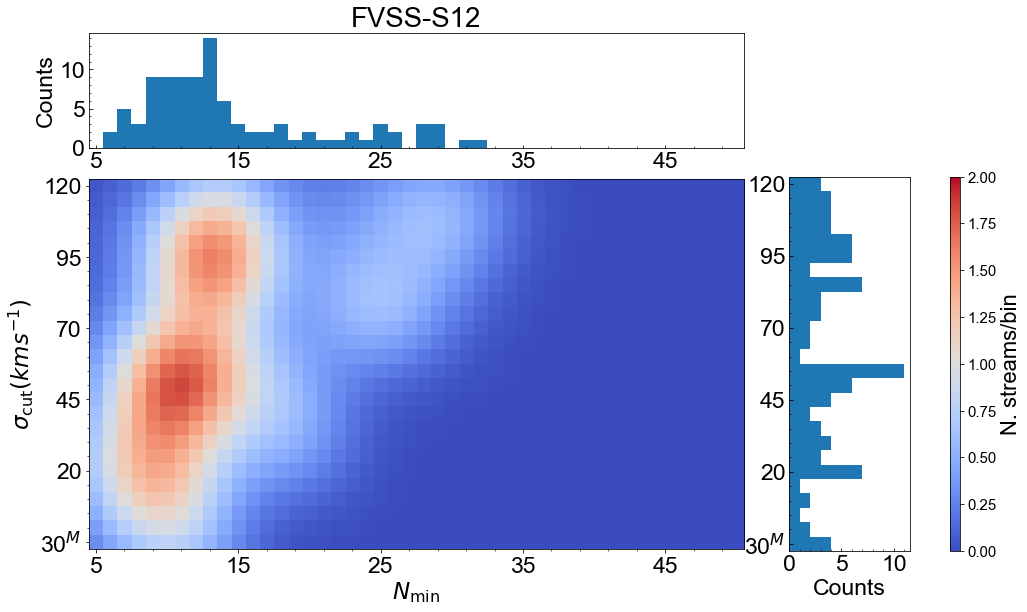}\\
    \includegraphics[scale=0.16]{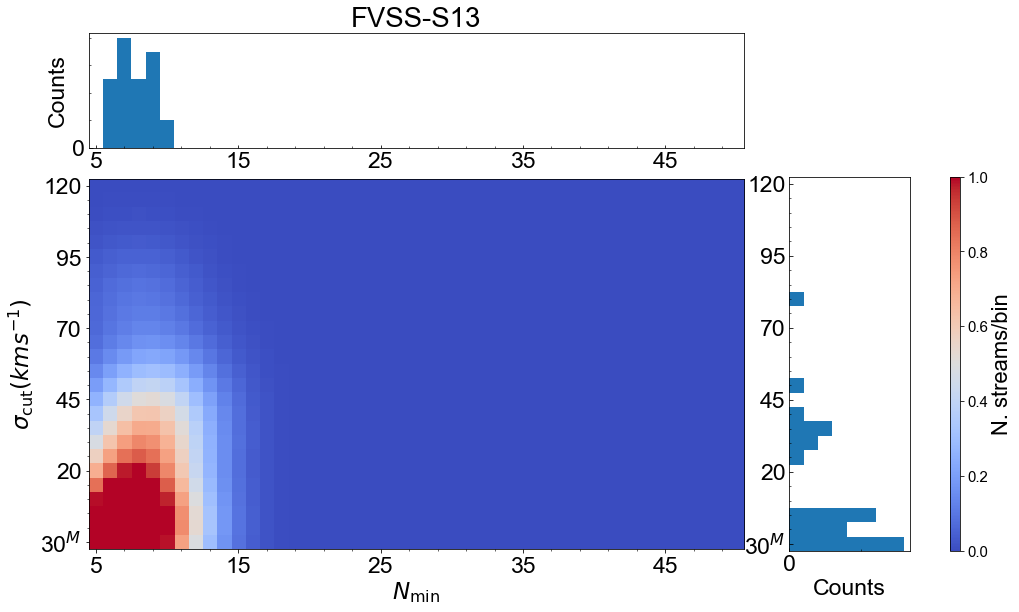}\\

    \caption{Density plot of the $\sigma_{\rm cut}$ as a function of $N_{\rm min}$ for each stream, color-coded by the number of times COSTA detected the stream with a given $\sigma_{\rm cut}$-$N_{\rm min}$ pair. Data were smoothed using a Gaussian kernel with a bandwidth equal to 3.}
    \label{fig:sigma_nmin}
\end{figure*}

\begin{table*}
    \tiny
	\caption{Statistics of the streams.}
	\label{tab:statistics_2}
	\hspace*{-1cm}
	\begin{tabular}{lccccccccccccccc} 
		\hline
		ID & RA & DEC & N & PNe & GCs & Rel. & Max Rel. & Occurr. & Vel. & $\sigma$ & Size & Distance & $\log(L)$ & SB\\
		& (J2000) & (J2000) & & & & & & & (kms$^{-1}$) & (kms$^{-1}$) & (arcmin) & (arcmin) & ($L_\textup{\sun,g}$) & (mag/arcsec$^2$)\\
		\hline
FVSS-S1 & 55.0119 & -35.4184 & $19_{-7}^{+11}$ & $14_{-10}^{+7}$ & $6_{-3}^{+7}$ & $76_{-18}^{+17}$ & 100 & 542 & $1500_{-30}^{+24}$ & $87_{-36}^{+21}$ & $6.7_{-2.7}^{+12.0}$ & 19.2 & $8.1_{-0.5}^{+0.2}$ & $30.0_{-0.4}^{+1.4}$\\
FVSS-S2 & 54.5819 & -35.5908 & $22_{-9}^{+12}$ & $7_{-4}^{+8}$ & $13_{-5}^{+11}$ & $74_{-17}^{+17}$ & 100 & 153 & $1563_{-34}^{+39}$ & $100_{-40}^{+14}$ & $4.2_{-2.0}^{+1.9}$ & 8.6 & $7.8_{-0.4}^{+0.3}$ & $29.7_{-0.8}^{+0.9}$\\
FVSS-S3 & 54.7895 & -35.5320 & $20_{-8}^{+9}$ & $8_{-3}^{+5}$ & $12_{-7}^{+6}$ & $75_{-16}^{+17}$ & 100 & 439 & $1392_{-68}^{+37}$ & $83_{-78}^{+22}$ & $5.2_{-1.3}^{+2.1}$ & 9.6 & $7.9_{-0.2}^{+0.2}$ & $30.0_{-0.5}^{+0.5}$\\
FVSS-S4 & 54.5123 & -35.4844 & $18_{-7}^{+9}$ & $9_{-3}^{+6}$ & $8_{-3}^{+4}$ & $85_{-22}^{+13}$ & 100 & 410 & $1473_{-29}^{+33}$ & $83_{-35}^{+27}$ & $1.9_{-0.6}^{+0.5}$ & 5.7 & $7.9_{-0.2}^{+0.2}$ & $27.7_{-0.5}^{+0.4}$\\
FVSS-S5 & 54.4688 & -35.5341 & $16_{-6}^{+8}$ & $7_{-2}^{+3}$ & $9_{-4}^{+6}$ & $68_{-14}^{+19}$ & 98 & 80 & $1351_{-19}^{+42}$ & $79_{-32}^{+20}$ & $2.9_{-0.9}^{+1.5}$ & 9.0 & $7.8_{-0.2}^{+0.1}$ & $28.9_{-0.4}^{+0.4}$\\
FVSS-S6 & 54.1540 & -35.0617 & $20_{-8}^{+7}$ & $0_{-0}^{+0}$ & $20_{-8}^{+7}$ & $85_{-21}^{+11}$ & 100 & 414 & $1795_{-25}^{+30}$ & $89_{-61}^{+19}$ & $20.2_{-7.4}^{+3.1}$ & 32.7 & - & -\\
FVSS-S7 & 54.3363 & -35.3508 & $19_{-7}^{+5}$ & $11_{-4}^{+4}$ & $7_{-2}^{+3}$ & $74_{-16}^{+17}$ & 99 & 146 & $1391_{-30}^{+31}$ & $74_{-50}^{+28}$ & $4.2_{-0.5}^{+1.2}$ & 15.2 & $8.0_{-0.2}^{+0.1}$ & $29.2_{-0.3}^{+0.5}$\\
FVSS-S8 & 54.3083 & -35.4652 & $15_{-2}^{+9}$ & $10_{-4}^{+6}$ & $7_{-2}^{+1}$ & $70_{-17}^{+11}$ & 96 & 27 & $1319_{-65}^{+30}$ & $84_{-47}^{+24}$ & $5.1_{-0.7}^{+0.8}$ & 15.3 & $7.9_{-0.2}^{+0.2}$ & $29.7_{-0.5}^{+0.6}$\\
FVSS-S9 & 54.5230 & -35.3838 & $14_{-4}^{+7}$ & $2_{-1}^{+4}$ & $11_{-3}^{+5}$ & $71_{-14}^{+19}$ & 99 & 97 & $1388_{-33}^{+24}$ & $46_{-38}^{+36}$ & $2.9_{-0.6}^{+1.9}$ & 6.2 & $7.2_{-0.3}^{+0.5}$ & $30.3_{-1.2}^{+0.8}$\\
FVSS-S10 & 54.8525 & -35.3793 & $19_{-6}^{+8}$ & $14_{-5}^{+3}$ & $5_{-2}^{+5}$ & $61_{-8}^{+16}$ & 97 & 45 & $1488_{-60}^{+41}$ & $98_{-53}^{+11}$ & $6.4_{-2.0}^{+1.9}$ & 12.1 & $8.1_{-0.2}^{+0.1}$ & $29.9_{-0.2}^{+0.5}$\\
FVSS-S11 & 54.6265 & -35.3504 & $14_{-2}^{+3}$ & $10_{-3}^{+3}$ & $4_{-2}^{+2}$ & $69_{-17}^{+22}$ & 98 & 23 & $1251_{-13}^{+102}$ & $49_{-11}^{+40}$ & $2.4_{-0.3}^{+0.3}$ & 6.0 & $7.9_{-0.2}^{+0.1}$ & $28.1_{-0.3}^{+0.4}$\\
FVSS-S12 & 54.6571 & -35.5594 & $14_{-4}^{+10}$ & $6_{-2}^{+1}$ & $8_{-2}^{+7}$ & $69_{-12}^{+13}$ & 99 & 97 & $1555_{-38}^{+32}$ & $55_{-36}^{+44}$ & $2.2_{-0.7}^{+1.5}$ & 6.8 & $7.7_{-0.2}^{+0.1}$ & $28.5_{-0.2}^{+0.4}$\\
FVSS-S13 & 54.6927 & -35.6170 & $9_{-1}^{+1}$ & $3_{-0}^{+1}$ & $5_{-1}^{+1}$ & $78_{-20}^{+14}$ & 99 & 27 & $2046_{-27}^{+13}$ & $35_{-11}^{+15}$ & $1.8_{-0.3}^{+0.4}$ & 10.6 & $7.4_{-0.2}^{+0.1}$ & $28.8_{-0.3}^{+0.4}$\\
\hline
	\end{tabular}
\begin{minipage}{180mm}
For each stream, we report identification, ID, RA and DEC coordinates, the average number of tracers, N, and the number of PNe and GCs. We also report the median reliability, Rel., the maximum reliability, Max Rel., the number of times COSTA detected the stream, Occurr., followed by their median radial velocity, Vel, velocity dispersion, $\sigma$, the size -- defined by the longer average distance among particles, and cluster-centric distance of their centroid. Last two columns report the total luminosity and surface brightness brightness as computed in \S\ref{sec:stream_lum}.\\
\textbf{Note: }FVSS-S13 dispersion is measured (with errors included) and not the intrinsic one.
\end{minipage}

\end{table*}

\subsection{Running COSTA on the combined GC+PN sample}
\label{sec:runningCOSTA}
Once the reliability map has been drawn from the WNS, we 
run COSTA on the detection sample (DS, see G+20), which is made of the real positions and radial velocities of the combined GC+PN sample, discussed in \S\ref{sec:gc+pn}. 
This produces a series of detections, consisting {of }
a number of candidates that recur in multiple set-up, every time with a slightly different number of particles but with a common bulk of members, as demonstrated in G+20.  
To define these 
``representative'' particles corresponding to a given substructure,
we decide to select the ones corresponding to the median parameters among all the allowed configurations, marginalizing over all the other parameters in the 4D parameter space  (see e.g. Fig. \ref{fig:rel_map}). Similarly, we can obtain all physical properties characterizing the stream as the median values among all configurations that select the representative particles.
For instance, some properties can be directly derived from the region of the parameter space with the higher density of detections, in particular in the space defined by the $N_{\rm min}-\sigma$ (as discussed in G+20, their \S~4 and Fig. 11). 

Finally, to increase the detection significance, we perturb the PN+GN velocity field.
We obtain 10 additional random realizations (i.e. 11 velocity fields in total), of the radial velocity field, each time assuming a Gaussian centered in the individual particle $V_{\rm rad}$ and with a mean velocity error of 37 km s$^{-1}$.
We use these random artificial velocity fields to check, for each stream detected in the unperturbed dataset, if additional detection configurations are allowed by the randomized data, and finally obtain the distribution of these configurations in the parameter space. This is shown in details in Appendix \ref{sec:reliability} where we report all configurations for the detected streams in the parameter space (see Fig.~\ref{fig:rel_maps_all_streams}). The detected streams derived with the above procedure are presented in the next section.

We remark here that we have excluded from our search the inner regions of the three galaxies as GCs and PNe catalogs are generally highly incomplete there because of the bright galaxy background. 
In particular, we have excluded the regions inside $5'$ from the NGC~1399 center and inside two effective radii from the other galaxies, i.e. $49''$ for NGC 1404 and $84''$ for NGC 1387.
Furthermore, we have also excluded all selected groups of particles that have one or more members overlapping these excluded area. 


\begin{figure*}
	\centering
	\hspace{-0.5cm}
    \includegraphics[width = \textwidth]{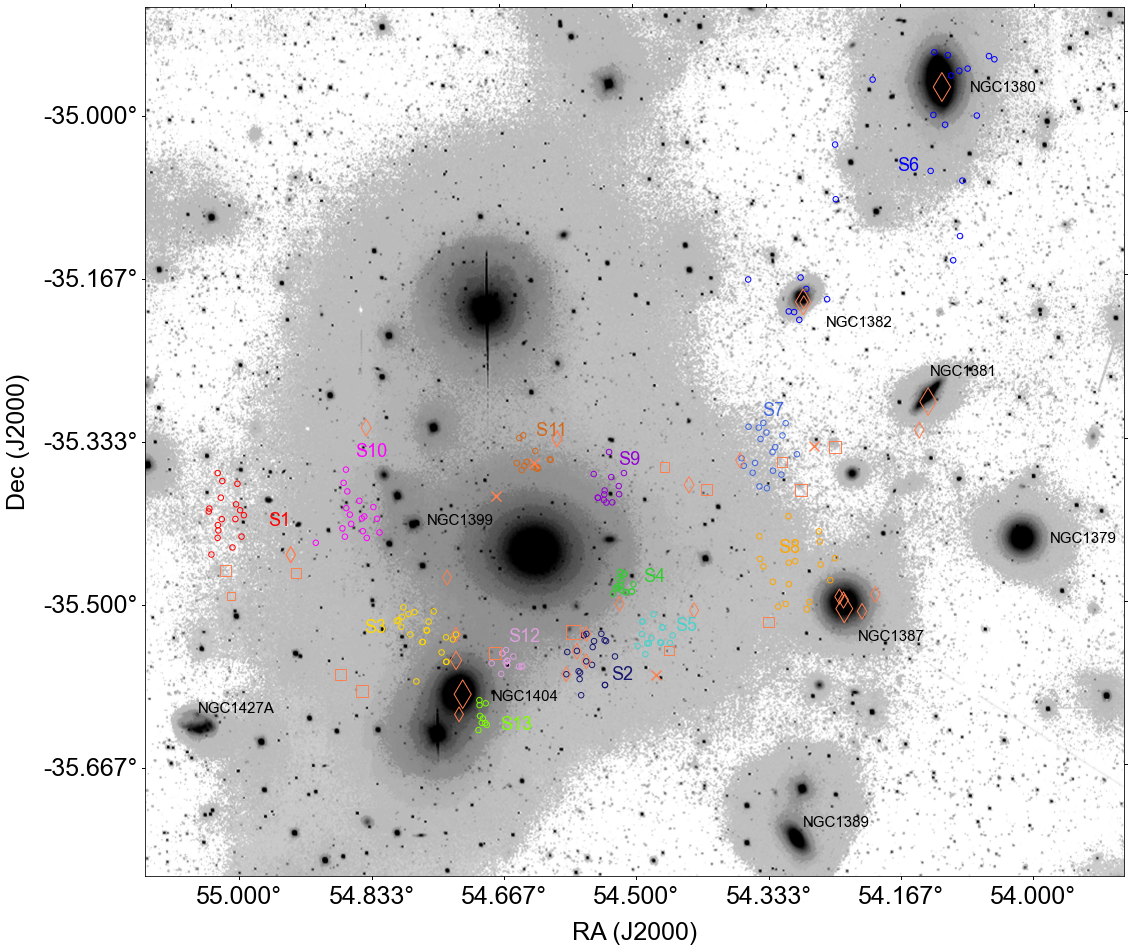}
    \caption{Image of the Fornax cluster core region. Representative particles of each cold substructure, found by COSTA {in the closest configuration to the median set-ups,} are 
    displayed, coloured accordingly {to } the {associated } stream. 
    Orange squares and diamonds indicate galaxies listed in Table~\ref{tab:spatial correlation} and Table~\ref{tab:velocity correlation} respectively,
    with the symbol size proportional to their $i-$band luminosities. Orange crosses represent galaxies {listed in the tables with no luminosity measurement. }
    }
    \label{fig:all_streams_image}
\end{figure*}

\section{Results}
\label{sec:Results}
In this section we present the candidate streams found in the {GC+PN} 
combined catalog. 
We have detected in total 13 stream candidates for which a detailed illustration of the COSTA parameters is given in Appendix \ref{sec:reliability}, where we show the different projections of the parameter space as compared with the reliability map. 

As anticipated in \S\ref{sec:runningCOSTA}, to visualize the most significant set-up configurations, we focus on a particular projection of the parameter space, the $N_{\rm min}-\sigma_{\rm cut}$, that we have {proven to be  }
representative of the stream physical quantities. Hence, we define the ``median'' configuration as the median of the parameters in the $N_{\rm min}-\sigma_{\rm cut}$ projected space, reported in Fig.~\ref{fig:sigma_nmin}, where we clearly see that set-ups {generally concentrate} 
in narrow regions in the projection space. 

\subsection{A catalog of cold substructures}
\label{sec:acatalog}


The stream candidates are reported in Table~\ref{tab:statistics_2}, where we list the stream ID, coordinates of the centroid of the stream, number of particles belonging to the stream (divided in GCs and PNe), {their} mean velocity, 
velocity dispersion, size (defined as the maximum distance among the particles) and distance from the cluster center. In the same Table, we also report the luminosity associated to the stream, based on the PN specific number density and the related surface brightness.
More details about the definition of each of these parameters will be given in the next Sections. 


Finally, in Table~\ref{tab:statistics_2} we also report the median and maximum reliability among all configurations 
in which COSTA detected the stream and the occurrence of the stream detection over the Montecarlo re-samplings of the GC+PN velocity field.
The reliability exceeds 70\% for 9 streams, and in particular two streams have a reliability greater than 85\%. 
The total number of particles ranges from 9 to 22, while their sizes are spread out over a quite large interval, which goes from 1.8 arcmin up to $\sim$20 arcmin.
Most of the streams have a comparable numbers of of GCs and PNe, while FVSS-S6 is around NGC~1380 where there is no spatial coverage of PNe (see Fig.~\ref{fig:gcs_pne}).

The position of the particles composing {each stream  are plotted in different colors }
in Fig.~\ref{fig:all_streams_image}, {overlapped on the deep $g-$band image of the core of Fornax cluster from FDS.} To visualize the stream particles we {choose }
the closest configuration to the median parameter set-up of each stream, as representative of the ``average stream''. 
In the same Figure we also report dwarf galaxies from literature \citep{munoz-2015, eigenthaler-2018}, which are in the vicinity of the streams and that can be likely associated to them (see details in \S\ref{sec:stream_descript}).
These are plotted as orange diamonds or squares (depending whether they have velocity measurements or not, respectively) with size proportional to the $i-$band total magnitude taken from \citet{cantiello-2020}. From this figure we see that streams are either very close to some of the detected dwarf galaxies (e.g. FVSS-S1, FVSS-S2, FVSS-S5, FVSS-S7, FVSS-S12) or live in the halo or are associated to larger galaxies in the Fornax core (e.g., FVSS-S3, FVSS-S4, FVSS-S6, FVSS-S8, FVSS-S13). {In addition, }
there are candidate streams that are particularly compact (e.g. S4, S12, S13) or rather diffuse (e.g. S8) or stretched (e.g. S6). A detailed discussion about the stream association with the galaxy population in the core of Fornax is reported in \S\ref{sec:discussion}.

\begin{figure*}
    \centering
    \includegraphics[scale=0.45]{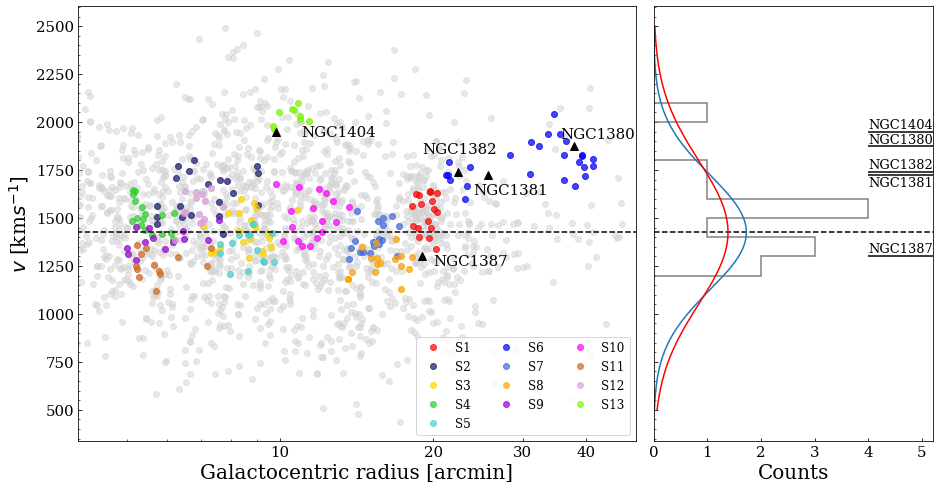}
    \caption{\emph{Left: }particles of each cold substructure, with same colours as in Fig.~\ref{fig:all_streams_image}, in the phase-space, overlapped to all particles used in this work (light gray). \emph{Right: }the blue and red solid lines are gaussians with mean = 1425 \kms, i.e. the NGC~1399 systemic velocity, and standard deviation of 300 \kms and 374 \kms, namely the velocity dispersion of the Fornax cluster and of the Fornax members analyzed by \citet{drinkwater-2001}. }
    \label{fig:phase_space_streams}
\end{figure*}

In Fig.~\ref{fig:phase_space_streams} we 
show the candidate streams in the phase space of all GCs and PNe in the Fornax core. 
The locations of the major galaxies in the area are also reported {in the same figure. Also here we can see notable}
differences in the stream typologies: 1) compact streams with small extensions (i.e. along the x-axis) and small dispersion, i.e. 
a compact distribution in radial velocity (y-axis), for instance FVSS-S4, FVSS-S11, FVSS-S12, FVSS-S13; 2) streams that are clearly related to individual large galaxies, e.g. S13 around NGC 1404 {and FVSS-S8 with NGC 1387}; 3) extended streams that seem to connect different large galaxies, e.g. S6 connecting NGC 1380, NGC 1381, NGC 1382; other streams still rather compact in velocity but more diffuse in radius, e.g. FVSS-S1, FVSS-S2, FVSS-S3, FVSS-S5, FVSS-S10.

Based on a visual inspection, there seem to be likely associations between the stream candidates and the position of galaxies of different size/luminosity (not only dwarf systems). In \S\ref{sec:discussion}, we will investigate in close details these possible associations, according to their vicinity and kinematical similarities. Here {we estimate important }
global physical properties of these candidate streams and {search for signatures} 
of their genuineness, {although we are aware that }
the final confirmation should come from direct observations with deep photometry to find evidence of 
stellar tails.

\subsection{Stream luminosity and surface brightness}
\label{sec:stream_lum}
For the streams including PNe, we can use some empirical formula to estimate the luminosity of the associated stellar population. In particular, as discussed in G+20, we can estimate the total light associated to the PN population using their specific number density within 2.5 mag from the cut-off magnitude of the luminosity function $\alpha_{2.5}=50\times10^{-9}$ PN$/L_\odot$ (\citealt{feldmeier2004}), which corresponds, on average, to the completeness limit of our sample (see Fig. \ref{fig:pnlf}). 
The total bolometric luminosity is obtained 
as $L_{\rm bol}=N_{\rm PN}/\alpha_{2.5}$. This can be converted in{to a} 
$g-$band luminosity, to compare the stream luminosity with the photometry from the Fornax core galaxies (e.g. I+16). 
We {estimate }
that, using a combination of F G and K stars, this amounts to about 1/3 of the flux in the optical range. Finally, considering an average 75\% completeness in the fraction of true recovered particles (see Fig.~14 of G+20) we can also correct the final luminosity by this factor. {The final, corrected 
$g-$band luminosities are listed in }
Table \ref{tab:statistics_2}, together with the corresponding surface brightness (SB), {computed} assuming a squared area with the side equal to the size {given} in the same table. 

The SB values we have obtained, always $>28$mag arcsec$^{-2}$, agree with typical predictions from cluster simulations: \citet[][C15 hereafter]{Cooper-2015a} found streams produced by disrupted galaxies covering a broad range of the galaxy luminosity function, including major galaxies.
However, typical surface mass densities of obvious features are in the range $10^5 -10^6$ M$_\odot$ kpc$^{-2}$, which correspond to $\sim27.5-30$ mag arcsec$^{-2}$ for the assumed mass-to-light ratios.
Such faint structures are hard to be seen even in very deep images. {As a matter of fact, }
a careful visual inspection of the deep images from FDS \citep[e.g.][]{venhola-2017} {and the }
Next Generation Fornax Survey \citep[e.g.][]{munoz-2015} has revealed no clear counterparts for our candidate streams, except in one case, the FVSS-S8 
({\it NGC 1387 stream} hereafter), which we have independently found with COSTA {but which} 
was previously 
claimed 
in I+16. As 
we will discuss later, this is the first kinematical confirmation of a photometric stream candidate. 

\subsection{Stream radial velocities as compared to other populations of the intracluster medium}
\label{sec:stream_ICpop}
Mean stream velocity and velocity dispersion {values} of all set-ups from different stream occurrences are computed respectively using {the following }
standard definition (see also P+18):
\begin{equation}
v_{\rm mean} = \frac{1}{N}  \sum v_i,~~~ 
\sigma_{I}^2 = \frac{1}{N-1}  \sum (v_i - v_{\rm sys})^2 - (\Delta v)^2,
\label{eq:std}
\end{equation}
where we {indicate } 
with $\sigma_I$ the intrinsic velocity dispersion, having subtracted the measurement errors $\Delta v=37$ kms$^{-1}$ in quadrature. The final median values of radial velocity and velocity dispersion are reported in 
Table~\ref{tab:statistics_2}. In this section we concentrate on the mean velocity estimates, while in the next section we will discuss the stream intrinsic velocity dispersion values.

The distribution of radial velocities of stream candidates is quite sparse and, in principle, can {give insights on }
the global kinematics of streams as member population of the cluster. {The mean velocity value of all the stream candidates is }
$1501\pm59$ kms$^{-1}$, which is consistent with the systemic velocity of NGC 1399 ($1425$ kms$^{-1}$), while their standard deviation is $213\pm42$ kms$^{-1}${. This value is }
smaller than the kinematics of other intracluster objects, which generally have a larger velocity dispersion, see  e.g.  cluster galaxies ($\sigma=374$ kms$^{-1}$, from \citealt{drinkwater-2001}), intracluster PNe ($\sigma=285$ kms$^{-1}$, from S+18) and GCs ($\sigma=285$ and $\sigma=304$ kms$^{-1}$ for red and blue ones respectively, from P+18), at $R\sim20'$\footnote{If we compare the velocity dispersion of the stream candidate population at their mean distance from the center, $13'\pm1.9'$, with the ones of the PNe and red and blue GCs at the same distance from S+18 and P+18, which are $312$ kms$^{-1}$, 312 kms$^{-1}$ and 361 kms$^{-1}$ respectively, we find the stream kinematics even more discrepant from the other intracluster populations.}. This indicates that, if real, streams possibly have a decoupled dynamics from other virialized populations. This would be the case if they have a peculiar density
or orbital distribution. One possibility is that these streams are produced by satellites that are preferentially placed on high eccentric orbits, hence possessing a strong radial anisotropy, different to other cluster member families. In this case streams should form in the pericenter of their trajectory around the central galaxy (see e.g. simple models from \citealt{longobardi-2018}) and, hence, have also a more centrally concentrated distribution with respect to other satellite systems (e.g. galaxies and intracluster GCs and PNe). 

Alternatively, the low velocity dispersion can also indicate that 
some 
stream candidates might be 
just a random extraction of the hot intracluster populations of tracers. 
In Appendix \ref{sec:spuriousness} we demonstrate that, if streams are a population of cluster members whose overall velocity dispersion should be comparable to that of other dynamical members (i.e. of the order of 300 \kms), the maximum number of spurious streams that might produce a ``dilution'' of their measured $\sigma$ down to $213$ \kms is $N_{\rm spur}\sim6$. 
In this worst case, more than half of the stream candidates (i.e. 7 over 13) are real, although in Appendix \ref{sec:spuriousness} we discuss why the fraction of real streams is very likely higher than that.


\subsection{Stream internal velocity dispersion}
{We focus here on the stream internal kinematics. The }velocity dispersion values show a wide distribution ranging from 35 kms$^{-1}$ (FVSS-S13) to 100 kms$^{-1}$ (FVSS-S2), while the mean value is 74 kms$^{-1}$. 
As seen in G+20, the accuracy of these estimates is hard to assess, as both incompleteness and contamination can alter the final {estimated velocity }dispersion value. 
{However, the estimates shown in }
Fig.~\ref{fig:sigma_nmin} fairly {take }
into account the statistical fluctuations as they come from stream detections from a large variety of set-up configurations (see \S\ref{sec:Results}). 
In G+20 we have also demonstrated that, even in case of significant contamination, the bias on the final stream dispersion estimates is confined within the statistical fluctuation. {This is because }
COSTA tends to collect only the particles that have a small scatter with respect to the intrinsic bulk kinematics of the stream (if the number of stream particles is dominant).
Also, the lowest velocity dispersion values are just nominal, as they are smaller than our measurement errors and have been obtained after subtracting the measurement errors in quadrature. Hence, for these ones we will assume 37 kms$^{-1}$ as an upper limit in the discussion hereafter.

In order to evaluate the dynamical range of the stream velocity and compare their internal kinematical structure with respect to their local cluster environment,
in Fig.~\ref{fig:phase_space_streams} 
we overplot the stream particles as reported in Fig.~\ref{fig:all_streams_image} to the total projected phase-space of the GC+PN system of the Fornax core.
First, stream particles show a velocity range which is rather colder (lower dispersion) than the underlying radial velocity distribution of the total PN+GC sample at the same radius. Also their mean velocities are confined well within the dynamical range allowed by the cluster potential. This is shown in the histogram reported in the same Fig.~\ref{fig:phase_space_streams},
where we also mark the systemic velocities of the large galaxies in the area and the velocity range corresponding to the maximum velocity dispersion measured by galaxies (red line) and the ICL (blue line), assuming a Gaussian distribution normalized to the number of streams. 
Some streams show a clear association to some of the giant galaxies (e.g. NGC 1404 and NGC 1387), in all other cases they should have some other association (see \S\ref{sec:stream_descript}).

We also do not see any chevron features, as one would expect from nearly shell-like orbits \citep[see e.g.][]{romanowsky-2012,longobardi-2015a}, 
but rather short sized substructures, as the ones detected in recent stripping events seen in hydrodynamical simulations (G+20). 
Overall, the kinematics of the streams shows that these structures are decoupled from the local potential (i.e. streams have a lower velocity dispersion with the respect to the particles at the same distance from the center), even though their mean velocities are well inside the dynamical range allowed by the cluster potential. This is compatible with the assumption that these candidate streams are tracing the kinematics of the interaction of parent dwarf galaxies and the overall cluster plus central galaxy potential, although they might not be in dynamical equilibrium in such potential (see \S\ref{sec:stream_ICpop}). 

\begin{figure*}
	\centering
	\hspace{-0.5cm}
    \includegraphics[width=18.5cm]{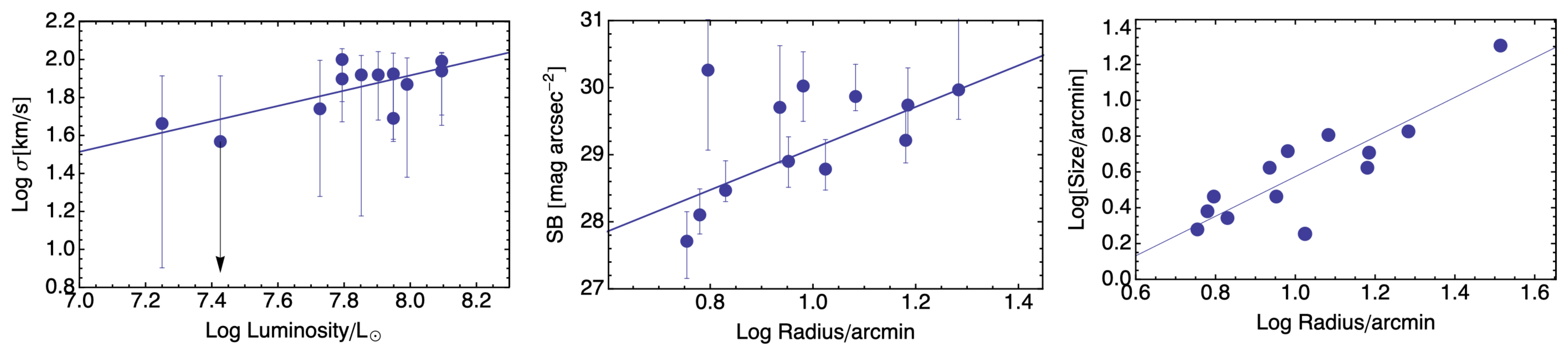}
    \caption{Correlation of the stream luminosity inferred by the PN specific number density with the estimated velocity dispersion (left) and the stream surface brightness (center) and the size of the stream (rigth) with distance from the cluster center.}
    \label{fig:stream_scaling}
\end{figure*}

\subsection{Correlations among Stream properties}
\label{sec:correlations}
Before describing in more details the properties of the individual streams and looking into their association with the dwarf population of the Fornax cluster, we discuss {here the correlations} 
among {some of } the parameters reported in Table~\ref{tab:statistics_2}. 

In {particular, in the left panel of } Fig.~\ref{fig:stream_scaling} 
we show the correlation {between } 
the stream luminosities {and the } 
inferred velocity dispersion. The (logarithm of {the}) velocity dispersion clearly increases with the stream (log) luminosity,
as also measured by the linear regression of the two correlations 
{shown as solid line} in the 
figure, for which we found a slope of $0.40\pm0.12$,
i.e. compatible with a non-zero correlation at $>3\sigma$ level, despite {the} large errorbars. This is confirmed by a high Spearman's rank correlation coefficient of $\rho = 0.76$, which corresponds to a significance of more than $99\%$.
We {find }
no significant correlation {between the stream surface brightness or the size, and its velocity dispersion. }
Similarly, we do not find a significant correlation between the stream luminosity and surface brightness and 
between luminosity 
and the distance from the cluster center. 
On the other hand, we find a significant correlation of both the stream surface brightness and stream size with the clustercentric distance, as shown in the {middle and right panels of } 
Fig.~\ref{fig:stream_scaling}{, respectively. } 
Here we also overplot the linear fit in log-log space for which {we obtain a }
slope of  
$3.09\pm1.19$ and $1.10\pm0.20$ respectively, i.e. both consistent with non-zero correlation at $>2.5\sigma$ significance. 
{We note that }
the correlation with the surface brightness seems weaker because of the presence of a larger scatter at small radii, especially because of stream 9. This has a $\mu_g=30.3$ mag/arcsec$^2$ {which is estimated on the basis of only }
2 PNe, making its value quite insecure (we will return on this stream later). The Spearman's rank correlation coefficient for the correlation between surface brightness and clustercentric distance is $\rho= 0.48$ and a significance $>90\%$; instead, if we remove stream 9, the correlation coefficient increases to $\rho = 0.73$, which corresponds to a significance of $>99\%$. For the correlation between stream size and clustercentric distance, $\rho = 0.74$ and it is significant at $> 99\%$.

These correlations, although based on stream candidates, 
suggest the presence of physical connections among the stream parameters.
However, we cannot {\it a priori} exclude that they can be the result of some selection effects. 
For instance, the correlations between the surface brightness and the size with the distance from the cluster center, can be the consequence of the decreasing density of tracers (see Fig. \ref{fig:histo_sample}). 
In particular, one can expect that streams of fewer particles would be more easily identified at larger distances from the center because of the lower density of the overall ICL population of tracers. This might explain the anti-correlation between the SB and the radius.
However, this cannot
explain why, at larger distances,
we do not find streams with larger particle numbers and smaller sizes, i.e. with higher SB, which would be even easier to find in a lower density environment.
This means that if no smaller sized, high-brightness streams are located in the bottom-right sides of both SB vs. Radius and Size vs. Radius panel in Fig. \ref{fig:stream_scaling}, it is likely because they do not exist. 

Another possible 
selection effect that could mimic a trend in the stream properties is the ``tracer completeness''. Due to the higher density of particles in the center, COSTA could collect a large number of contaminants than at larger distances. This might produce overall brighter streams in the centers (i.e. more contaminants per unit of real stream particle), eventually with a higher SB then the ones at larger distances, which is compatible what we see in the SB vs. Radius relation in Fig. \ref{fig:stream_scaling}. 
To be more quantitative, in G+20 (Fig. 14) we have conducted a series of tests to estimate, for simulated streams of different shape/size, the ``observed'' and ``true'' completeness (OC and TC respectively). The former is defined as the ratio of true stream members ($N_{\rm memb}$) over the total number of particles selected as stream ($N_{\rm tot}=N_{\rm memb}+N_{\rm cont}$) -- including contaminant particles, $N_{\rm cont}$ -- and the latter is defined as true selected members ($N_{\rm memb}$) over total true members ($N_{\rm str}$). With a little of algebra one can easily relate the total numbers of particles recovered by COSTA, $N_{\rm tot}$, to the OC and TC, by $N_{\rm tot}=$TC/OC$\times N_{\rm str}$. From Fig. 14 in G+20 we can see that TC/OC is strongly varying as a function of the parameter and the distance from the cluster center, but for 
$N_{\rm min}=15-25$, which are typical values for most of the 
the real streams (see Fig. \ref{fig:sigma_nmin}), 
the TC/OC goes from 0.9/0.7 in the inner bin, to 0.9/0.85 in the outermost bin, corresponding to a factor $\sim1.2$ ratio of the detected numbers from the center bin to the outer bin.
This means that 1) streams are detected at all distances, although 2) the number of the associate members can indeed change from the center to the outskirts (mainly due to a different contamination, see Fig. 13 of G+20). However the overall variation does not exceed a factor 1.2, considering the $N_{\rm min}$ involved (but can reach 1.6 if a wider $N_{\rm min}$ range is allowed). This is not sufficient to justify the 2 mag arcsec$^{-2}$ variation in SB seen in the middle panel of Fig. \ref{fig:stream_scaling}, where one would expect a factor $>6$ in numbers, if the overall size of streams is not increased significantly by the contaminants.


According to the arguments above, we can conclude that the correlations in Fig. \ref{fig:stream_scaling} are likely to be real. 
If so, we need to understand what is the physics behind them, in particular, if these correlations are compatible with being produced in 
the interaction between dwarf/intermediate galaxy systems and the environment (either larger galaxies or the cluster potential).
For instance, about the luminosity-velocity dispersion correlation, one can expect that the more massive a dwarf galaxy is, the more light is stripped during an interaction and the larger is the velocity dispersion of the particles that possibly still hold the memory of the dwarf internal kinematics. Note that this effect would be less efficient for more massive galaxies, which tend to loose less stellar mass by tidal stripping than dwarf-like systems (see \citealt{rudick-2009}).

On the other hand, the correlation of the surface brightness with the distance from the cluster center can be interpreted as the effect of dynamical friction. 
We can qualitatively understand that the higher is the density of the medium, the higher is the dynamical friction a satellite experiences. 
This can be illustrated by using a simplified formula for the dynamical drag: $F_d\propto G^2 M^2 \rho/v_M^2$ \citep{carroll-1996}. Here $G$ is the gravitational constant, $M$ is the mass of the satellite, $\rho$ is the density of the stellar medium the satellite is entering, and $v_M$ is the velocity of the satellite. This equation shows that a higher $\rho$ produces a stronger dynamical pull behind the intruder. This would produce more compact streams from satellite closer to the cluster center than further satellites of the same mass, hence impacting both on size and surface brightness of the candidate streams. $F_d$ also depends on $v_M$, which is larger for systems falling toward the centers, while it can be small at the pericenter, where $F_d\sim0$, hence generating more diffuse tails. 
For this reason we should expect some scatter added to the correlation by the proximity of the stream to the pericenter of the dwarf orbits, which might be larger toward the center, due to the intrinsic higher compactness of the streams. 
Since there is no clear variation of the luminosity function of dwarf galaxies as a function of the distance from the Fornax center \citep[see e.g.][]{venhola-2018}, then {\it we can expect that the correlation between stream size and distance should directly reflect the statistical effect of the dynamical friction as a function of the clustercentric radius}.

Is there some other physical mechanism able to make similar predictions?
Beside standard tidally stripped streams, C15 discussed the clumpy 2D distribution of stars associated with `sub-resolution' haloes that survive in the semi-analytic part of their simulations but not in the N-body part (their Fig. 2 panel 6). These latter possibly represents the remnants of disrupted galaxies that cannot be resolved with low-resolution dark-matter particles. These fragments, similar to the patchy distribution of some candidates we see in Fig. \ref{fig:all_streams_image}, are likely the product of the violent relaxation that might have involved progenitors of the bright central cluster galaxy (BCG). According to C15, these progenitors can have any mass, but more likely they are associated to low-mass dark haloes that are easily stripped below a total mass of 20 particles, corresponding to a $\sim2\times10^8 M_\odot$. 
Unfortunately, in C15 there is no radial velocity information, nor internal velocity dispersion of these clumps to check against the correlations we show in Fig. \ref{fig:stream_scaling}. From C15 Fig. 2 (panel 6) we notice that the densest knots are present at $R>100$ kpc from the center, possibly suggesting an anti-correlation of the SB with the radius. However, given the ``semi-analytic'' nature of their stellar particles we do not to want to over-interpret the projected distribution of these orphan stars. In fact, according to C15, the most conservative  interpretation is that they have to belong to the BCG halo component, with no detailed information of their actual geometry and properties. 

To conclude, we cannot exclude that some of our streams are the product of a violent relaxation involving massive progenitors of the BCG (see also \S\ref{sec:discussion}), although we should probably expect for these ones to show a larger velocity dispersion than the dwarf-like galaxies we are intrinsically selecting with COSTA ($<120$\kms). Hence, to finally test this scenario, we need more detailed predictions about the size, internal kinematics, morphology and frequency of these surviving structures, possibly from simulations more closely reproducing Fornax (e.g. the clusters in C15 are all about ten times richer or more).

\section{Discussion}
\label{sec:discussion}
In this section {we }
discuss the stream properties in more details and {investigate whether a connection exists }
between these properties 
and the ones of the galaxy population, in particular the dwarf-like systems, in their vicinity. This can help us to get more insight on the mechanisms that are contributing to the building-up of the intracluster stellar population in the Fornax cluster, as a prototype of a rather evolved galaxy cluster system with still ongoing galaxy transformation (see e.g. \citealt{2019A&A...628A...4R,2020A&A...640A.137R}). 

Especially in its core, recent observations have revealed signatures of interactions between the cD, NGC 1399, and other bright galaxies, e.g. overdensities in the photometrically selected GCs, \citet{dabrusco-2016}, C+20, \citet{Chaturvedi-2021}; faint $\mu_\textup{r} \sim 28-29$ mag/arcsec$^2$ and diffuse intra-cluster patches of light (\citealt{Iodice-2017}), which were predicted in earlier dynamical studies (\citealt{napolitano-2002}) and mirrored by asymmetries in the X-ray halo emission of NGC~1399 \citep{paolillo-2002,su-2017}. 



Tidal stripping of stars (including PNe) and GCs from the outskirt of galaxies in close passages through the cluster core is, indeed, predicted from N-body simulations \citep{rudick-2009}  
to provide the main mechanism 
for the origin of the ICL in Fornax.
However evidence of the tidal stripping origin of the ICL are all indirect and circumstantial. E.g.:  

1) the similarity between the fraction of the luminosity in the ICL with respect to the total light of cD ($\sim 5 \%$) and the fraction of blue GCs {over }
the total population of 
GCs ($\sim 4-6 \%$) in the same region of the ICL {seems to suggest that blue GCs are an }
intracluster population in the Fornax core (\citealt{Iodice-2017} but see also \citealt{bassino-2003} for further evidence of ICL in Fornax);

2) a lower GC specific frequency ($S_\textup{N} \sim 2$) with respect to typical values for cluster massive ellipticals ($S_\textup{N} \sim 5$) for the GC population of NGC 1404 supports the scenario of a tidal stripping of GCs in a close passage to the cD \citep[e.g.][]{bekki-2003}. 
Evidence of such interactions were found in \citet{napolitano-2002} using the velocity structure of PNe as kinematical tracers;

3) the presence of extended structure found {especially }
in the GC population (DA+16, C+20, but see also \citealt{spiniello-2018} for evidence of a cold structure in the PN population)
are some examples of indirect evidence supporting this scenario, 
but they do not provide a ``smocking gun'' proof.\\

Correlating the stream candidates found by COSTA with the parent galaxies from which they might have formed, would both provide direct evidence of tidal stripping and a proof that this is still ongoing and producing streams in the Fornax cluster.
To do that, we will take advantage of new collections of Fornax dwarf candidates 
\citep{munoz-2015, venhola-2017,eigenthaler-2018}, and recent spectroscopic compilation of objects (including normal galaxies, dwarf galaxies and UCDs) {in the }
Fornax cluster from \citet{maddox-2019}. This latter includes previous literature from \citet{ferguson-1989,Hilker-1999,drinkwater-2000a,drinkwater-2001,Mieske-2004,bergond-2007,Firth-2007,Firth-2008,Gregg-2009,schuberth-2010,Huchra-2012}. We finally crossmatch the catalog of new UCDs already classified in P+18 with the catalog from \citet{Saifollahi-2021} as a further check of the correct classification of the former.  
Associating the candidate streams to the properties of the dwarf galaxies is important: 1) because this is part of the validation of the stream candidates, since we have assumed that the COSTA algorithm can detect tails of streams particles recently lost by their parent dwarf systems, 2) because we expect that these recently stripped systems shall save the record of the kinematical properties of the parent galaxy (see e.g. \citealt{gatto-2020}) and, hence, tell us something about their internal dynamics\footnote{This latter aspect is beyond the purpose of the current paper, but will be the topic of forthcoming developments of this project.}, 3) because we can understand both the origin of the streams and the fate of the stripped (parent) systems.

\subsection{Close inspection of the stream candidates}
\label{sec:stream_descript}
We start by taking a closer look to the individual streams that are overplotted on the Fornax core image in Fig.~\ref{fig:all_streams_image} and try to relate them 
to the properties of the known galaxies around them. As mentioned above, our aim is to check whether there is a realistic association between the candidate streams and dwarf galaxies or any other galaxy sample in the Fornax core, to validate our working hypothesis. 

\textbf{FVSS-S1:}
It is the most recurrent stream (it is found with 542 set-ups) and has also a relatively high median reliability (76\%), with some set-ups {even} reaching the 100\%.
FVSS-S1 is made of 19 particles (i.e. by averaging the number of particles over all set-ups), {14 of which are PNe. }
This is also among the most luminous streams ($\log L/L_\sun=8.1$, $g-$band) but with a quite low SB ($\sim30$ mag/arcsec$^2$) due to its large size ($6.7'$).
{FVSS-S1 }
is placed at about 20\arcmin~ East from NGC~1399, in a region where \citet{Ordenes-Briceno2018} have identified a dwarf galaxy overdensity.
From the cross-match with \citet{maddox-2019}, {we find }
an ultra-compact dwarf (UCD, ID=F11747, with absolute magnitude in $i-$band, $M_i=-11.81$, see Table \ref{tab:velocity correlation}) 4.4$'$ away, with a systemic velocity of $1448$ \kms,  within $\sim2\sigma$ from the mean stream radial velocity (1500\kms). It is also closer ($2.8'$) to a slightly fainter dwarf system (NGFS 034003-352754 from \citealt{munoz-2015}, $M_i=-11.57$, see Table \ref{tab:spatial correlation}), which has an effective radius of 9.6$''$ but no systemic velocity measurement from literature to confirm {the }
association. 
{Both galaxies }
are located along the direction of the maximum elongation of the stream, suggesting {that } this can be the tail of one of the galaxies moving southward on the sky plane. From Fig.~\ref{fig:all_streams_image} the candidate stream seems to lie on a region of light excess, although this is just at the detection limit of the FDS images.
%
%
%

\textbf{FVSS-S2:}
This structure is located South-West of NGC~1399. It has an average median reliability of 74\%, with a maximum reliability of 100\%. It is made of $\sim$20 particles, of which 7 PNe, corresponding to a luminosity of $\log L/L_\sun=7.79$.
It has a relatively high velocity dispersion (100 \kms), while its radial velocity is more than 100 \kms\ higher than that of the BCG.  
An object classified as a bright ($M_i=-11.66$) GC from the \citet{maddox-2019} catalog (see Table~\ref{tab:velocity correlation}), located at less than 3$'$, has a systemic velocity (1490 \kms) within $\sim$ 2$\sigma$ from its mean velocity (1563 \kms). This is also classified as a UCD in Fornax in \citet{Gregg-2009}. If the stream is associated to this galaxy, the morphology seen in Fig.~\ref{fig:all_streams_image} is consistent with a stream stretched from a dwarf galaxy slowly falling toward the cluster center and seems to coincide with one of the overdensities found by \citet{cantiello-2020}. {We also see from Table 4 and Fig.~\ref{fig:all_streams_image}  that 2 GCs from the P+18 sample are also classified {as }
UCDs, although they are just close to the lower limit separating them from bright GCs.}
FVSS-SS2 has also two bright dwarfs at less than 5\arcmin~(see Table~\ref{tab:spatial correlation}) {away}, NGFS 033819-353151 with $M_i = -14.79$ and NGFS 033842-353308 with $M_i = -12.23$,
but the former has a $cz$=1725\kms, taken from the Simbad database\footnote{http://simbad.u-strasbg.fr/}, which makes it inconsistent with S2. Hence we keep NGFS 033842-353308 as a possible association in Table~\ref{tab:spatial correlation}.

\begin{figure*}
    \includegraphics[width=18.5cm]{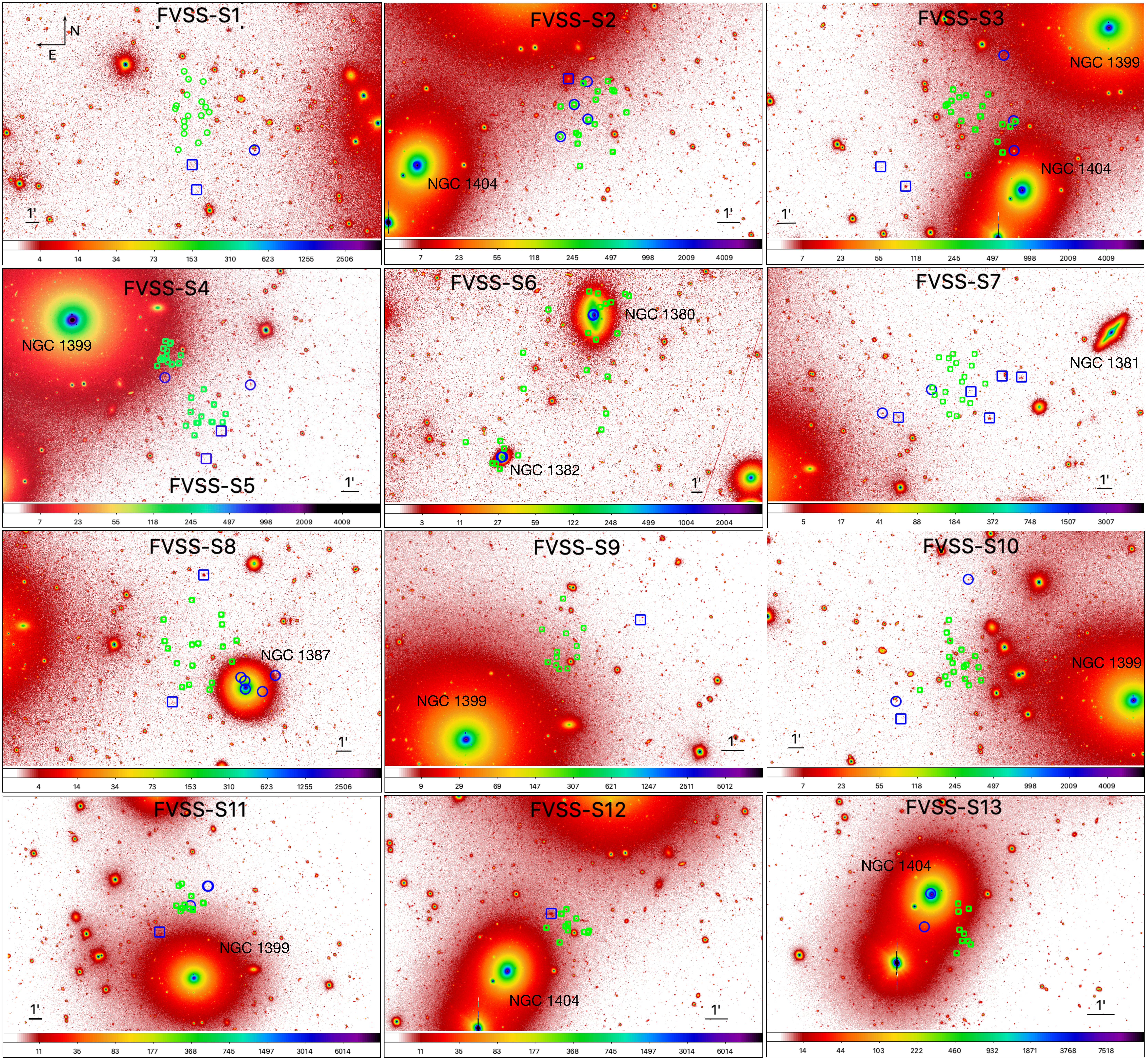}
    \hspace*{-1.cm}
    \caption{Zoom images of each cold substructure. Green circles indicate particles of the streams, whereas blue squares and circles represent all galaxies listed in Table~\ref{tab:spatial correlation} and Table~\ref{tab:velocity correlation}, respectively.}
    \label{fig:zoom_streams}
\end{figure*}

\textbf{FVSS-S3:}
This stream is the most recurrent after FVSS-S1, and it has a median reliability of 75\% with its maximum at 100\%.
It is composed by 20 particles, 8 of which are PNe, giving it a luminosity of $\log L/L_\sun=7.85$ and a SB of $\sim30$ mag/arcsec$^2$. The low SB does not make it possible to observe any counterpart in the deep imaging of the Fornax core. It is located very close to NGC~1404, but because of its mean radial velocity we exclude the possibility that it is associated to it (1392 \kms vs 1947 \kms).
{FVSS-S3 }
has two dwarfs within $\sim$ 5$'$ from its median centroid (see Table~\ref{tab:spatial correlation}), but none of them has a measured systemic velocity, and both look too displaced to be a convincing association. Other more likely associations are 2 UCDs from P+18 and 1 from S+21 within 4$'$ from the stream centroid,  and having radial velocities with 100 \kms\ from the stream mean velocity. Among these, one is rather bright ($I = 13.66$ $mag$, see Table~\ref{tab:velocity correlation}) and represents a convincing association. Fig.~\ref{fig:zoom_streams} (top panel) displays a zoom in the surrounding of FVSS-S3 overlapped also with nearby dwarf listed in Tables~\ref{tab:spatial correlation} and \ref{tab:velocity correlation} (yellow squares). The brightest UCD is right above NGC 1404 and is clearly visible in the image. This actually is the first UCD ever discovered (\citealt{Hilker-1999}, \citealt{drinkwater-2000a}), and is still the most massive confirmed UCD of the Fornax cluster (\citealt{Hilker-2007}). It has a small stellar envelope of ~100 pc size (\citealt{Evstigneeva-2008}) 
and very likely a SMBH (\citealt{Afanasiev-2018}).

\textbf{FVSS-S4:}
This structure has the highest median reliability, together with S6, and it is also very recurrent. 
Moreover, it is the closest to NGC~1399 and also very compact, resulting in a SB of 27.7 mag/arcsec$^2$. Despite this it is the candidate with the brightest surface brightness, its vicinity to NGC~1399 and NGC~1396 makes its luminous counterpart undetected also in the $g-$band image where the SB profile of the BGC has been subtracted.
The only possible association we have found is a faint UCD ($i-$mag$=20.4$, Table~\ref{tab:spatial correlation}) from P+18 with radial velocity $\sim1544$ \kms, which is consistent within $2\sigma$ with the mean velocity of the stream.

\textbf{FVSS-S5:}
This candidate stream occurs 80 times and has a median reliability of 68\% and a maximum reliability of 98\%.
It is located close to NGFS 033750-353302 \citep{munoz-2015}, which has a measured effective radius, ($R_\textup{e} = 3.2''$), a total magnitude of $M_i=-9.75$, but no measured systemic velocity in literature (Table~\ref{tab:spatial correlation}). 
It has also another dwarf placed at 2.5\arcmin (NGFS 033754-353429), but with no information about the size and luminosity 
(see also Table~\ref{tab:spatial correlation}). 
Another possible association is a UCD from P+18 at 2.6$'$ distance from the stream centroid, with $M_i=-11.3$ and radial velocity 1319\kms (Table~\ref{tab:velocity correlation}), very close to the one of the stream (1351\kms).

\textbf{FVSS-S6: discovery of a giant, undetected stream connecting NGC~1380, NGC~1381 and NGC~1382.}
This is a very recurrent structure with the highest median reliability of 85\%.  
It is made entirely by GCs, because there is no coverage of PNE observations, and  
shows a semi-ellipsoidal shape connecting NGC~1380 with NGC~1382 on one side, and possibly with NGC~1381 on the opposite side, suggesting that this stream might have been originated from one of these galaxies. 
Indeed, its mean radial velocity (1795 \kms) is considerably higher than that of
NGC~1399 (i.e. 1425 \kms) and very similar to the radial velocities of the three connected galaxies (1877 \kms, 1724 \kms\ and 1740 \kms, respectively for NGC~1380, NGC~1381 and NGC~1382, \citep[][]{vanderbeke-2011,drinkwater-2001,donofrio-1995}.
Since some of the GCs look still ``attached'' to the halo of NGC~1382, we argue this might be the progenitor. This is possibly supported by the fact that the stream has a velocity dispersion of $89_{-61}^{+19}$ \kms, which is fully compatible with the one of this galaxy (i.e. $77$ \kms, \citealt{wegner-2003}). 
We finally notice that the West side of the stream (connecting NGC~1382 and NGC~1380) has a mean velocity $\sim 1748 \kms$ 
(i.e. consistent with the systemic velocity of NGC~1382), while the ones on the East side has mean velocity $\sim 1849\kms$. Finally, the stream GCs  
around NGC~1380 have mean velocity $\sim 1786\kms$, hence showing a clear gradient compatible with an hyperbolic trajectory with the pericenter being located somewhere around NGC~1380 itself.

In the same region, there are other dwarf galaxies having a systemic velocity within 3$\sigma$ from the mean velocity of the stream (F08831, F08944, F16830, F16832, F15915, see Tab.~\ref{tab:spatial correlation}), making this region highly populated by systems with a significant velocity offset with respect to the cluster barycenter. 
However, the association with these systems is not as significant as the one with NGC~1382. 

\textbf{FVSS-S7:}
This structure is located north-west of NGC~1399 {and recurs 146 times.} 
As for S1, this stream also seems to lie on a region with light excess with respect to the surrounding regions. 
It has 5 dwarfs from the \citet{munoz-2015} sample at less than 5$'$ from its centre, of which the closest one is about 1$'$ away and has a low magnitude in the $I-$band $=-10.48$ -- see Table~\ref{tab:spatial correlation}. There are 
two systems with a comparable systemic velocity from the \citet{maddox-2019} catalog. The closest of these two  
(F15141) has been re-observed in S+21, while the other one (F10101) is classified as UCD from \citet{Gregg-2009} and it is also brighter ($I-$mag$=-12.24$) -- see Table~\ref{tab:velocity correlation}.

\textbf{FVSS-S8: kinematical confirmation of a photometric stream candidate.} 
This stream has a low occurrence (27) and 
a median reliability of $70\%$, however it overlaps to the photometric stream previously reported by \citet{dabrusco-2016} and \citet{Iodice-2016}. 
Hence, the independent finding from COSTA represents the first kinematical confirmation of one of the newly found substructures in the deep photomerty of Fornax cluster by FDS survey. 
{From the PNe luminosity function, we } estimate a luminosity of $\log L/L_\sun=7.9$ and a SB of $\sim29.7$ mag/arcsec$^2$, i.e. fully compatible with the estimates from deep FDS photometry (29$\leq \mu_g \leq 31$ mag arcsec$^{-2}$, \citealt{Iodice-2016}). 
Kinematically, the measurement of the mean stream velocity, 1319 kms$^{-1}$, is comparable with the systemic velocity of NGC~1387 (1302 kms$^{-1}$).  
However, the measured velocity dispersion is $84$ kms$^{-1}$, i.e. about half the central values found for NGC~1387 ($\sigma=170$\kms$^{-1}$, \citealt{wegner-2003}).  
{A possible explanation would be a steeply decreasing velocity dispersion profile towards the outskirts of the galaxy due to a low concentration dark matter halo \citep[see e.g.][]{napolitano-2009}.}
However, we cannot exclude that {the stream originates from a different }
dwarf-like system. {Indeed, there are }
2 catalogued systems in \citet{munoz-2015} within 5$'$, three matches with the \citet{maddox-2019} catalog within $6'$ and 4 UCDs from P+18, with compatible radial velocities (see Table~\ref{tab:velocity correlation}).
This could indeed explain the inconsistency of the stream colour with {these of }
NGC~1399 and NGC~1387, both redder than the average $g-i\sim0.9$ measured along the stream (FDS, \citealt{Iodice-2016}). Two {of the } UCDs in the area have {instead} $g-i\sim0.9$. 
If these latter have interacted with the halo of  NGC~1387, then the stream colour from the FDS images can be a mix of the stripped stars from NGC~1387 and the two UCDs.  

\textbf{FVSS-S9:}
This structure is one of the closest to NGC~1399. It is rather compact and it is made by an average of 14 members, almost all GCs (11 on average). 
FVSS-S9 has a low velocity dispersion (46 \kms), making it one of the coldest {stream candidates } of the COSTA catalogue.  
There is one object from \citet{munoz-2015} at about 3\arcmin~with a relative low absolute luminosity, i.e. $M_i=$-9.99, and classified as LSB galaxy, {but } with no systemic velocity measured. {FVSS-S9 }
is also in the proximity of the largest of the dwarf galaxies, NGC~1396, in the halo of NGC~1399, which {however} has a {much } 
smaller systemic velocity ($\sim 808$ \kms, \citealt{drinkwater-2001}) than the one estimated for FVSS-S9 ($\sim$1388 \kms), hence making {an association} unlikely. 
The mean velocity of this stream candidate, compatible with NGC~1399, and the vicinity to the central galaxy both suggest that this candidate stream might be produced by some yet undetected dwarf galaxy falling toward the cluster center (similarly to what seen for FVSS-S4). Its very low surface brightness ($\sim$ 30.3 mag/arcsec$^2$) makes this diffuse stream hard to detect in the deep images of the halo of NGC~1399.

\textbf{FVSS-S10:}
This stream has the lowest median reliability (61\%) and it occurs {only } 45 times. 
It is made of about 20 particles, {of which }
75\% are PNe. 
{FVSS-S10 } 
has a mean estimated velocity higher than the one of NGC~1399 (i.e. 1488 \kms) and its velocity dispersion is about 100 \kms. 
There is one catalogued dwarf at about 3$'$ (see Table~\ref{tab:spatial correlation}) with no information about its systemic velocity. 
The associations to this is rather insecure and the visual inspection of the FDS images shows no evident counterpart because of the intrinsic faint SB of the stream ($\sim29.9$ mag/arcsec$^2$, see Table \ref{tab:statistics_2}) and the presence of bright stars in the West direction.   
On the other hand, there are 2 convincing associations to the \citet{maddox-2019} catalog (Table~\ref{tab:velocity correlation}) with compatible velocities within $2\sigma$,  and located at the opposite sides of the stream. One of the two (F11747) is classified as UCD, the other is a bright GC ($I-$mag$ =-12.48$, likely a misclassified UCD).

\textbf{FVSS-S11:}
This is another stream inhabiting the halo regions of NGC~1399, but occurring {only} 
few times (23). 
Like other similar substructures close to the cluster centre (e.g. S4 and S9), it is very compact. {FVSS-S11} 
has an average radial velocity about 200 \kms~lower than NGC~1399. It has two compact objects with a systemic velocity within 2$\sigma$ (Table~\ref{tab:velocity correlation}), of which the most convincing one has no luminosity information, and a further dwarf at about 3$'$ with no velocity and no luminosity information (Table~\ref{tab:spatial correlation}).
Akin the other compact structures in the strict halo of the central galaxy, these streams possibly form a family whose progenitors have been gone through a more dramatic transformation, {maybe }
because caught in the final passages before being fully merged in the BCG. The possible association with UCD systems for this stream, and other ones of the same kind, is rather interesting and suggests that these can be the last phase of the disruption of former dwarf galaxies having left the compact core only surviving (see \S\ref{sec:streams_UCDs}).

\textbf{FVSS-S12:}
This candidate stream is spatially very close to NGC~1404, is very compact (2.2$'$) and relativ{ely} cold (55 \kms). {FVSS-S12 }
is made of 14 particles almost {equally } split between GCs (8) and PNe (6), and it has a median reliability of 69\%. Being the difference of the mean radial velocity of its particles about 400 \kms lower than the systemic velocity of NGC~1404, but rather close to the one of NGC~1399, {FVSS-S12 }
is more likely associated to this latter. For its compactness and position, this stream seems to join the family of the halo streams including also S4, S9 and S11, but has no clear associations.
We finally remark that this stream is very close in position and radial velocity to FVSS-S12, with which for some parameter combinations it has a few particles in common. This hints for a connection between the two substructures that, however, COSTA did not find to be significant (in the step-{v} described in \S\ref{sec:steps})

\textbf{FVSS-S13:}
This is the most compact stream, placed right to the West of NGC~1404{, but only occurs 27 times}. 
The systemic velocity suggests a connection with NGC~1404 (2046 \kms vs 1947 \kms). 
No dwarfs have been discovered in its proximity. However,  there is a UCD from P+18 (confirmed by S+21) at {a distance of} $1.4'$, {with a similarly high 
radial velocity (2000 \kms),}  
which is possibly interacting with the NGC~1404 halo and getting material stripped.
\subsection{Summary of the stream-galaxy associations}
We conclude this section by summarizing the results on the stream-galaxy associations. As mentioned before, the difficulty to find a clear luminous counterpart of the stream candidates found by COSTA in the Fornax core makes the stream catalog in Table~\ref{tab:statistics_2} unsecure. On the other hand, convincing indications that streams have associations with progenitor candidates allow us to make a step forward {in } 
understanding 
their origin. {This can also trigger }
further investigations {and new observations }(e.g. dedicated deeper photometry) to look for the integrated light signature of these streams. 

About the origin of the streams and the connection with Fornax systems we have found, in the previous section, a series of evidence that we summarize here. 
We stress though that a more physical interpretation of these associations is beyond the scopes of this paper and will be addressed in a forthcoming work (Napolitano et al. in preparation).

\subsubsection{Streams associated to normal galaxies} 
FVSS-S6 is likely connected to NGC~1382 while FVSS-S8 is connected to NGC~1387. 
In particular, FVSS-S6 looks similar to the coherent streams of tidally stripped material from large galaxies found in C15 (Fig. 2 panel 5).
For the first time, we bring evidence of a
kinematical link of these streams with parent galaxies, due to the similarity of the mean velocity of the stream tracers and {their associated systems. }
However, while the velocity dispersion of FVSS-S6 ($\sigma=89 \kms$) is fully compatible with the one of the NGC 1382 from stellar spectroscopy (77 \kms, average from Simbad database),
for FVSS-S8,  
the stream {velocity dispersion value, }$\sigma=84 \kms$, is {much smaller }
that the one of the internal galaxy kinematics ($\sigma_{*}=167 \kms$). We have also discussed, in \S\ref{sec:stream_descript}, that being the $g-i$ colour ($=0.9$) of the stream bluer than the NGC~1387 
outskirts,  there is a chance that the stream GC+PN population is a mix of particles coming from NGC~1387 and the bluer UCD (F09174 and F09244 in Table \ref{tab:velocity correlation}) found in the proximity of the stream.  

We conclude this section by noting that, in general, the velocity dispersion values estimated for the streams in Table \ref{tab:statistics_2}, i.e. of the order of $70-90$ \kms (after the subtraction of the measurement errors), are typical of galaxies of absolute magnitude of the order of $V=-19.5$, and total luminosity $L_{\rm T}\sim5\times10^9L_{\sun,V}$, according to the Faber--Jackson relation (FJR, e.g. \citealt{2011MNRAS.417.1787F}). Looking at the typical luminosities of the streams also reported in Table \ref{tab:statistics_2}, we see that these latter would correspond on average to $\sim$3-10\% of the $L_T$ expected by the Faber-Jackson. In other world, most of the streams would be compatible with the stripping of a minimal amount of stars from these intermediate luminosity systems (e.g. \citealt{rudick-2009, Cooper-2015a}). However we could not find these associations for most of the streams above, not even at reasonably large distances. This leaves us with the only option that these streams might be still produced by dwarf-like systems with mid-sized velocity dispersions, but that these have heavily transformed and did not survive as recognizable dwarf systems (see also \S\ref{sec:streams_no_assoc}).    

 
\subsubsection{Streams associated to UCDs}
\label{sec:streams_UCDs}
Streams FVSS-S1, FVSS-S2, FVSS-S3, FVSS-S4, FVSS-S5, FVSS-S7, FVSS-S10, FVSS-S11, and FVSS-S13 have, instead, a kinematical link (i.e. radial velocity within $\sim2\sigma$) to systems classified as compact objects/UCDs (see Table~\ref{tab:velocity correlation}). 
For all these streams (except S11) the estimated velocity dispersion,  
$\sim70-90$ \kms, is higher than the one expected from a typical FJR of UCDs \citep[see e.g.][]{mieske-2008,penny-2014}. {However,} 
the errorbars are so large that all of them 
are consistent within $1\sigma$ with the typical velocity dispersions of UCDs on the FJR ($\sim40$ \kms at $M_i=-12$). 

\subsubsection{Streams with no clear associations}
\label{sec:streams_no_assoc}
There are two remaining streams, FVSS-S9, and FVSS-S12 that miss a clear kinematical association (see Table \ref{tab:velocity correlation}).
As such, these streams may remain dubious and listed among the possible spurious streams foreseen in \S\ref{sec:stream_ICpop}. 
However, as discussed in \S\ref{sec:correlations}, clumps of stripped stars apparently dissociated from visible stellar systems are predicted from simulations (C15) as a result of violent relaxation due to the interaction of  galaxies merging in the halo region of the BCG. Among the stellar streams in Table \ref{tab:statistics_2}, the two streams with no clear associations are the most likely candidates to belong to this family. This does not exclude that some other streams associated to UCDs may be remnants of disrupted galaxies of different mass. For instance, looking at Fig.~\ref{fig:zoom_streams}, FVSS-S10 which is not that close to its tentative UCD association or FVSS-S13, in the halo of NGC 1404, could be good examples. The former has a velocity dispersion of $\sim90$ \kms, after the measurement errors subtraction, the latter 35 \kms, likely produced by the disruption of a rather bright galaxy, $V\sim-19.5$, and a faint dwarf, $V\sim-16.0$,  respectively.

\begin{table*}
    \centering
    \caption{Distance between candidate streams and their closest dwarfs taken from \citet{munoz-2015} and \citet{eigenthaler-2018}.}
    \label{tab:spatial correlation}
    \begin{tabular}{c|c|c|c|c|c|c|c}
      ID & ID dwarf & RA & Dec & Dist. & $i-$mag & $I-$mag &  Comments\\
         & (NGFS) & (J2000) & (J2000) & ($\arcmin$) & (mag) & (mag) & \\
    \hline
FVSS-S1 & 034002-352930 & 55.0072 & -35.4916 & 4.4 & 22.80 & -8.71 &   \\
FVSS-S1 & 034003-352754 & 55.0130 & -35.4651 & 2.8 & 19.94 & -11.57 & 1 (LSB)\\
FVSS-S2 & 033819-353151 & 54.5782 & -35.5309 & 3.6 & 16.72 & -14.79 & $cz=$1725; LEDA 74806; LSB\\
FVSS-S3 & 033922-353524 & 54.8436 & -35.5900 & 4.4 & 19.33 & -12.18  & 2 (LSB)\\
FVSS-S3 & 033929-353421 & 54.8705 & -35.5726 & 4.6 & 19.92 & -11.59  & 1 (LSB)\\
FVSS-S5 & 033750-353302 & 54.4578 & -35.5504 & 1.1 & 21.76 & -9.75  &  2 (LSB)\\
FVSS-S5 & 033754-353429 & 54.4744 & -35.5748 & 2.5 &  & \\ 
FVSS-S7 & 033700-352035 & 54.2490 & -35.3430 & 4.3 & 18.48 & -13.03 &  1 (LSB) \\
FVSS-S7 & 033706-352031 & 54.2750 & -35.3420 & 3.0 & & & 2 (LSB) \\
FVSS-S7 & 033710-352312 & 54.2915 & -35.3867 & 3.1 & 18.65 & -12.86 &  FCC 191; 1 (LSB) \\
FVSS-S7 & 033716-352130 & 54.3154 & -35.3582 & 1.1 & 21.03 & -10.48 &  2 (LSB) \\
FVSS-S7 & 033738-352308 & 54.4104 & -35.3855 & 4.2 & 20.00 & -11.51 &  2 (LSB) \\
FVSS-S8 & 033710-352312 & 54.2915 & -35.3867 & 4.8 & 18.65 & -12.86 & FCC 191; 1 (LSB) \\
FVSS-S8 & 033720-353118 & 54.3328 & -35.5216 & 3.6 & 20.63 & -10.88 &  \\
FVSS-S9 & 033751-352146 & 54.4621 & -35.3629 & 3.2 & 21.52 & -9.99 & 2 (LSB) \\
FVSS-S10 & 033942-352806 & 54.9250 & -35.4684 & 6.4 & 20.87 & -10.64 & 2 (LSB)  \\
FVSS-S11 & 033842-352329 & 54.6737 & -35.3915 & 3.4 & & & 2 (LSB) \\
FVSS-S12 & 033842-353308 & 54.6759 & -35.5521 & 1.0 & 19.28 & -12.23 & FCC B1281; LSB\\

\hline
    
    \end{tabular}
\vspace{0.5cm}    

\begin{minipage}[b]{180mm}
    \textbf{References -}: 
    \textbf{1: } \citet{venhola-2017}
    \textbf{2: } \citet{Mieske-2007}
\end{minipage}

\end{table*}

\begin{table*}
\centering
    \caption{Crossmatch with \citet{maddox-2019}, \citet{pota-2018} and \citet{Saifollahi-2021}. }
    \label{tab:velocity correlation}
    \hspace*{-1cm}

    \begin{tabular}{c|c|c|c|c|c|c|c|c|c}
      ID & ID dwarf & RA & Dec & Dist. & Vel. D & Vel. S & $i-$mag & $I-$mag & References (Comments)\\
        & & (J2000) & (J2000) & ($\arcmin$) & (\kms) & (\kms) & (mag) & (mag) & \\
    \hline
FVSS-S1 & F11747 & 54.9315 & -35.4499 & 4.4 & 1448 & $1500_{-30}^{+24}$ & 19.70 & -11.81 & 1, 4 (UC), 9 (UCD)\\
FVSS-S2 & F10648 & 54.5734 & -35.5508 & 2.4 & 1490 & $1563_{-34}^{+39}$ & 19.85 & -11.66 & 1, 7 (UCD), 8, 9 (UCD)\\
FVSS-S2 & & 54.5613 & -35.533 & 3.6 & 1504 & $1563_{-34}^{+39}$ & 20.47 & -11.04 & 6 (UCD)\\
FVSS-S2 & & 54.5868 & -35.5741 & 1.0 & 1524 & $1563_{-34}^{+39}$ & 20.44 & -11.07 & 6 (UCD)\\
FVSS-S2 & & 54.5617 & -35.561 & 2.0 & 1611 & $1563_{-34}^{+39}$ &  20.43 & -11.08 & 6 (UCD)\\
FVSS-S3 & & 54.7251 & -35.5327 & 3.1 & 1487 & $1392_{-68}^{+37}$ & 20.47 & -11.04 & 6 (UCD)\\
FVSS-S3 & & 54.7357 & -35.4746 & 4.3 & 1382 & $1392_{-68}^{+37}$ & 20.47 & -11.04 & 7 (UCD)\\ 
FVSS-S3 & & 54.7251 & -35.5592 & 3.5 & 1491 & $1392_{-68}^{+37}$ & 17.85 & -13.66 & 6 (UCD)\\
FVSS-S4 & & 54.5194 & -35.5023 & 1.1 & 1544 & $1473_{-29}^{+33}$ & 20.4 & -11.11 & 6 (UCD)\\
FVSS-S5 & & 54.426 & -35.5094 & 2.6 & 1319 & $1351_{-19}^{+42}$ & 20.2 & -11.31 & 6 (UCD)\\
FVSS-S6 & F16825 & 54.1146 & -34.9753 & 5.5 & 1827 & $1795_{-25}^{+30}$ & 10.55 & -20.96 & 3, Simbad (NGC1380 - 1877 \kms) \\
FVSS-S6 & F15896 & 54.1321 & -35.2967 & 14.1 & 1763 & $1795_{-25}^{+30}$ & 11.45 & -20.06 & 2, Simbad (NGC1381 - 1724 \kms)\\
FVSS-S6 & F08944 & 54.1432 & -35.3257 & 15.9 & 1817 & $1795_{-25}^{+30}$ & 19.78 & -11.73 & 1\\
FVSS-S6 & F15915 & 54.2875 & -35.1950 & 10.3 & 1740 & $1795_{-25}^{+30}$ & 21.68 & -9.83 & 2, Simbad (NGC1382 - 1740 \kms)\\
FVSS-S6 & F09556 & 54.2882 & -35.1946 & 10.3 & 1771 & $1795_{-25}^{+30}$ & 12.75 & -18.76 & 1 \\
FVSS-S7 & F10101 & 54.4317 & -35.3811 & 5.0 & 1326 & $1391_{-30}^{+31}$ & 19.27 & -12.24 & 1, 8\\
FVSS-S7 & & 54.3675 & -35.3563 & 1.6 & 1373 & $1391_{-30}^{+31}$ & 20.11 & -11.40 & 7 (UCD)\\
FVSS-S8 & F09174 & 54.1985 & -35.4936 & 5.6 & 1373 & $1319_{-65}^{+30}$ & 19.77 & -11.74 & 1\\
FVSS-S8 & F09244 & 54.2153 & -35.5108 & 5.3 & 1375 & $1319_{-65}^{+30}$ & 19.90 & -11.61 & 1 \\
FVSS-S8 & F15912 & 54.2375 & -35.5081 & 4.3 & 1257 & $1319_{-65}^{+30}$ & 11.00  & -20.51 & 2, Simbad (NGC1387 - 1302 \kms)\\
FVSS-S8 & F16897 & 54.2437 & -35.4958 & 3.6 & 1379 & $1319_{-65}^{+30}$ & 20.35 & -11.16 & 4 (UC)\\
FVSS-S8 & & 54.2385 & -35.4992 & 4.0 & 1246 & $1319_{-65}^{+30}$ & 19.38 & -12.13 & 6 (UCD)\\
FVSS-S10 & F11453 & 54.8357 & -35.3207 & 3.6 & 1420 & $1488_{-60}^{+41}$ & 19.03 & -12.48 & 1, 10\\
FVSS-S10 & F11747 & 54.9315 & -35.4499 & 5.7 & 1448 & $1488_{-60}^{+41}$ & 19.70 & -11.81 & 1, 4 (UC), 9 (UCD)\\
FVSS-S11 & F10736 & 54.5970 & -35.3336 & 1.8 & 1370 & $1251_{-13}^{+102}$ & 19.54 & -11.97 & 1, 10 (Bright GC)\\
FVSS-S11 & F17231 & 54.6258 & -35.3583 & 0.5 & 1402 & $1251_{-13}^{+102}$ & & & 5, 9 (UCD)\\
FVSS-S11 & & 54.5968 & -35.3335 & 1.8 & 1309 & $1251_{-13}^{+102}$ & 19.55 & -11.96 & 6 (UCD)\\
FVSS-S13 & & 54.7219 & -35.6144 & 1.4 & 2000 & $2046_{-27}^{+13}$ & 20.49 & -11.02 & 6 (UCD)\\
FVSS-S13 &  & 54.7171 & -35.5938 & 1.8 & 1948 & $2046_{-27}^{+13}$ & 10.75 & -20.76 & Simbad (NGC1404 - 1942 \kms)\\
\hline

    \end{tabular}

\vspace{0.5cm}    
\begin{minipage}[b]{180mm}
    \textbf{Note}:
    for \citet{maddox-2019} we retrieve 
    all sources with a radial estimated velocity within 3$\sigma$ the average stream velocity. For \citet{pota-2018} and \citet{Saifollahi-2021} we took only sources with a radial estimated velocity within 50 \kms~the average stream velocity. We adopted as a maximum distance $Dist =$ {\sc SIZE} + 1\arcmin~ from the stream centre coordinates for all the three crossmatches. 
    Magnitudes in the {\emph i-}band come from 
    \citet{cantiello-2020}.\\
    \textbf{References -}
    \textbf{1: } Drinkwater 2dF. 
    \textbf{2: } Drinkwater FLAIR II. 
    \textbf{3: } \citet{ferguson-1989}.
    \textbf{4: } \citet{bergond-2007}. 
    \textbf{5: } \citet{Mieske-2004}.
    \textbf{6: } P+18. 
    \textbf{7: } S+2021. 
    \textbf{8: } D+2021.
    \textbf{9: } G+2009. 
    \textbf{10: } S+10. Brigh GC
\end{minipage}

    
\end{table*}

\section{Conclusions}
\label{sec:conclusions}
The census of streams around the brightest cluster galaxies is a key diagnostic to understand the origin and assembly of the large galaxy haloes around galaxies in  dense environment and the building-up of the intracluster stellar population. 
Streams have been predicted in simulations and lately {also} directly detected in nearby universe \citep{napolitano-2003,cooper-2010,romanowsky-2012,foster-2014,longobardi-2015a}. 
However, without the kinematical information it is hard to connect the streams with the galaxy formation mechanisms and, ultimately, fully confirm the predictions about their evolution. {For instance, }
one firm prediction of the galaxy formation scenario is that dwarf galaxies, orbiting around the cluster centers, experience tidal interaction with the potential of the cluster and the galaxies orbiting in it. After many passages, these dwarf galaxies loose most of their stellar mass that is accreted in the intracluster medium to form a diffuse component \citep{rudick-2009}. 
Remnants of these disruption processes can possibly produce some family of observed galaxy populations in the cluster cores, like UCDs that have lately been recognized to be an abundant population in the cluster cores \citep[see e.g.][]{Janssens-2019}.  
Hence, streams {carry a wealth of information }
about the transformation of the dwarf systems and how much of {their baryoninc material }
can be fully disrupted, hence affecting the luminosity function in cluster cores, with respect to the lower density environments.


In order to explore all these avenues, the first step is to find a robust approach to detect such substrucures. Due to their low surface brightness (of the order of 29 mag arcsec$^{-2}$), streams are hard to be detected with deep photometry \citep{cooper-2010},
although 
bright streams have been seen in imaging surveys \citep[see e.g.][]{Mihos-2005,Martinez-Delgado-2010,Iodice-2017,Montes-2019}. 
The next, should be their kinematical characterization.
There have been few works being able to kinematically characterize faint streams \citep[e.g.][]{merrett-2003},
including some preliminary experiments 
making use of GCs \citep{romanowsky-2012} and PNe \citep{longobardi-2015a} in galaxy haloes.

In this paper, we have used a sample of about 2000 objects accurately selected from the Fornax VLT Spectroscopical Survey (FVSS) from \citet{pota-2018} and \citet{spiniello-2018} in order to have a comparable sample in terms of number density and depth from the two types of tracers. We have also checked that their velocities and spatial distributions were statistically similar (at more than 5\% significance, see \S\ref{sec:datasets}), to be used together to extract candidate streams with our new COld STream finding Algorithm \citep[COSTA,][]{gatto-2020}. 

We have tested the algorithm of a simulated reduced phase space of the Fornax core where there is the major contribution of at least three large galaxies, NGC~1399, NGC~1387 and NGC~1404, which, with their local velocity field, complicate the overall phase space structure. 
We have 
obtained the first catalog of 13 stellar stream {candidates},  selected in the intracluster regions of the Fornax cluster (Table \ref{tab:statistics_2}).  For these streams we provided mean centroid position (RA and DEC), number of PN and GC particles forming them, reliability, radial velocity (i.e. mean velocity of the PN+GC particles composing them), velocity dispersion (as the standard deviation of the PN+GC particles composing them), size, total luminosity (as derived by the number of PNe multiplied by an average plausible PN specific number density, see \S\ref{sec:stream_lum}) and surface brightness. 
Based on statistical arguments (see \S\ref{sec:stream_ICpop}) we have predicted that more than half of these are likely to be real. This is the main result of the paper, which aims at providing a first attempt to select cold substructures in a cluster environment. 
Here below the main conclusions we drew from this catalog:
\begin{itemize}
    \item the detected streams have been found at different distance from the cluster center, with a radial velocity distribution which follows the one of other cluster members (e.g. galaxies and IC PNe and GCs) but with  a smaller velocity dispersion. This has been explained either with the possible presence of spurious streams or with a non-equilibrium of the stream population within the cluster;
    
    \item the estimates of the streams surface brightness vary between $\sim28$ and $\sim30$ mag arcsec$^{-1}$, too faint to be deblended from the diffuse light distribution in the core of the Fornax cluster with current imaging data. Deeper, more targeted observations are needed to obtain a photometric confirmation;
    
    \item the streams internal velocity dispersion values vary between $\sim35$ \kms and $\sim100$ \kms, in line with typical velocity dispersions of dwarf-like galaxies, very likely their main progenitor systems;
    
    \item based on the correlations between their luminosity and velocity dispersion and the size and surface brightness as a function of the distance from the cluster center, we have discussed in \S\ref{sec:correlations} that streams show signatures of dynamical friction, i.e. they are produced, as expected, by tidal stripping. However we could not exclude that other physical mechanisms are also involved in the formation of (a part of) the observed streams. For instance, violent relaxation of galaxies, up to intermediate luminosity, that have merged into the halo of the central galaxy (see e.g. C15);
    
    \item we have discovered a new giant stream (FVSS-S6) {likely} connecting NGC~1380
    and NGC~1382. In particular,  we {have suggested that this stream is }
    produced by the galaxy NGC~1382 (with which it shares a compatible velocity dispersion) having performed an hyperbolic orbit around NGC~1380;
    
    \item we have independently detected the stream FVSS-S8, {photometrically observed in }
    \citet{Iodice-2016} and associated to NGC~1387, and estimated for this a velocity dispersion which is about half of the one of the galaxy. Due to the compatibility of the radial velocity of this stream with two blue ($g-i\sim0.9$) UCDs in the proximity of NGC~1387, we suggested that this stream might be 
    formed by a mix of particles stripped by both NGC~1387 and the two satellite systems;
    
    \item  
we have demonstrated kinematical links of most of the streams with UCDs inhabiting the Fornax cluster core. Within the large uncertainties of their velocity dispersion estimates, most of the streams are compatible with the velocity dispersion {values } expected for the UCDs they are seemingly connected with.  However, part of the excess of velocity dispersion can be reconciled with a dwarf--like Faber-Jackson relation if we assume that these UCDs are the remnants of more luminous galaxies having lost a significant fraction of their initial luminous mass and, for that, have been moved away from the original FJR. 
This would be compatible with the formation scenario of the UCD galaxies as remnants of nucleated dwarf spheroidal (see e.g. \citealt{2013MNRAS.433.1997P}). 
More evidence of a connection of the physical mechanism (the tidal stripping) and the transient effect (the stream) producing the transformation of the dwarf galaxies into UCDs will be discussed in a forthcoming paper (Napolitano et al. in preparation).

\item Two stream candidates are lacking a clear close physical connection (either in position or velocity) with dwarf-like systems. As such, they can be either spurious detections or, more interstingly, they can represent a class of streams from disrupted progenitors merged in the halo of the central galaxies. These are expected to produce orphan clumps of stellar tracers, with no clear parent systems, very similar to the one we have observed (e.g. FVSS-S9 and FVSS-S12). However, more accurate predictions on the morphology, photometry, and kinematics of these family of streams are needed to draw a clear comparison with our observational findings. In this respect, higher resolution N-body+tagging particles or hydrodynamical simulations would be very desirable to better resolve and characterize these substructures.


\end{itemize}

In a forthcoming analysis, we plan to look more in the last two families of streams described here above. In particular we will investigate in detail the FJR of the streams for which we have a clear galaxy association with UCDs and draw more quantitative conclusions about their origin. Furthermore, we plan to search for more candidates of these ``orphan clumps'' predicted in C15. If some of these clumps come from large galaxies and follow the FJR, they might have a larger velocity dispersion than the upper limit we have imposed for dwarf-like systems (i.e. 120 \kms) but possibly not too large to be detected as “cold substructures” by COSTA.

\begin{acknowledgements}
We thank the referee, Dr. A. Cooper, for the careful review of the paper that allowed us to improve the interpretation and discussion of the paper results. NRN acknowledges financial support from the “One hundred top talent program of Sun Yat-sen University” grant N. 71000-18841229, and from the European Union Horizon 2020 research and innovation programme under the Marie Skodowska-Curie grant agreement n. 721463 to the SUNDIAL ITN network.
CS is supported by an `Hintze Fellow' at the Oxford Centre for Astrophysical Surveys, which is funded through generous support from the Hintze Family Charitable Foundation.  
\end{acknowledgements}


\bibliographystyle{aa}
\bibliography{mybibliography} 



\begin{appendix}

\section{K-S test for red and blue GCs vs. PNe}
\label{sec:red_blue_pne}
\begin{figure}
	\includegraphics[width=9.3cm]{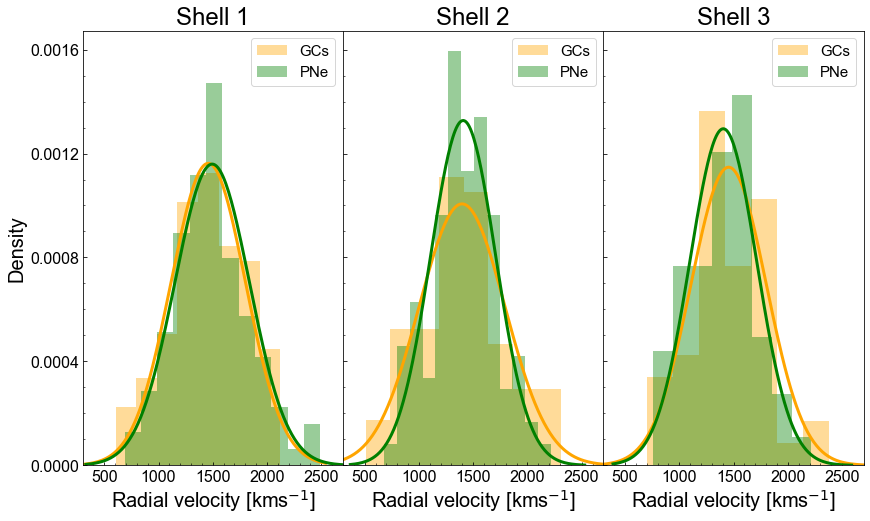}\\
	\includegraphics[width=9.3cm]{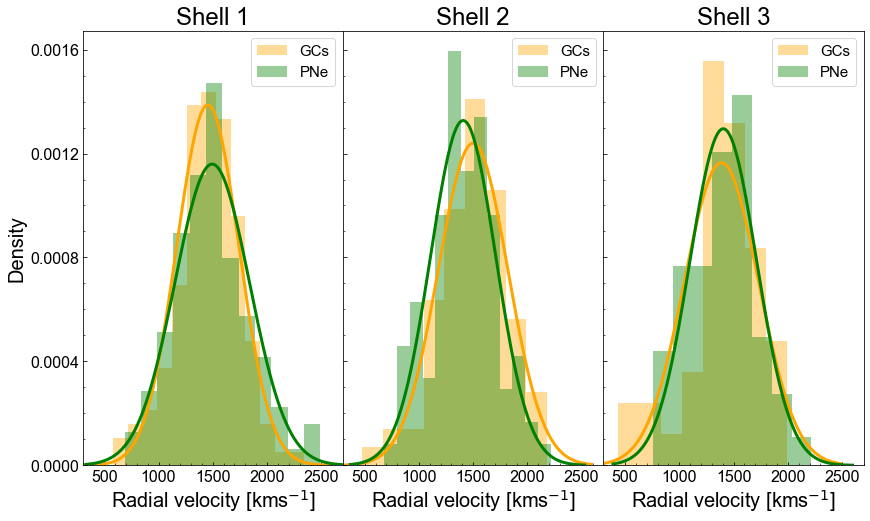}\\
	\includegraphics[width=9.3cm]{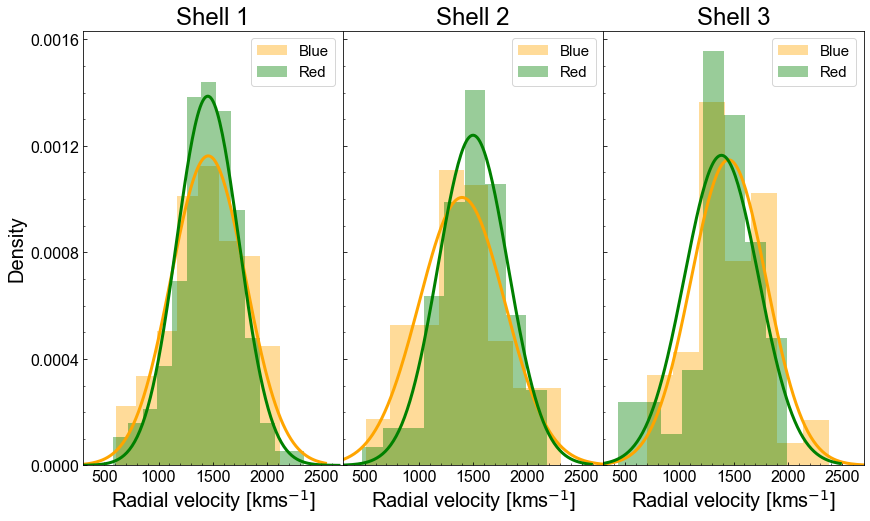}\\
    \caption{Outcomes of the K-S test, in the three shells at different radial distances. \emph{Top:} blue GCs and PNe. \emph{Middle:} red GCs and PNe. \emph{Bottom:} red GCs and PNe.}
	\label{fig:K-S_test_different_pop}
\end{figure}

{In this section we test the kinematical consistency of the PNe with the separated population of red and blue GCs. In \S\ref{sec:gc+pn} we have commented that the PNe and GCs are generally dynamically decoupled, but that PNe and red GCs usually follow closer the properties of the old stellar population of galaxies. Hence the use of the GC population as a whole can produce some bias, if the kinematical properties are significantly different. 
For this reason we have adopted the same color value used by P+18, namely $g-i = 0.85$~mag, to separate blue and red GCs, 
In Fig.~\ref{fig:K-S_test_different_pop} we displayed the K-S test performed on the three populations paired in turn: blue GCs and PNe (top panels), red GCs and PNe (middle panels) and finally with blue and red GCs (bottom panels), for each of the three shells defined in Table~\ref{tab:K-S test}. The results for all cases are reported in Table~\ref{tab:K-S_test_different_pop}. Overall, the p-values are higher than the significance level of 5\%, with the exception of one case, i.e. red GCs and PNe in the Shell 2. However, in all other cases the derived $p-$value are even larger that the ones obtained for the GC as a single population. In particular, we do not find statistical differences of the blue GC population with either the red GC and PNe, which was our primary concern. Hence we conclude that due to the general agreement found in all other shells, we can reasonably assume that the three populations do not show difference across the cluster that can introduce biases. On the contrary, stressing the diversity of tracers could introduce a selection bias in the search for streams, which we wanted to prevent because this is the first time cold streams are tentatively isolated and our understanding of these objects has just started.  
}

\begin{table}
	\centering
	\caption{Parameters and p-values of the Kolmogorov-Smirnov test.}
	\label{tab:K-S_test_different_pop}
	\hspace*{-0.5cm}
	\begin{tabular}{l c c c c c}
		\hline
		Case & PNe & red GCs & blue GCs & Shell & p-value\\
		\hline
		blue GCs - PNe & 208 & - & 94 & 1 & 0.95\\
		blue GCs - PNe & 202 & - & 76 & 2 & 0.60\\
		blue GCs - PNe & 100 & - & 49 & 3 & 0.44\\
		red GCs - PNe & 208 & 139 & - & 1 & 0.52\\
		red GCs - PNe & 202 & 75 & - & 2 & 0.03\\
		red GCs - PNe & 100 & 43 & - & 3 & 0.81\\
		red GCs - blue GCs & - & 139 & 94 & 1 & 0.45\\
		red GCs - blue GCs & - & 75 & 76 & 2 & 0.10\\
		red GCs - blue GCs & - & 43 & 49 & 3 & 0.71\\
		\hline

	\end{tabular}
\end{table}

\section{Stream reliability maps}
\label{sec:reliability}
\begin{figure*}
    \hspace{-0.5cm}
    \includegraphics[scale=0.16]{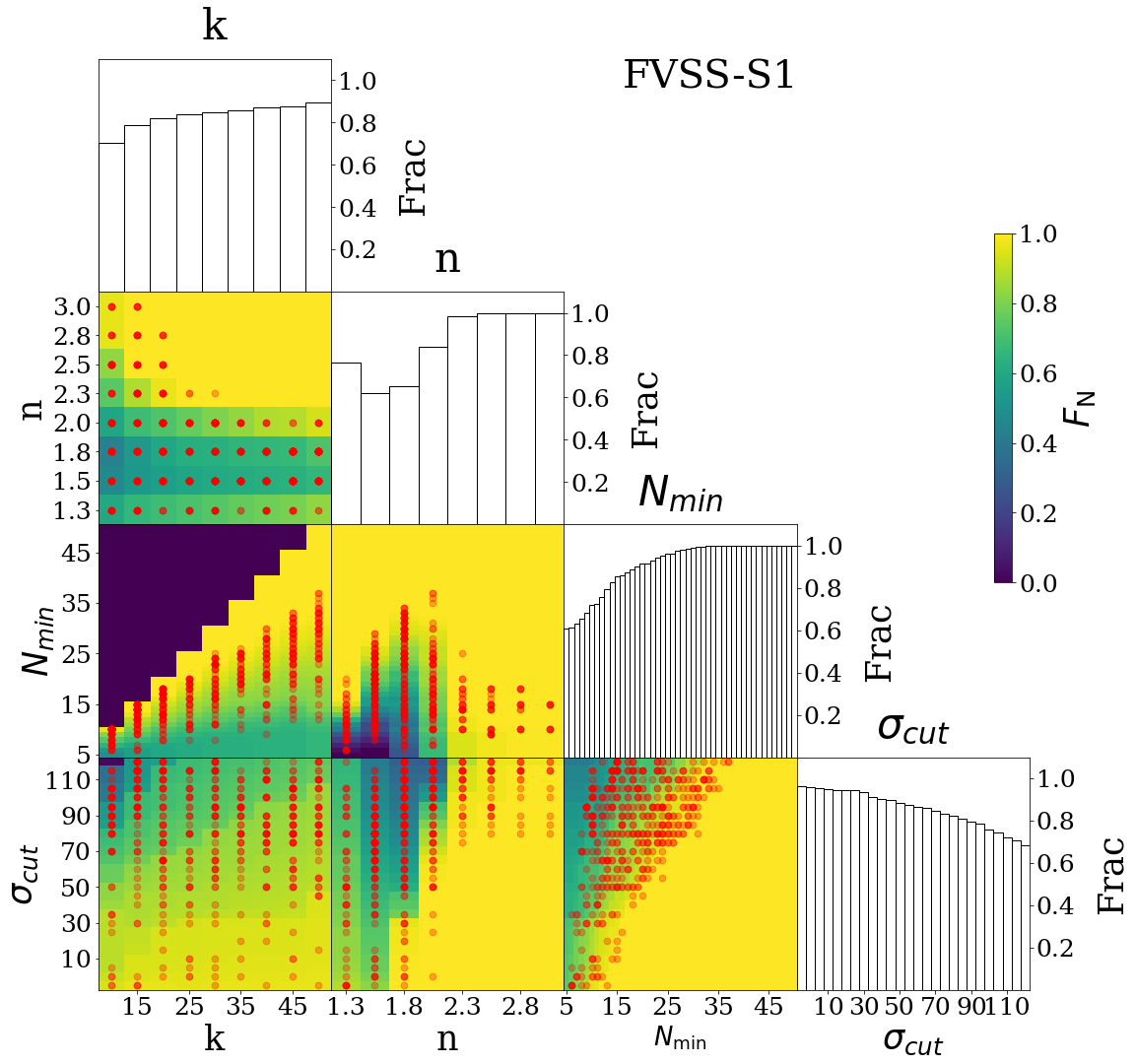}
    \includegraphics[scale=0.16]{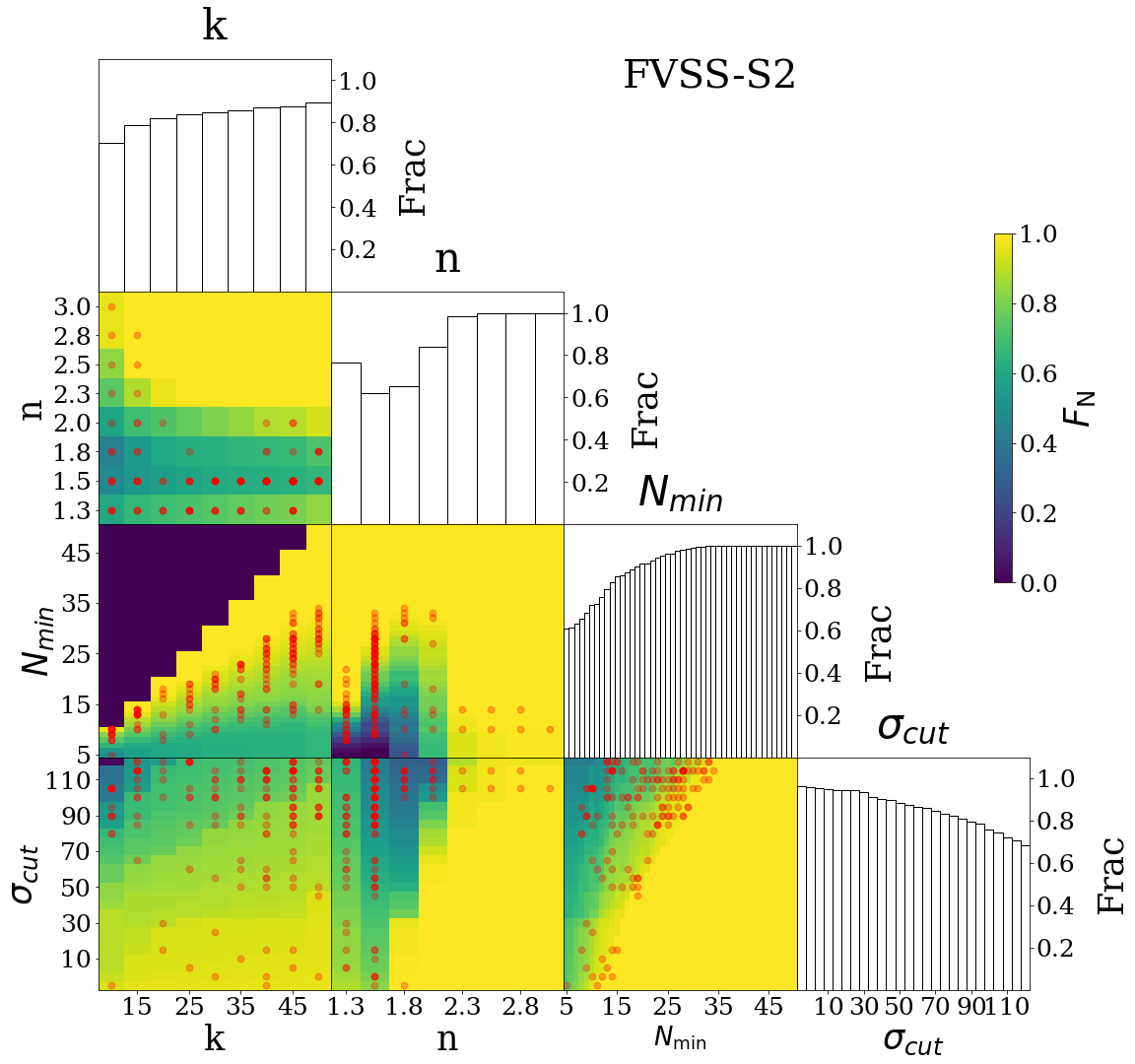}
    \includegraphics[scale=0.16]{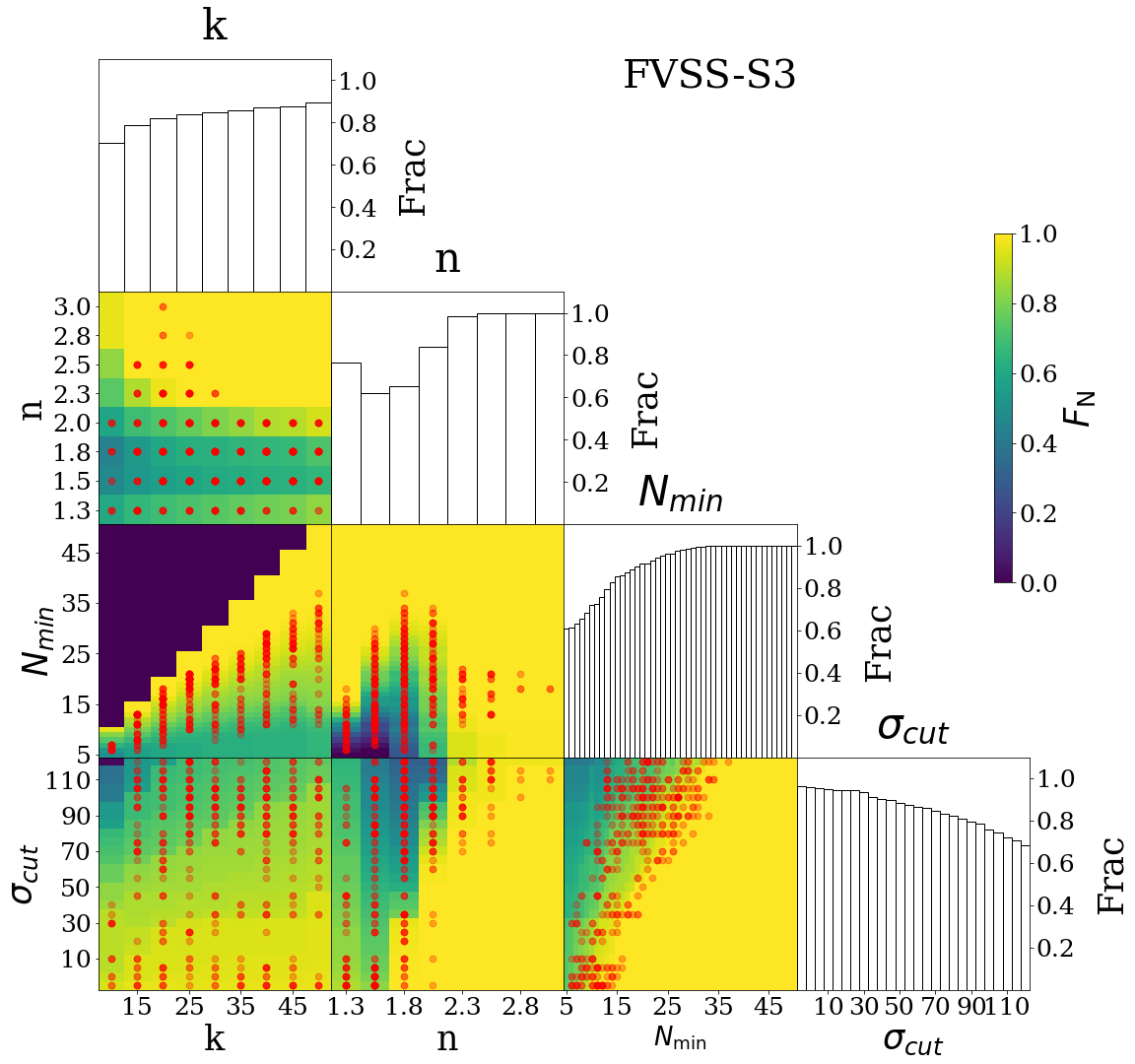}\\
    \hspace*{-0.5cm}
    \includegraphics[scale=0.16]{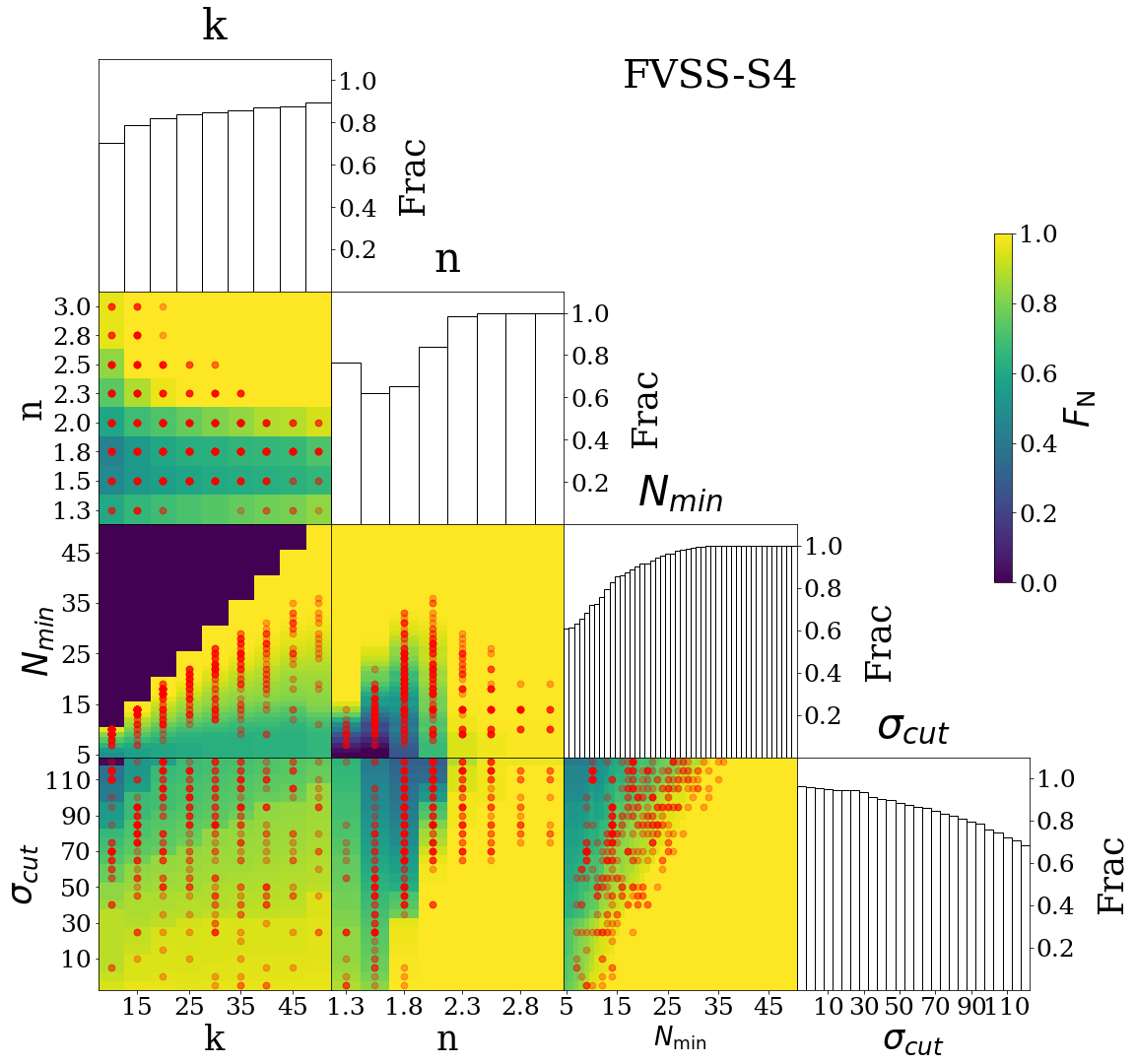}
     \includegraphics[scale=0.16]{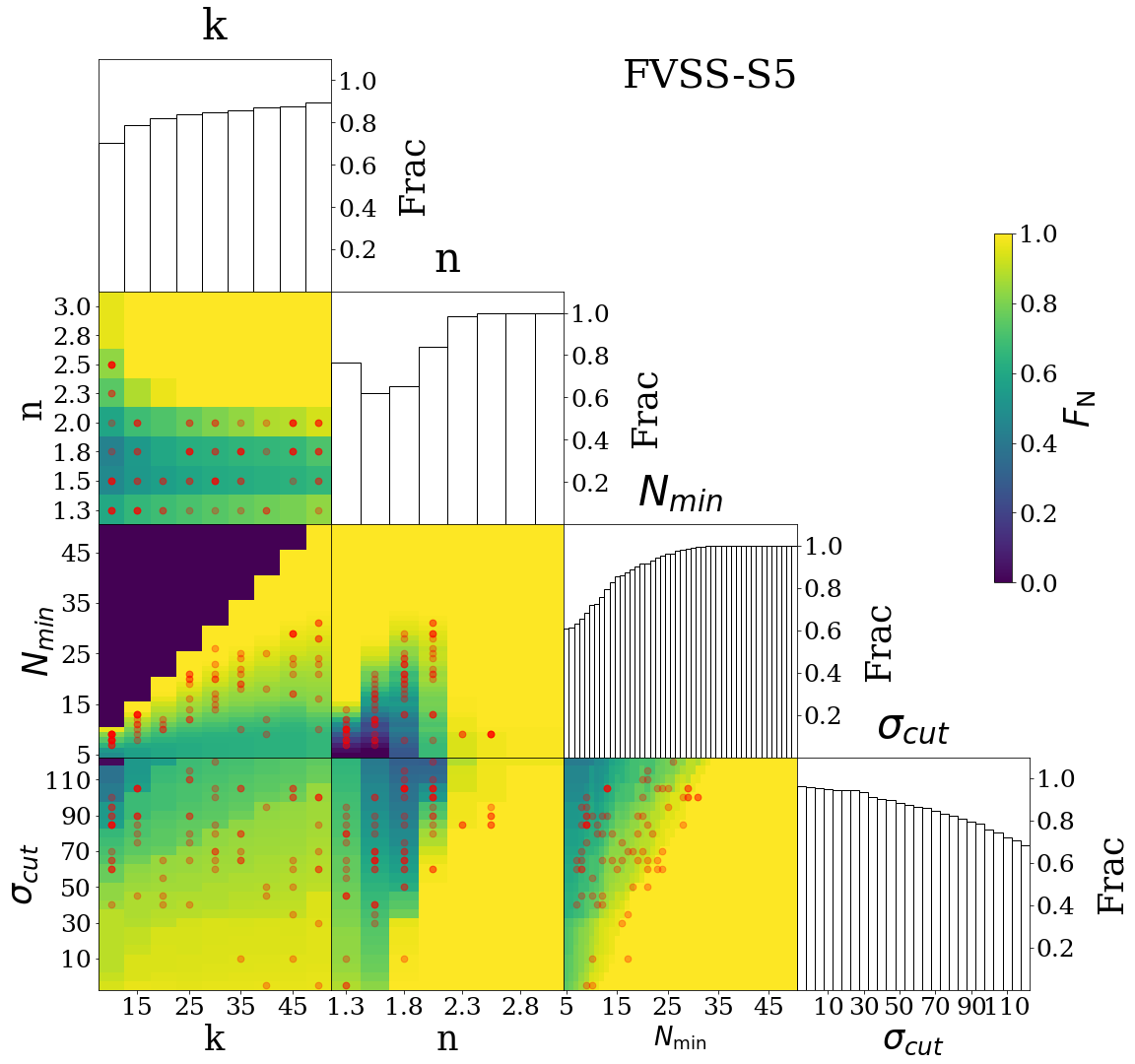}
    \includegraphics[scale=0.16]{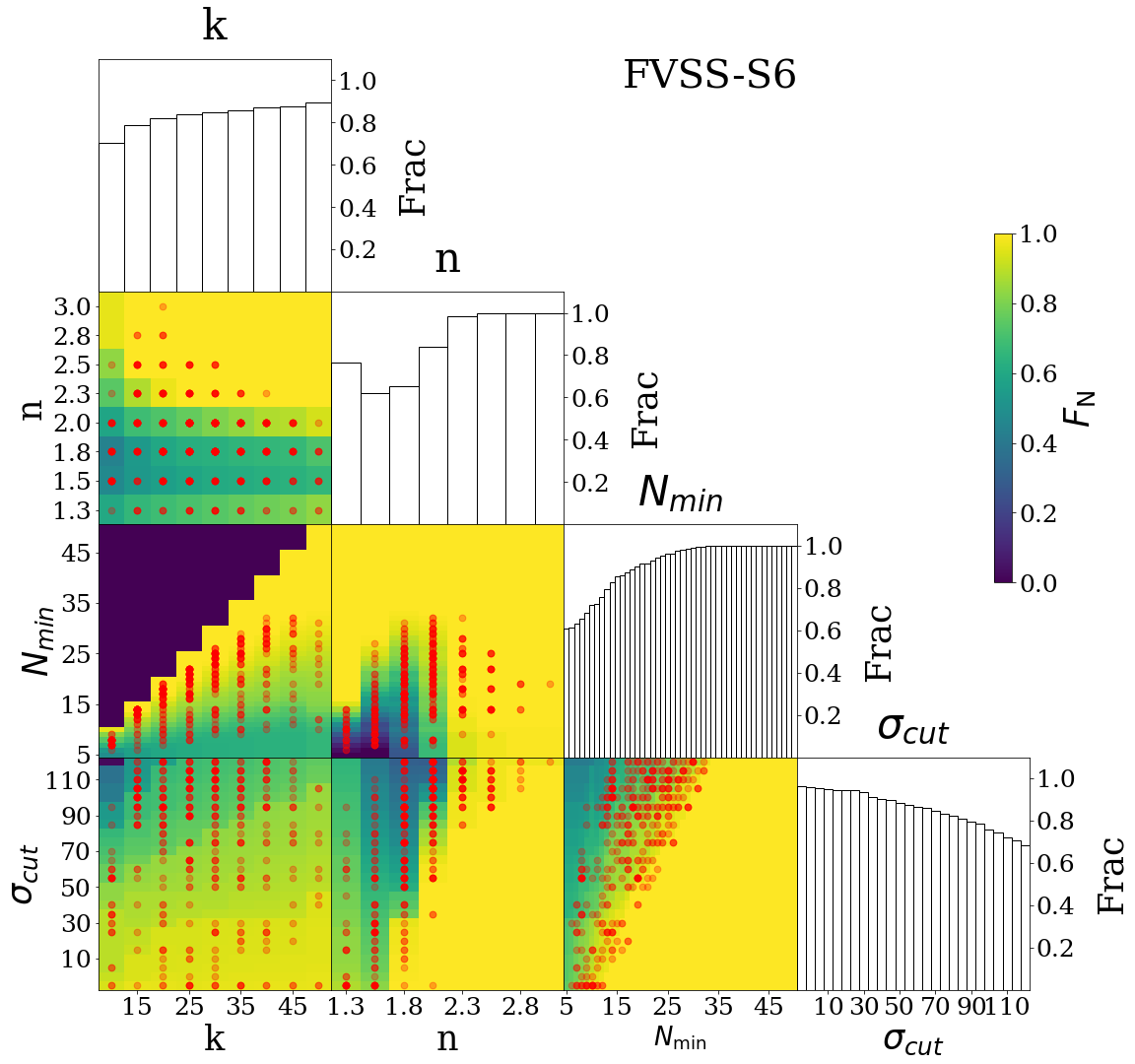}\\
    \hspace*{-0.5cm}
    \includegraphics[scale=0.16]{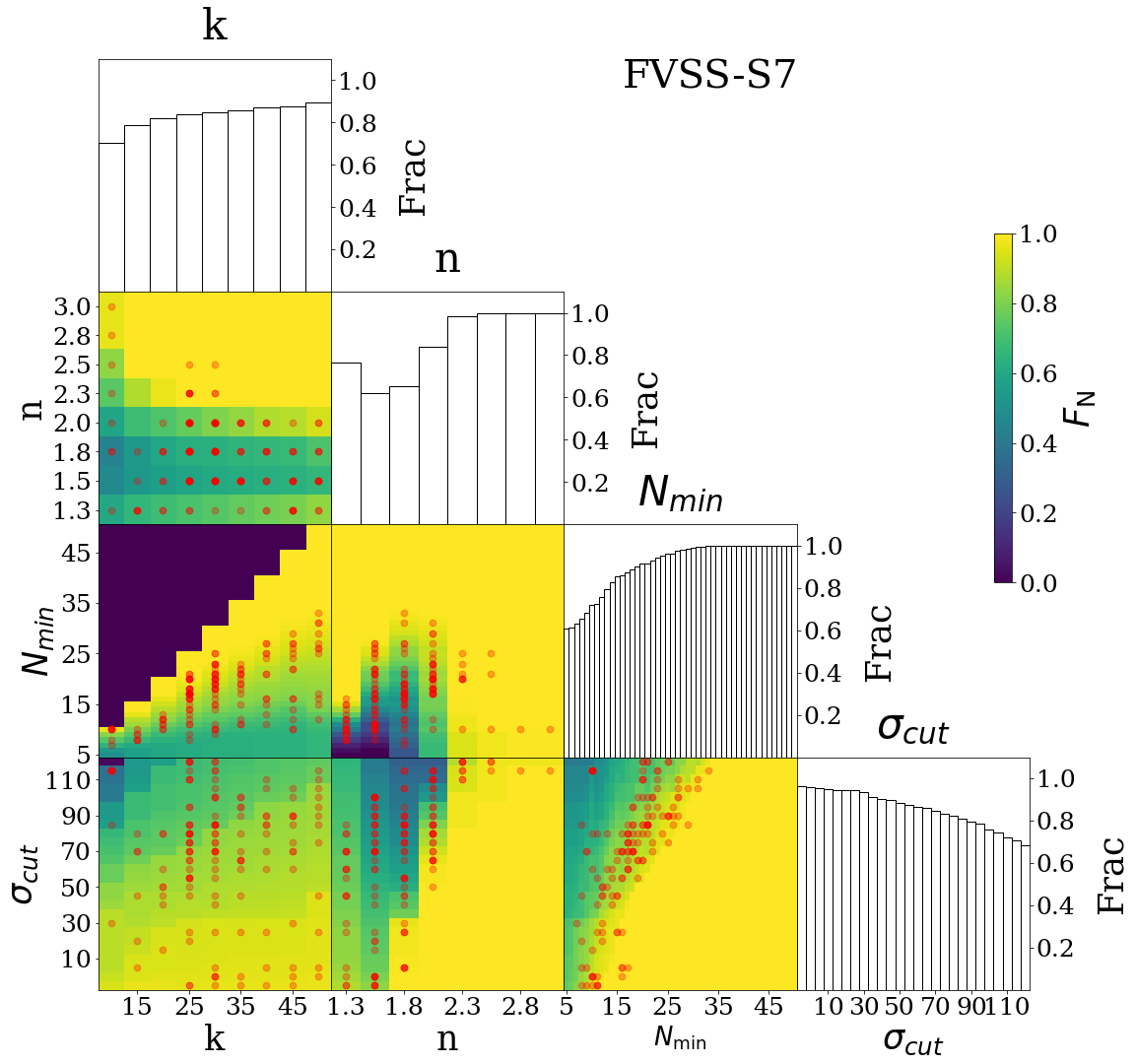}
    \includegraphics[scale=0.16]{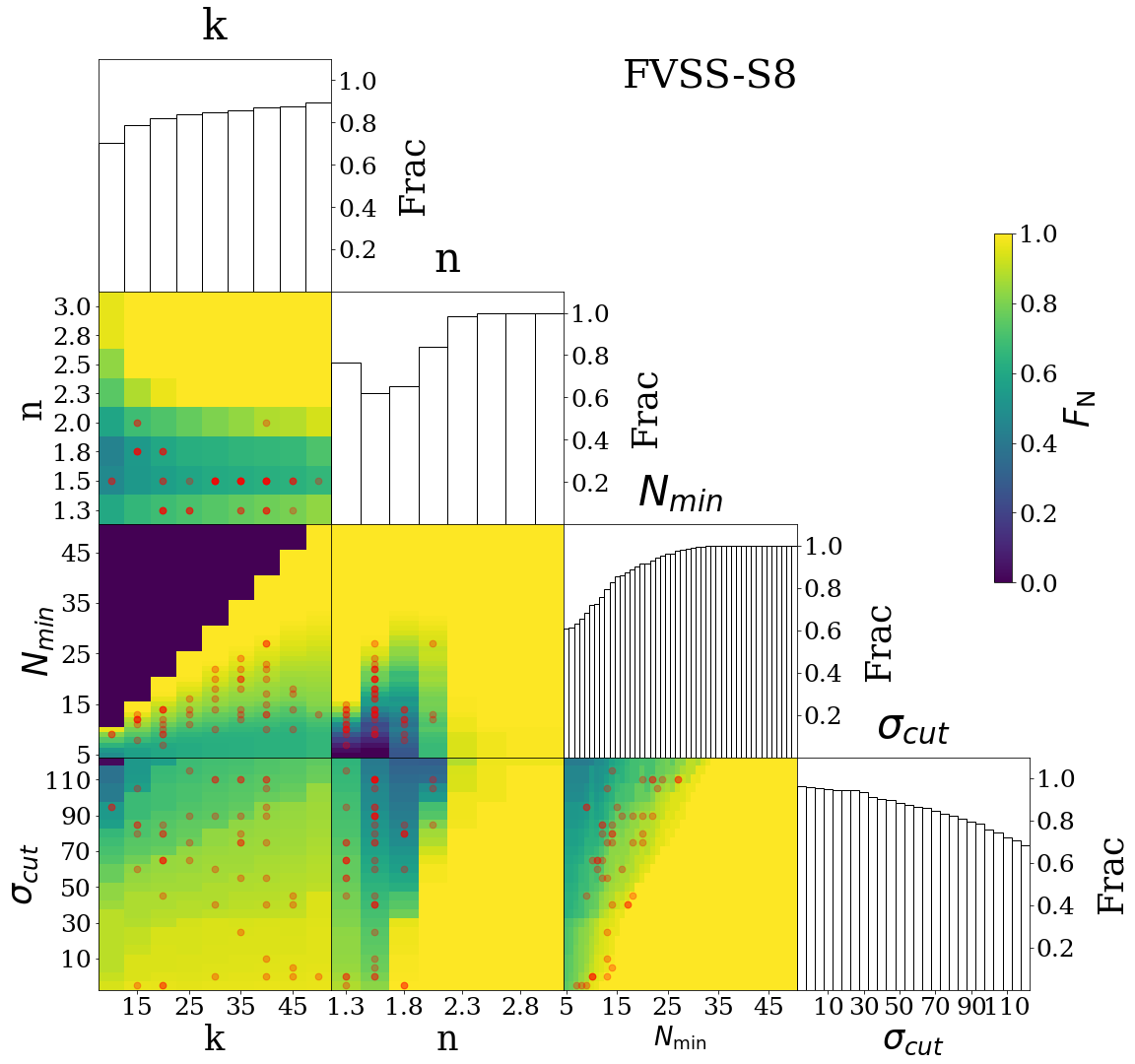}
    \includegraphics[scale=0.16]{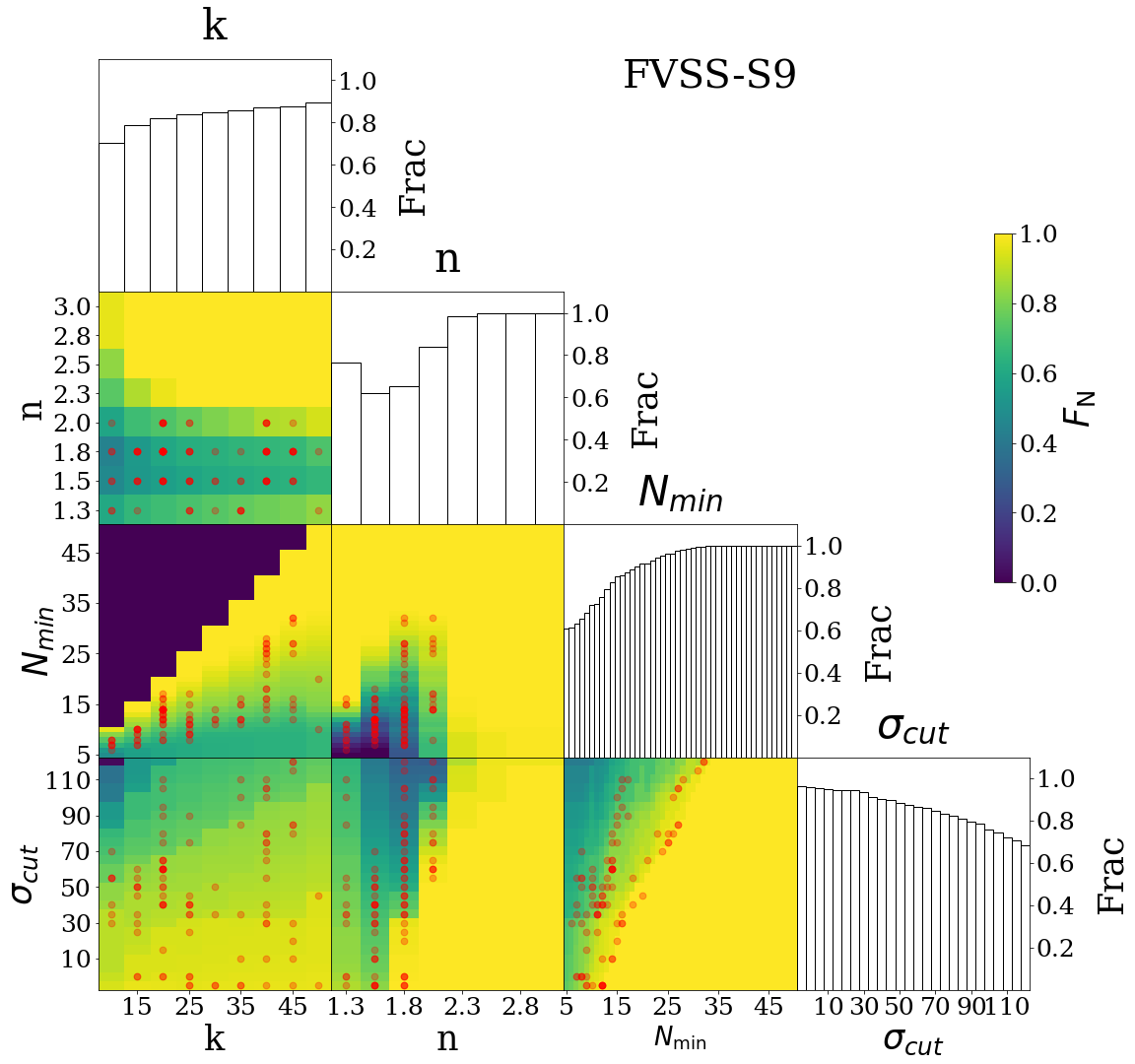}\\
\caption{Reliability map for the Fornax cluster obtained with a reliability threshold of 50\%. For every stream detected by COSTA, we mark as red points, the set-ups where COSTA revealed them.}
    \label{fig:rel_maps_all_streams}
\end{figure*}

In Fig~\ref{fig:rel_maps_all_streams} we displayed, for each of the 13 cold substructures found with COSTA, the reliability maps of the Fornax cluster core obtained by using only the WNS as described in \S\ref{sec:simulation of the cluster}, overlapped with the set-ups in which COSTA detected the streams.
\begin{figure*}
    \hspace{-0.5cm}
    \includegraphics[scale=0.16]{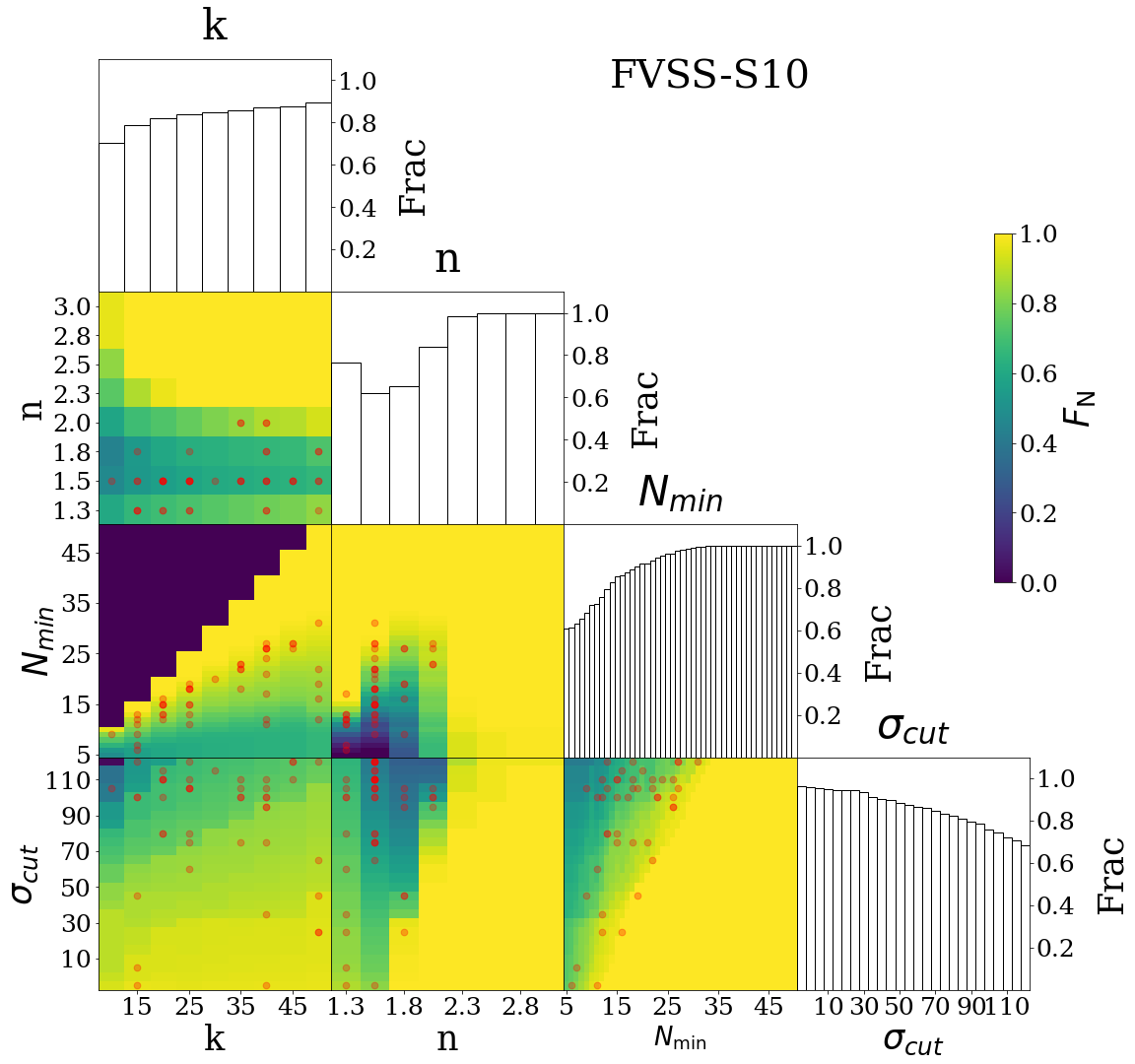}
    \includegraphics[scale=0.16]{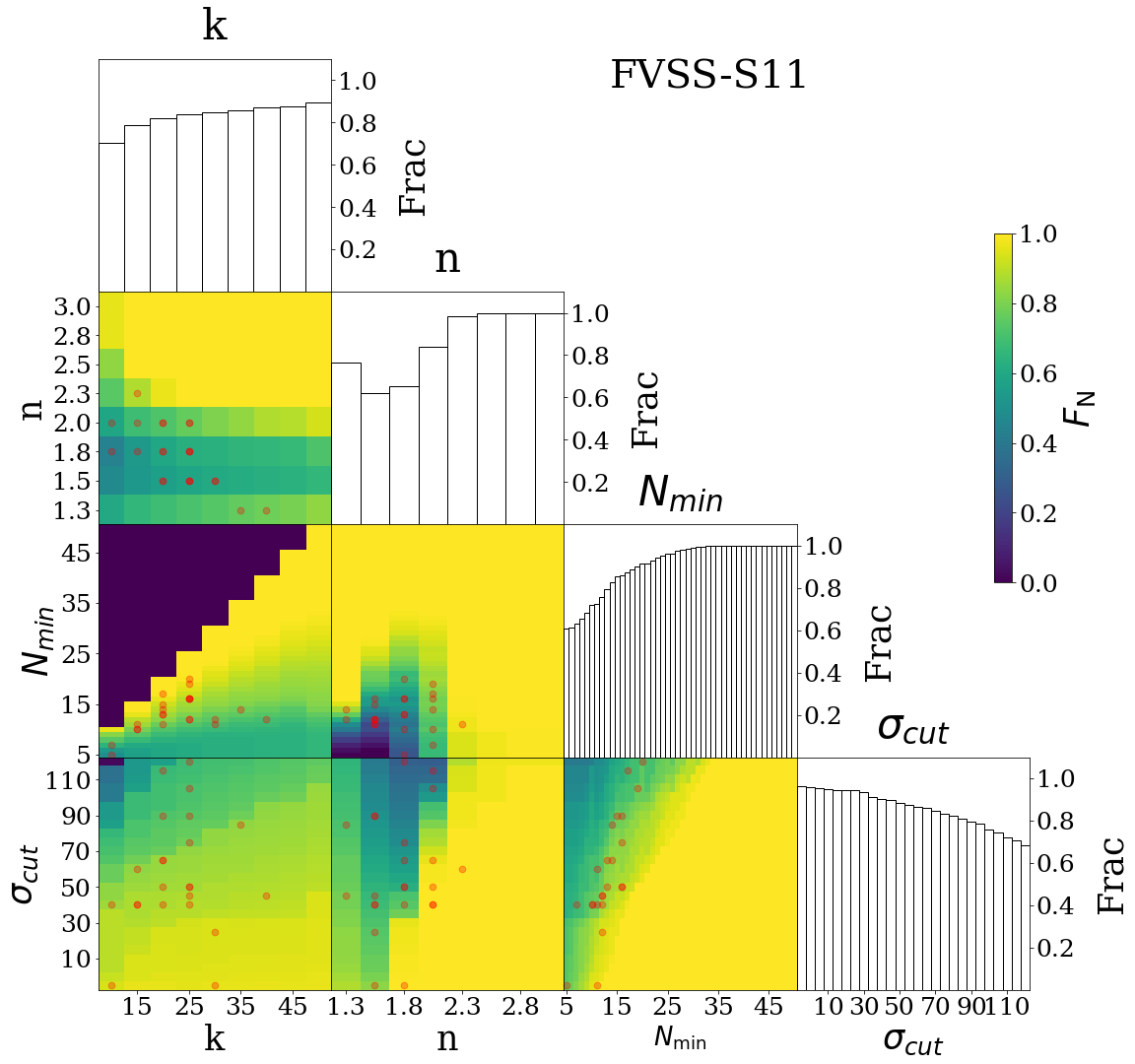}
    \includegraphics[scale=0.16]{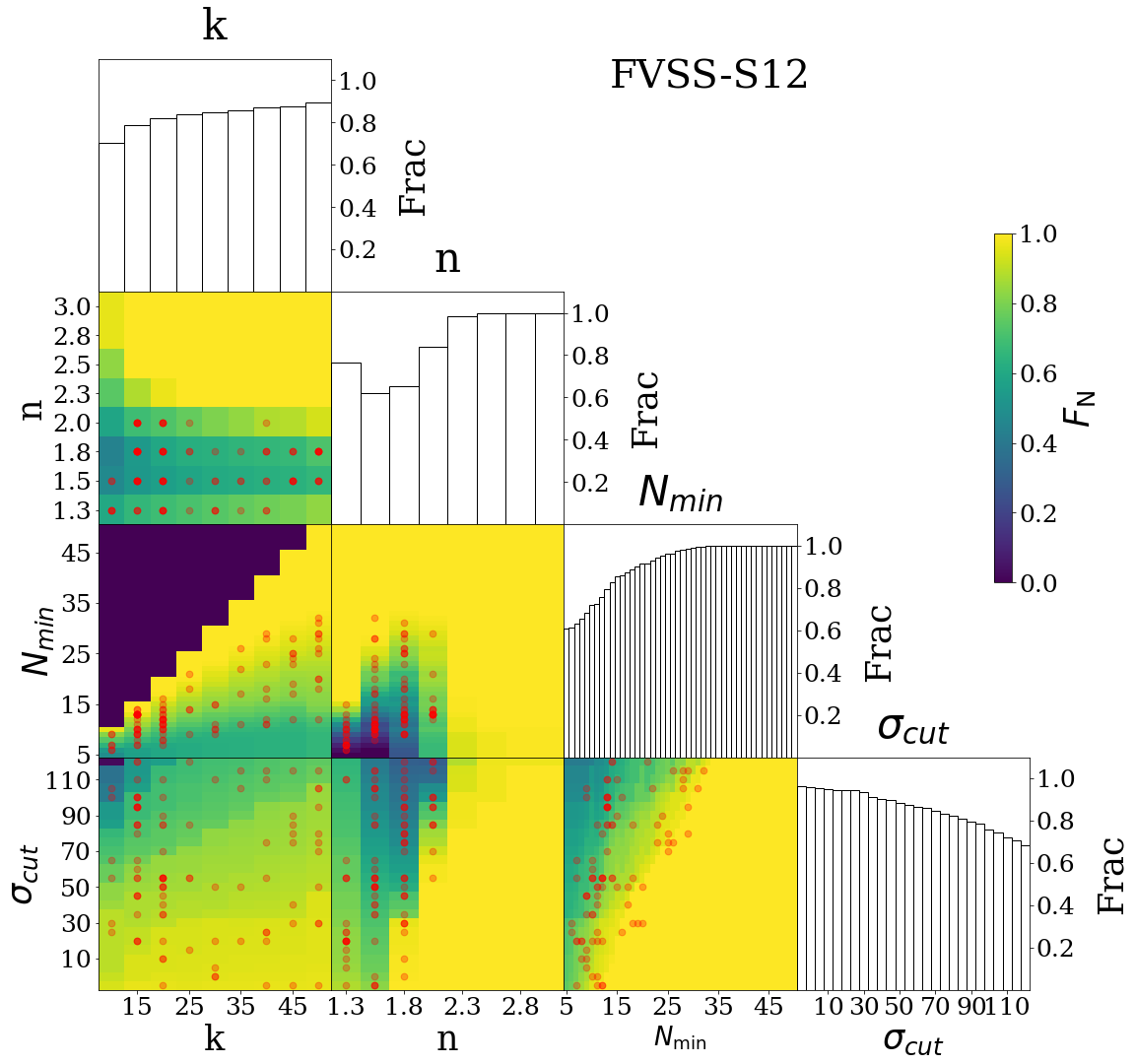}\\
    \hspace*{-0.5cm}
    \includegraphics[scale=0.16]{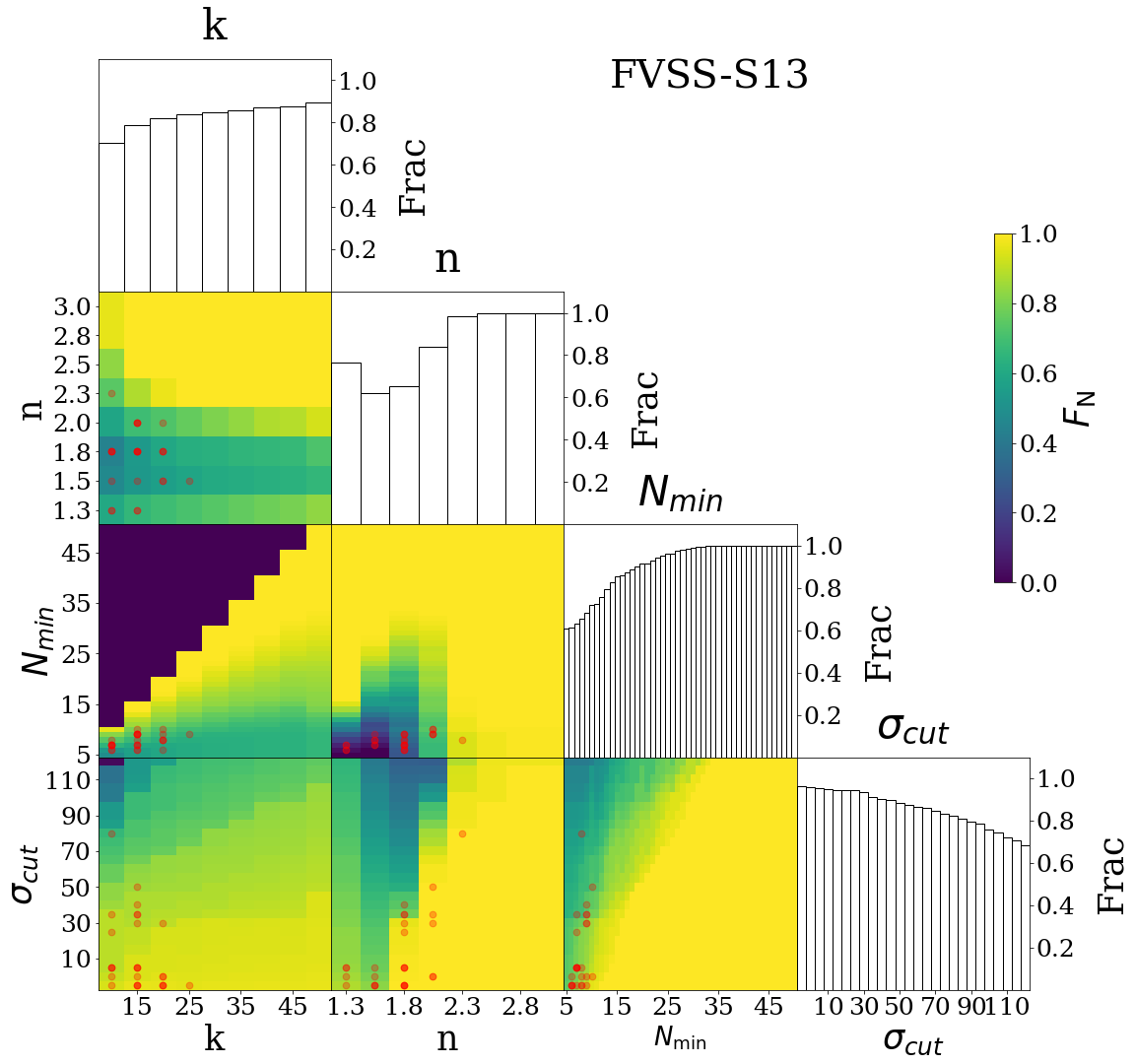}\\
\end{figure*}

\section{Spuriousness of streams candidates}
\label{sec:spuriousness}
 In \S\ref{sec:stream_ICpop} we have mentioned the possibility that some candidate streams can be just a random extraction of ``cold tails'' of a generally hotter population of tracers.
{In order to investigate this latter possibility, we follow a very simple statistical argument and suppose that the stream particles are randomly extracted by a population being in equilibrium with the cluster potential. We can then assume that they have a Normal velocity distribution with mean equal to the velocity of NGC~1399 and standard deviation $\sigma\sim300$ kms$^{-1}$. Then the mean of these groups of particles should have a standard deviation equal to $\sigma/\sqrt{N}$ where $N$ is the number of streams, i.e. typically 300$/\sqrt{13}=83$ kms$^{-1}$. }

To confirm this, we can use the Montecarlo simulations above, with no added artificial streams{. We select }
13 groups of 20 random particles (i.e. the typical number size of one of the detected streams) in each simulation and compute their mean velocity (the mean group velocities). {We then use these } 
to determine the average velocity and the standard deviation, finding that the mean was compatible with the systemic velocity assumed for the cluster and the standard deviation of the random groups was $80\pm17$ kms$^{-1}$ over the 100 simulations{, }
smaller than the one {we measure }
from the mean velocities of the streams in Table~\ref{tab:statistics_2}.  This suggests that the overall kinematics of the stream candidates is incompatible with random particles extracted from an equilibrium population. 
Then, why do not they share the kinematics of other intracluster objects (i.e. galaxies, PNe and GCs)? 
Do they posses a decoupled dynamics from other equilibrium populations
or are some of the candidate streams spurious? 
If spurious, we have seen that they should contribute to the overall velocity dispersion of the ``real stream''
with a low velocity dispersion (of the order of $\sim80$ kms$^{-1}$, as just discussed) which would reduce the intrinsic velocity dispersion of the stream population. 

{Under the assumption that streams have been lately stripped by the parent galaxy,  it is fair to believe that they should still record the kinematics of the parent galaxy in the cluster potential. If this is the case, some pseudo-equilibrium should still hold and the mean velocities of real streams should follow the dynamics of the cluster as their parent galaxies do.
This, in turns, allows us to estimate what fraction of spurious streams would produce an observed factor $\sim213/300$ between the scatter of the stream radial velocities and the typical cluster velocity dispersion value in the area, by simply assuming that all spurious streams have contributed with a $\sigma_{\rm spur}\sim83$ kms$^{-1}$ (see above).  
}
Hence we can write $N_{\rm Eq}\sigma_{\rm Eq}^2=N_{\rm obs}\sigma_{\rm obs}^2+N_{\rm spur}\sigma_{\rm spur}^2$, where $N_{\rm obs}$ and $\sigma_{\rm obs}$ ($=213$ kms$^{-1}$) are the number of stream candidates and the measured dispersion for them respectively, $N_{\rm spur}$ and $\sigma_{\rm spur}$ are the possible spurious streams and their individual velocity dispersion (see above) and $N_{\rm Eq}$ and $\sigma_{\rm Eq}$ ($=300$ kms$^{-1}$) are the true number of stream and the equilibrium velocity dispersion. Being, by definition, $N_{\rm Eq}=N_{\rm obs}-N_{\rm spur}$, with a little of algebra, we obtain $N_{\rm spur}=N_{\rm obs}(\sigma_{\rm Eq}^2-\sigma_{\rm obs}^2)/(\sigma_{\rm Eq}^2+\sigma_{\rm spur}^2)$, which returns $N_{\rm spur}\sim6$.
Thus, in case streams are a population of dynamical components which is in equilibrium with the cluster potential, we expect that our catalog might contain up to 6 spurious detections. This simple estimate does not take into account the density slope of the streams and the anisotropy of their orbital distribution, which are 
fully unknown and can both affect the expected velocity dispersion of them as an equilibrium population. On the other hand, it is likely that the streams are not a canonical equilibrium population and the $N_{\rm spur}\sim6$ has to be taken as an upper limit. Indeed, as we will see in \S\ref{sec:correlations}, the stream population shows signature of dynamical friction, which suggests the presence of ongoing dissipative processes. This would reasonably drive loss of kinetic energy and consequently cause a 
reduction of the velocity dispersion of the streams as an intracluster population {making }
the actual $N_{\rm spur}$ smaller than {that } estimated above. 
As a bottom line of this digression, we can conclude that, even {in the worst case } 
($N_{\rm spur}\sim6$), at least half of the stream candidates (i.e. 7 over 13) is real.  

\section{On the connection od FVSS-S2 and FVSS-S12}
\label{sec:s2-s12}
In \S\ref{sec:stream_descript} we have mentioned that FVSS-S2 and FVSS-S12 are, among all streams in Table \ref{tab:statistics_2}, the ones having the most similar radial velocity also being close in position.
This might indicate a link between the two substructures that has been either missed or considered not significant by COSTA. This latter possibility cannot be excluded if the size of a stream is particularly large (as discussed in Sect. 3.4.2 of G+20). We have {\it a posteriori} checked the presence of overlapping particles among the different parameter combinations, e.g. the ones shown in Fig. \ref{fig:sigma_nmin} and Fig. \ref{fig:rel_maps_all_streams}. Indeed, we have found a few particles in common for some parameter set-ups which favor the detection of large streams (large $k$, $N_{\rm min}$ and $\sigma_{\rm cut}$). However, in the majority of the parameter combinations, COSTA does detect them as two separate structures with a velocity dispersion values that are only marginally compatible (within 2$\sigma$, see Table \ref{tab:statistics_2}, but with a very large error on the FVSS-S12 velocity dispersion value). 
Looking at the $\sigma-N_{\rm min}$ diagrams in Fig. \ref{fig:sigma_nmin}, we can see that the parameter spaces of the two streams, barely overlap, hence demonstrating that the numbers of configurations that return common particle is rather small. On the other hand, if joined together, the parameter spaces of the two streams would cover a very large volume, much wider than a medium sized cold substructure (see e.g. G20 Fig. 12). 

\begin{figure}
	\includegraphics[width=9.3cm]{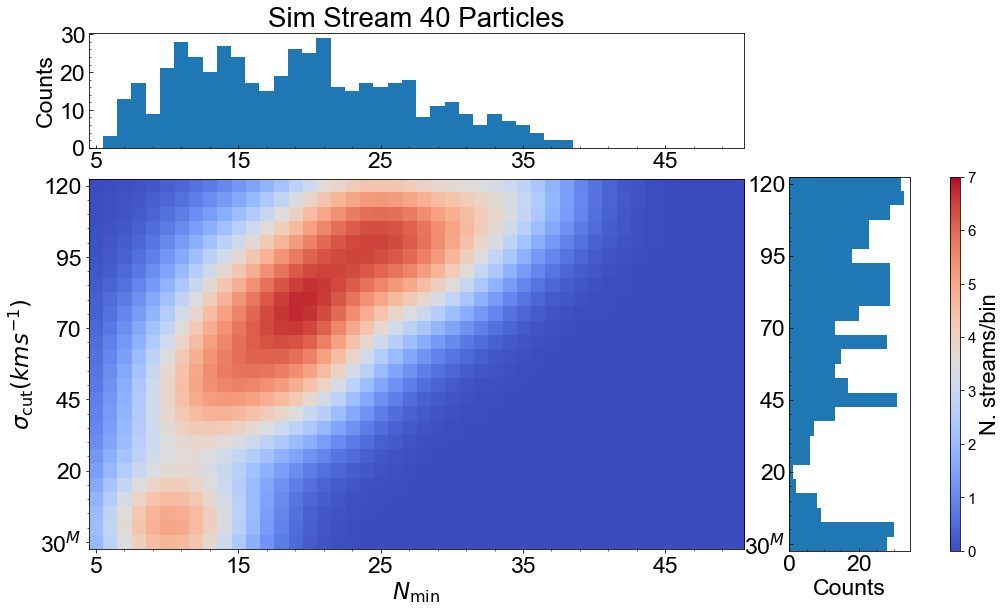}
    \caption{Density plot of the $\sigma_{\rm cut}$ as a function of $N_{\rm min}$ for a simulated stream of stream 40 particles, $10'\times 5'$ size, with an intrinsic velocity dispersion of 90 \kms, mimicking a large stream as expected if FVSS-S2 and FVSS-S12 are part of a single substructure. Color code and smoothing are defined as the real streams in Fig. \ref{fig:sigma_nmin}.}
	\label{fig:sigma_nmin_large}
\end{figure}

We have tried to mimic the presence of a much larger structure, by adding a large artificial stream ($10'\times 5'$) made of 40 particles, with an intrinsic velocity dispersion of 90 \kms, in the combined GC+PN sample of the Fornax core, and then applying COSTA to recover it (following the procedure adopted for real streams described in \S\ref{sec:runningCOSTA}). This simulated stream should reasonably reproduce the properties of the FVSS-S2 and S12 combined together, assuming that FVSS-S2 is a fragment of a bigger stream with higher velocity dispersion, which upper limit is given by the $\sigma$ estimated for FVSS-S12 (see Table \ref{tab:statistics_2}). In particular, we have derived the $\sigma-N_{\rm min}$ diagram, which is reported in Fig. \ref{fig:sigma_nmin_large}. This shows two main features. 
First, the parameter space covered by the allowed configurations is distributed along a banana shaped region revealing a parameter degeneracy. This can be a consequence of an increasing fraction of contaminants in larger $N_{\rm min}$, producing larger velocity dispersion configurations. 
Second, the distribution of parameters is multi-variate with three peaks at $(\sigma;~N_{\rm min})\simeq(10,~10),~(20,~80),~(25,~110)$. 
These might correspond to some subgroups that, if not continuously connected, could be selected as independent units. 

This test shows that it is possible that FVSS-S2 and -S12 can be part of a larger stream, whose main body is made by the latter. This would possibly support the origin suggested for FVSS-S12 in \S\ref{sec:stream_descript}, as a stream produced by the disruption of a larger galaxy via violent-relaxation, being similar to the clumpy remnants found in C15. 

\label{lastpage}
\end{appendix}
\end{document}